\documentclass[aps,prd,singlecolumn,superscriptaddress,floatfix,nofootinbib]{revtex4-2}

\usepackage{graphicx,natbib,amssymb,amsmath,stmaryrd,makecell}
\usepackage[utf8]{inputenc}
\UseRawInputEncoding
\usepackage{amsmath}
\usepackage{bm}
\usepackage{xcolor}
\usepackage{natbib}
\usepackage[title]{appendix}
\usepackage{graphicx}
\usepackage{float}
\usepackage[normalem]{ulem}
\usepackage{aas_macros}

\usepackage{scalerel}
\usepackage{tikz}
\usetikzlibrary{svg.path}
\definecolor{orcidlogocol}{HTML}{A6CE39}
\tikzset{
  orcidlogo/.pic={
    \fill[orcidlogocol] svg{M256,128c0,70.7-57.3,128-128,128C57.3,256,0,198.7,0,128C0,57.3,57.3,0,128,0C198.7,0,256,57.3,256,128z};
    \fill[white] svg{M86.3,186.2H70.9V79.1h15.4v48.4V186.2z}
                 svg{M108.9,79.1h41.6c39.6,0,57,28.3,57,53.6c0,27.5-21.5,53.6-56.8,53.6h-41.8V79.1z M124.3,172.4h24.5c34.9,0,42.9-26.5,42.9-39.7c0-21.5-13.7-39.7-43.7-39.7h-23.7V172.4z}
                 svg{M88.7,56.8c0,5.5-4.5,10.1-10.1,10.1c-5.6,0-10.1-4.6-10.1-10.1c0-5.6,4.5-10.1,10.1-10.1C84.2,46.7,88.7,51.3,88.7,56.8z};
  }
}

\newcommand\orcidlink[1]{\href{https://orcid.org/#1}{\mbox{\scalerel*{
\begin{tikzpicture}[yscale=-1,transform shape]
\pic{orcidlogo};
\end{tikzpicture}
}{|}}}}

\newcommand\orcidicon[1]{\href{https://orcid.org/#1}{\usebox{\ORCIDlogo}}}

\usepackage{hyperref}
\hypersetup{colorlinks}
\defcitealias{Kusiak_2022}{K22}

\begin{document}
\title{Enhancing Measurements of the CMB Blackbody Temperature Power Spectrum by Removing CIB and Thermal Sunyaev-Zel'dovich Contamination Using External Galaxy Catalogs}

\author{Aleksandra Kusiak\,\orcidlink{0000-0002-1048-7970}}
\thanks{{\bf these authors contributed equally to this work}}
\email{akk2175@columbia.edu}
\affiliation{Department of Physics, Columbia University, New York, NY 10027, USA}

\author{Kristen M.~Surrao\,\orcidlink{0000-0002-7611-6179}}
\thanks{{\bf these authors contributed equally to this work}}
\email{k.surrao@columbia.edu}
\affiliation{Department of Physics, Columbia University, New York, NY 10027, USA}

\author{J.~Colin Hill\,\orcidlink{0000-0002-9539-0835}}
\affiliation{Department of Physics, Columbia University, New York, NY 10027, USA}

\date{\today}

\begin{abstract}
Extracting the CMB blackbody temperature power spectrum --- which is dominated by the primary CMB signal and the kinematic Sunyaev-Zel'dovich (kSZ) effect --- from mm-wave sky maps requires cleaning other sky components. In this work, we develop new methods to use large-scale structure (LSS) tracers to remove cosmic infrared background (CIB) and thermal Sunyaev-Zel'dovich (tSZ) contamination in such measurements. Our methods rely on the fact that LSS tracers are correlated with the CIB and tSZ signals, but their two-point correlations with the CMB and kSZ signals vanish on small scales, thus leaving the CMB blackbody power spectrum unbiased after cleaning. We develop methods analogous to delensing (\textit{de-CIB} or \textit{de-(CIB+tSZ)}) to clean CIB and tSZ contaminants using these tracers. We compare these methods to internal linear combination (ILC) methods, including novel approaches that incorporate the tracer maps in the ILC procedure itself, without requiring exact assumptions about the CIB SED. As a concrete example, we use the \emph{unWISE} galaxy samples as tracers. We provide calculations for a combined Simons Observatory and \emph{Planck}-like experiment, with our simulated sky model comprising eight frequencies from 93 to 353 GHz. Using \emph{unWISE} tracers, improvements with our methods over current approaches are already non-negligible: we find improvements up to 20\% in the kSZ power spectrum signal-to-noise ratio (SNR) when applying the de-CIB method to a tSZ-deprojected ILC map. These gains could be more significant when using additional LSS tracers from current surveys, and will become even larger with future LSS surveys, with improvements in the kSZ power spectrum SNR up to 50\%. For the total CMB blackbody power spectrum, these improvements stand at 4\% and 7\%, respectively.  Our code is publicly available in \verb|deCIBing|.\footnote{\url{https://github.com/olakusiak/deCIBing}}
\end{abstract}

\maketitle

\section{Introduction}
\label{sec:intro}

The cosmic microwave background (CMB) blackbody temperature power spectrum is dominated by the primary CMB signal on large and moderate angular scales and the kinematic Sunyaev-Zel'dovich (kSZ) effect on small angular scales. The kSZ effect --- the Compton scattering of CMB photons off electrons moving with non-zero line-of-sight velocity --- is a unique cosmological and astrophysical probe. It provides information on the distribution of electrons in galaxies, groups, and clusters, as well as probes the cosmological velocity field \cite{SZ_1972, SZ_1980, Ostriker1986}.

However, detecting the power spectrum of the kSZ effect with CMB data has been extremely challenging, as it preserves the blackbody spectral energy distribution (SED) of the primary CMB. Furthermore, measuring the kSZ auto-power spectrum requires cleaning other sky components to high precision, particularly the cosmic infrared background (CIB) and thermal Sunyaev-Zel'dovich (tSZ) effect, as well as other foregrounds. Fortunately, the tSZ effect can be robustly removed using the constrained internal linear combination (ILC) method due to its unique spectral dependence~\cite{Eriksen2004, Remazeilles_2010}, albeit at the cost of increased noise in the final map due to the tSZ deprojection.  The CIB, however, is more challenging, and its deprojection in a constrained ILC relies on the use of an analytic model for the effective CIB SED (\textit{e.g.},~\cite{SO2019, Madhavacheril_2020}), which is unlikely to hold at high precision and also neglects decorrelation of the CIB across frequencies. Nevertheless, evidence ($\approx 3\sigma$) of the kSZ power spectrum has recently been found in data from the South Pole Telescope~\cite{Reichardt_2021}.  In addition, the kSZ effect has been detected in combination with large-scale structure (LSS) data using various methods, including the mean pairwise momentum method \cite{Ferreira_99, Hand_2012, Calafut_2021}, velocity-weighted stacking \cite{Ho2009, Shao_2011, Schaan_2016, schaan2020act}, or the projected-fields method \cite{Dore2004, DeDeo, Hill2016, Ferraro2016, Kusiak_2021}. Furthermore, another method, the large-scale velocity reconstruction \cite{Deutsch_2018, Munchmeyer:2018eey, giri_20} has been recently proposed. We refer the reader to Ref.~\cite{Smith_2018} for a detailed review of the kSZ-LSS detection methods, where the authors showed that, apart from the projected-fields estimator, all other methods are equivalently measuring the bispectrum of a form $\langle ggT \rangle$, where $g$ is an LSS tracer and $T$ is the kSZ field. With the upcoming CMB and LSS experiments, the kSZ effect will be measured with increasing signal-to-noise ratio (SNR), highly surpassing the few-$\sigma$ cross-correlations detected so far. 

In this work, we build new methods to clean CIB and tSZ contamination from blackbody CMB temperature maps, thus enhancing detection prospects for both the kSZ and primary CMB signals.  Our methods rely on the fact that LSS tracers are correlated with both the CIB and tSZ signals, but not with the primary CMB (apart from the integrated Sachs-Wolfe effect on very large scales) or with the kSZ signal --- the latter two-point correlation vanishes on small scales due to the equal likelihood of positive and negative line-of-sight velocities. We consider various methods to use these LSS tracers to remove the CIB and tSZ contaminants while leaving the CMB blackbody signal unbiased. As a concrete example, for our tracer maps we use the \emph{unWISE} galaxy samples~\cite{krolewski_2020}, a catalog comprising over 500 million objects on the full sky, spanning redshifts $0 \lesssim z \lesssim 2$.

Our first approach draws motivation from delensing of the CMB.  Delensing is the subtraction of the lensing B-mode component from the CMB, so as to enhance detection prospects for B-modes generated by primordial gravitational waves \cite{Kesden:2002ku, Knox:2002pe}. Tracers such as LSS data and the CIB have been used for this purpose \cite{smith2012delensing, Sherwin:2015baa}, and combinations of multiple tracers can improve the delensing performance \cite{Yu_2017}. Two of our methods of interest, \textit{de-CIBing} and \textit{de-(CIB+tSZ)ing}, are directly analogous to delensing.  These methods use the LSS tracers to clean the CIB and tSZ contaminants without requiring exact assumptions about the CIB SED.

Our second approach involves extensions of the widely used internal linear combination (ILC) technique~\cite{Eriksen2004, Remazeilles_2010}.  We consider ILC calculations with LSS tracer maps included as additional input maps in the ILC (along with the CMB and other mm-wave frequency maps), and we also consider the use of additional constraints involving the tracer maps in the ILC, in which we require the final ILC map to have zero cross-correlation with the tracers.  We compare this approach to standard ILC results (including constrained ILC with tSZ and/or CIB deprojection), as well as to the de-CIBing or de-(CIB+tSZ)ing methods described above. All of these methods can be applied directly to data without necessitating specific theoretical models for the  correlations between sky components and tracers (with a small exception when trying to optimally combine several tracers, further discussed in \S \ref{sec:deCIB}). We model the theoretical correlations in this paper primarily to provide forecasts of the methods without the use of full numerical simulations.

We model the microwave sky as consisting of the primary lensed CMB, kSZ, tSZ, CIB, and radio source signals, as well as detector and atmospheric noise for a combined Simons Observatory (SO) and \emph{Planck}-like experiment.  In total we consider auto- and cross-spectra at eight frequencies ranging from 93 to 353 GHz. In calculating the sky component power spectra, we assume the flat $\Lambda$CDM \emph{Planck} 2018 cosmology \cite{Planck2018}: $\omega_{cdm}= 0.11933$, $\omega_b=0.02242$, $H_0 = 67.66$ km/s/Mpc, $\ln(10^{10}A_s)=3.047$, and $n_s=0.9665$ with $k_\mathrm{pivot}=0.05\,\mathrm{Mpc}^{-1}$, and $\tau_\mathrm{reio}=0.0561$ (best-fit parameter values from the last column of Table~II of Ref.~\cite{Planck2018}). We use a halo model approach to compute the tSZ, CIB, and \emph{unWISE} galaxy correlations.  Throughout this work, we also assume the standard Tinker \textit{et al.}~(2008) halo mass function \cite{Tinker_2008}, Navarro-Frenk-White (NFW) halo density profiles \cite{Navarro_1997}, and the concentration-mass relation defined in Ref.~\cite{Bhattacharya_2013}. We adopt the $M_{200c}$ halo mass definition for all calculations. All masses are in units of  $M_\odot/h$, unless stated otherwise, and all error bars, unless stated otherwise, are 1$\sigma$. 

We compare forecasts of the different methods for CIB and tSZ contaminant removal using various evaluation metrics. One of our primary results is the calculation of auto-spectra of the cleaned maps resulting from each of our methods; this result is shown in Fig.~\ref{fig:Tclean_auto_spectra}. We also compute correlation coefficients of the cleaned maps with the CIB and tSZ signals to assess residual contamination, shown in Fig.~\ref{fig:TcleanxCIB_corr} and \ref{fig:TcleanxtSZ}, respectively. Our novel CIB and tSZ signal removal methods yield non-negligible improvements toward detecting the kSZ auto-power spectrum and total CMB blackbody temperature power spectrum using SO and \emph{Planck}.  With our de-CIB method using \emph{unWISE} galaxies, we project improvements in the SNR of the kSZ power spectrum of up to 20\%, depending on the details of the CIB and \emph{unWISE} modeling. For future LSS surveys, such as \emph{Euclid} \cite{Euclid2011, Euclid:2021} or \emph{Roman} \cite{Spergel2013wfirst, Akeson2019, Yung:2022}, we expect further improvements up to 50\%, as those surveys will probe 2-3 times more galaxies over a similar redshift range. For the total CMB blackbody temperature power spectrum, those numbers stand at approximately 4\% and 7\%, respectively. Note that we can also improve current forecasts by adding more external LSS tracers that correlate with both the CIB and tSZ fields, \textit{e.g.}, the Dark Energy Survey (DES) \cite{Rodr_guez_Monroy_2022}, 2MASS \cite{2mass_06}, or BOSS/eBOSS \cite{BOSSgal_2015, eBOSS_dr14_2017} galaxy catalogs, or even galaxy lensing or CMB lensing data.

The remainder of this paper is organized as follows. In \S \ref{sec.ILC} we review standard and constrained harmonic ILC foreground-removal methods.  In \S \ref{sec:deCIB} through \S \ref{sec:method1} we introduce our new methods for removing CIB and tSZ contamination using LSS tracers: de-CIB and de-(CIB+tSZ) in \S \ref{sec:deCIB}, an ILC with tracer maps as additional input ``frequency" maps in \S \ref{sec.method2}, and an ILC with an additional constraint requiring zero correlation of the ILC map with the tracer maps in \S \ref{sec:method1}. In \S \ref{sec:modeling} we describe the \emph{unWISE} galaxy catalogs used as LSS tracers for our demonstrations as well as some of the modeling choices for the component power spectra and comparisons of the theoretical curves to data. In \S \ref{sec:results} we present the results from the different methods in a combined SO \cite{SO2019} and \emph{Planck}-like \cite{Planck:2018} experiment, and in \S \ref{sec:forecasts} we provide forecasts for the methods using future LSS surveys. We discuss these results and forecasts in \S \ref{sec:Discussion}. We provide several details in the appendices. In Appendix \ref{sec:spectra} we describe theoretical models of the component auto- and cross-spectra in the halo model, and in Appendix \ref{sec:models} we provide details of the modeling choices and comparisons to data for the \emph{unWISE} galaxy HOD, CIB models, and tSZ model. In Appendices \ref{app:additional_plots} and \ref{app:planck_results}, we provide additional plots of the results. We also discuss the \emph{unWISE} pixel window function treatment, CIB -- galaxy cross-correlation, CIB -- CIB correlation coefficients, and CIB SED as a modified blackbody in Appendices \ref{app:pwf}, \ref{app:plots}, \ref{app:corr_coeff}, and \ref{app:mbb}, respectively.

\section{Harmonic ILC}
\label{sec.ILC}

\subsection{Standard ILC}
\label{subsec.standard_ILC}
The standard ILC \cite{Bennett2003, Tegmark2003, Eriksen2004, Delabrouille2009} is a method that constructs a map of a signal of interest by finding the minimum-variance linear combination of the observed maps at different frequencies that simultaneously satisfies the constraint of unit response to the signal of interest. We can express an ILC map as a linear combination of $N$ temperature maps at different frequencies. For an ILC performed in harmonic space,\footnote{For simplicity, we consider only harmonic-domain ILC calculations in this work.  For applications to real data, it will be advantageous to consider needlet ILC~\cite{Delabrouille2009}.} we have
\begin{equation}
    \label{eq.ILC_sum}
    \hat{T}^{\rm ILC}_{\ell m} = w^i_{\ell} T^i_{\ell m},
\end{equation}
where $T^i_{\ell m}$ are the harmonic transforms of the temperature maps at different frequencies indexed by $i$. In Eq.~\eqref{eq.ILC_sum} and throughout this section, we use the convention that repeated indices (excluding $\ell$) are summed over. 

To find the optimal weights, we minimize the variance of the ILC map:
\begin{equation}
    \label{eq:Rij}
    \sigma^2_{\hat{T}^{\mathrm{ILC}}_{\ell m} \hat{T}^{\mathrm{ILC}}_{\ell m}} = w^i_{\ell} w^j_{\ell} \left(\hat{R}_{\ell} \right)_{ij} \qquad \text{with} \qquad \left(\hat{R}_{\ell} \right)_{ij}= \sum_{\ell' = \ell-\Delta \ell /2}^{\ell+\Delta \ell /2} \frac{2\ell'+1}{4\pi} C_{\ell'}^{ij} \,, 
\end{equation}
 representing the empirical frequency-frequency covariance matrix of the data. The multipole bin width $\Delta \ell$ must be large enough to mitigate the ``ILC bias" that results from computing the covariances for ILC weights using a small number of modes \cite{Delabrouille2009}. 
 
 In a standard ILC this minimization is subject solely to the constraint of unit response to the signal of interest: 
\begin{equation}
    \label{eq.ILC_preserved_component}
    w^i_{\ell} a_i = 1 \,,
\end{equation}
where $a_i$ is the spectral response of the signal of interest at the $i$th frequency (\textit{e.g.}, unity for the CMB or kSZ signal, assuming the frequency maps are in blackbody temperature units). This constraint ensures signal preservation in the final ILC result at each $\ell$. The weights satisfying the optimization problem are found via Lagrange multipliers to be \cite{Eriksen2004} 
\begin{equation}
    \label{eq.ILCweightsstandard}
    w^i_{\ell} = \frac{ \left( \hat{R}_{\ell}^{-1} \right)_{ij} a_j }{\left( \hat{R}_{\ell}^{-1} \right)_{km} a_k a_m } 
\end{equation}
for $i,j,k,m \in \{1,...,N\}$.  Throughout this work, we focus solely on the construction of blackbody CMB+kSZ maps, such that $\textbf{\textit{a}}$ is a vector of ones of length $N$.

\subsection{Constrained ILC}
\label{subsec.constrained_ILC}

\subsubsection{One Deprojected Component}
\label{subsubsec.one_deproj_comp}
Now suppose we want to explicitly deproject some component from the final ILC map, \textit{i.e.}, require that the ILC weights have zero response to a contaminant with some specified SED. This gives the constraint
\begin{equation}
    \label{eq.ilc_deproj_constraint}
     w^i_{\ell} b_i = 0,
\end{equation}
where $b_i$ is the deprojected component's spectral response at the $i$th frequency channel. The minimization with the additional constraint gives the weights \cite{Remazeilles_2010}
\begin{equation}
\label{eq.cilc}
    w^j_{\ell} = \frac{ \left(b_k (\hat{R}_{\ell}^{-1})_{kl} b_l \right) a_i (\hat{R}_{\ell}^{-1})_{ij} - \left(a_k (\hat{R}_{\ell}^{-1})_{kl} b_l \right) b_i (\hat{R}_{\ell}^{-1})_{ij}}{\left(a_k (\hat{R}_{\ell}^{-1})_{kl} a_l\right) \left(b_m (\hat{R}_{\ell}^{-1})_{mn} b_n \right) - \left(a_k (\hat{R}_{\ell}^{-1})_{kl} b_l \right)^2}
\end{equation}
for $j \in \{1,...,N\}$.

\subsubsection{Two Deprojected Components}
\label{subsubsec.two_deproj_comps}
If we want to deproject two components from the final ILC map, we have the constraints 
\begin{equation}
    \label{eq.ilc_deproj_two_constraint}
    w^i_{\ell} b_i = 0 = w^i_{\ell} c_i \,,
\end{equation}
where $b_i$ and $c_i$ are the first and second deprojected components' spectral response at the $i$th frequency channel, respectively. The solution for the weights is 
\begin{equation}
    \label{eq.ILC_weights_two_deproj_comps}
    w^j_{\ell} = \left(\hat{R}_{\ell}^{-1}\right)_{ij} \frac{1}{Q_{\ell}} \left( (B_{\ell}C_{\ell}-F_{\ell}^2)a_i + (E_{\ell}F_{\ell}-C_{\ell}D_{\ell})b_i + (D_{\ell}F_{\ell}-B_{\ell}E_{\ell})c_i \right),
\end{equation}
where
\begin{align}
    A_{\ell} &= \left( \hat{R}_{\ell}^{-1} \right)_{ij} a_i a_j \nonumber \\
    B_{\ell} &= \left( \hat{R}_{\ell}^{-1} \right)_{ij} b_i b_j \nonumber \\
    C_{\ell} &= \left( \hat{R}_{\ell}^{-1} \right)_{ij} c_i c_j \nonumber \\
    D_{\ell} &= \left( \hat{R}_{\ell}^{-1} \right)_{ij} a_i b_j \nonumber \\
    E_{\ell} &= \left( \hat{R}_{\ell}^{-1} \right)_{ij} a_i c_j \nonumber \\
    F_{\ell} &= \left( \hat{R}_{\ell}^{-1} \right)_{ij} b_i c_j \nonumber \\
    Q_{\ell} &= A_{\ell}B_{\ell}C_{\ell} + 2D_{\ell}E_{\ell}F_{\ell} - A_{\ell}F_{\ell}^2 -B_{\ell}E_{\ell}^2 - C_{\ell}D_{\ell}^2 \,.
\end{align}

\subsubsection{Arbitrary Number of Deprojected Components (Multiply Constrained ILC)}
\label{subsubsec.mcilc}
Finally, we can generalize these results to deprojecting an arbitrary number of components. Suppose we want to deproject $N_f$ components. Using indices $\alpha$ and $\beta$ for the $N_f+1$ components (of which $N_f$ are to be removed and one is to be preserved), we define $\mathcal{A}_{i \alpha}$ to be the $\alpha$th emissive component's SED integrated over the $i$th channel's bandpass (\textit{i.e.}, the ``mixing matrix"). The weights are then given by 
\begin{equation}
    w^i_\ell = \frac{1}{2} \left( \hat{R}_{\ell}^{-1} \right)_{ij} \Lambda_\alpha \mathcal{A}_{j \alpha},
\end{equation}
where $\Lambda_\alpha$ is a vector of Lagrange multipliers. We have $N_f+1$ constraints: $w^i_\ell \mathcal{A}_{i0}=1$ and $w^i_\ell \mathcal{A}_{i \beta}=0$ for $\beta \neq 0$. Defining the $(N_f+1) \times (N_f+1)$ symmetric matrix $\mathcal{Q}^\ell_{\alpha \beta} = \left( \hat{R}_{\ell}^{-1} \right)_{ij} \mathcal{A}_{i \alpha} \mathcal{A}_{j \beta}$ (which has $(N_f+1)(N_f+2)/2$ independent entries), the solution for the weights is then 
\begin{align}
    \label{eq.mcilc_weights_long}
    w^j_\ell &= \left( \hat{R}_{\ell}^{-1} \right)_{ij} \frac{1}{\mathrm{det} (\mathcal{Q}^\ell)} \Bigg(\mathrm{det}(\mathcal{Q^\ell}_{1,2,...,N_f;1,2,...,N_f}) \mathcal{A}_{i0} - \mathrm{det}(\mathcal{Q^\ell}_{0,2,...,N_f;1,2,...,N_f}) \mathcal{A}_{i1} \nonumber
    \\& \qquad + \mathrm{det}(\mathcal{Q^\ell}_{0,1,3,...,N_f;1,2,...,N_f}) \mathcal{A}_{i2} - \mathrm{det}(\mathcal{Q^\ell}_{0,1,2,4,,...,N_f;1,2,...,N_f}) \mathcal{A}_{i3} + ... \Bigg),
\end{align}
where $\mathcal{Q^\ell}_{1,2,...,N_f;1,2,...,N_f}$ refers to the $N_f \times N_f$ sub-matrix of $\mathcal{Q^\ell}$ left after removing the zeroth row and zeroth column, $\mathcal{Q^\ell}_{0,2,...,N_f;1,2,...,N_f}$ refers to the $N_f \times N_f$ sub-matrix of $\mathcal{Q^\ell}$ left after removing the first row and zeroth column, and so on. Defining $\mathcal{Q}^{\ell ,S}_{\alpha}$ to be the sub-matrix of $\mathcal{Q}^\ell$ left after removing the $\alpha$th row and zeroth column, we can write Eq.~\eqref{eq.mcilc_weights_long} compactly as 
\begin{equation}
    \label{eq.mcilc_weights_compact}
    w^j_\ell = \left( \hat{R}_{\ell}^{-1} \right)_{ij} \frac{1}{\mathrm{det} (\mathcal{Q}^\ell)} \sum_{\alpha=0}^{N_f} (-1)^{\alpha} \mathrm{det} (\mathcal{Q}^{\ell ,S}_{\alpha}) \mathcal{A}_{i \alpha}
\end{equation}
We refer to this approach as the ``multiply constrained ILC"; see Ref.~\cite{Remazeilles:2021} for alternate versions of these expressions.

While deprojecting components in the ILC is useful for certain purposes, as it allows one to robustly guarantee that a contaminant with some SED is removed, there is a trade-off: deprojecting a component (adding a constraint to the ILC) increases the noise in the resulting ILC map since the feasible region allowed by the constraints is smaller \cite{Remazeilles_2010, Abylkairov2021}. This generally results in a lower SNR for the signal of interest in the final map.  Moreover, we may not know with certainty the SED of a component we seek to deproject, as in the case of the CIB emission. We thus consider alternatives to this explicit deprojection.


\section{Multitracer de-CIB and de-(CIB+tSZ)}
\label{sec:deCIB}

\subsection{Modification of Standard ILC: de-(CIB+tSZ)}
\label{sec:de_CIBplustSZ}

Suppose that we are given an external catalog of LSS tracers (\textit{e.g.}, galaxies, quasars, lensing convergence, etc.), which are correlated with the CIB, tSZ, and other signals in the mm-wave sky.  Our goal here is to build a method that combines this external catalog with the mm-wave frequency maps so as to remove the CIB, tSZ, and/or other contaminants.  We can think of cleaning the CIB and tSZ from our map --- \emph{de-(CIB+tSZ)ing} --- in an analogous way to delensing of the CMB, as we now describe.

The first step is to build a combined LSS tracer map that is optimally correlated with the CIB and tSZ fields.  Following Ref.~\cite{Yu_2017}, the linear combination of tracer samples that is most highly correlated with the combined CIB+tSZ signal at each frequency channel can be expressed as 
\begin{equation}
    \label{eq.decib_gi}
    g^i_{\ell m} = \sum_a c_{a,\ell}^i \, g^a_{\ell m} \,,
\end{equation}
where $g^a$ labels each tracer sample and $g^i$ is the optimal linear combination of these samples in terms of correlation with the (CIB+tSZ) signal at the $i$th frequency. The coefficients $c^i_{a,\ell}$ are given by (analogous to Eq.~8 in Ref.~\cite{Yu_2017})
\begin{equation}
    \label{eq.decib_coeffs}
    c^i_{a,\ell} = \sum_b (\rho^{\ell})^{-1}_{ab} \rho^{\ell}_{b,\mathrm{(CIB+tSZ)}_i} \sqrt{\frac{C_{\ell}^{\mathrm{(CIB+tSZ)}_i\mathrm{(CIB+tSZ)}_i}}{C_{\ell}^{g_a g_a}}},
\end{equation}
with $\rho^{\ell}_{ab}$ representing the correlation matrix of two tracer samples at a given $\ell$, defined as 
\begin{equation}
    \rho^\ell_{ab} = \frac{C_{\ell}^{g_ag_b}}{\sqrt{C_{\ell}^{g_ag_a} C_{\ell}^{g_bg_b}}} \,,
\end{equation}
and $\rho^{\ell}_{b,\mathrm{(CIB+tSZ)}_i}$ representing the correlation matrix of the tracer samples with the CIB and tSZ at the $i$th frequency at a given $\ell$:
\begin{equation}
    \rho^{\ell}_{b,\mathrm{(CIB+tSZ)}_i} = \frac{C_{\ell}^{g_b \mathrm{(CIB+tSZ)}_i}}{\sqrt{C_{\ell}^{g_bg_b} C_{\ell}^{\mathrm{(CIB+tSZ)}_i\mathrm{(CIB+tSZ)}_i}}} \,,
\end{equation}
where $C_{\ell}^{\mathrm{(CIB+tSZ)}_i\mathrm{(CIB+tSZ)}_i} $ is the auto-spectrum of the joint (CIB+tSZ) signal at the $i$th frequency, $C_{\ell}^{g_a g_a}$ is the auto-spectrum of tracer $g_a$, $C_{\ell}^{g_a g_b}$ is the cross-spectrum of two tracer samples $g_a$ and $g_b$, and $C_{\ell}^{\mathrm{(CIB+tSZ)}_i, g_a}$ is the cross-spectrum of the (CIB+tSZ) at the $i$th frequency with the tracer sample $g_a$.

The first step is then to remove the fraction of the tracer maps that is contained in the (CIB+tSZ) portion of the temperature map at each frequency, i.e.,
\begin{equation}
    (T^i_{\ell m})' = T^i_{\ell m}-f^i_\ell g^i_{\ell m}
\end{equation}
Then we can modify the standard ILC procedure to minimize the variance of the linear combination of \textit{modified} frequency maps subject to the constraint of unit response to the signal of interest. Thus, the frequency-frequency covariance matrix is now
\begin{equation}
    \label{eq.Rlij_decibtsz}
    \left(\hat{R}_{\ell} \right)_{ij} = \sum_{\ell' = \ell-\Delta \ell /2}^{\ell+\Delta \ell /2} \frac{2\ell'+1}{4\pi} \left( C_{\ell'}^{ij} -f^i_{\ell'} C_{\ell'}^{jg^i} - f^j_{\ell'} C_{\ell'}^{ig^j} +  f^i_{\ell'}f^j_{\ell'}C_{\ell'}^{g^ig^j} \right) \, ,
\end{equation}
where $C_{\ell'}^{jg^i}$ is the cross-power spectrum of the original temperature map at frequency $j$ and the linear combination of tracers $g^i$.

Finding this fraction $f^i_\ell$ amounts to finding the fraction of $g^i$ contained in the joint (CIB+tSZ) signal at each frequency: 
\begin{equation}
    f^i_{\ell} \equiv \frac{C_{\ell}^{\mathrm{(CIB+tSZ)}_i, g^i}}{C_{\ell}^{g^i,g^i}} = \frac{\sum_a c^i_{a,\ell} C_{\ell}^{\mathrm{(CIB+tSZ)}_i, g_a}}{\sum_{a,b} c^i_{a,\ell} c^i_{b,\ell} C_{\ell}^{g_ag_b}} \,.
\end{equation}

As described in, \textit{e.g.}, Ref.~\cite{Lizancos:2023}, $f^i_\ell$ is an ``optimal filter" that comes from minimizing the variance of the cleaned map at each $\ell$. For the specific form of this filter given here, it is only optimal under the hypothesis that the only components in the frequency maps that are correlated with the tracer density maps are the CIB and tSZ fields.

The ILC weights are then given by the usual standard ILC weights, but with the frequency-frequency covariance matrix replaced with that from Eq.~\eqref{eq.Rlij_decibtsz}. Then the power spectrum of $T_{\rm clean}$ is 
\begin{equation}
    C_{\ell}^{T^{\rm clean} T^{\rm clean}} = C_{\ell}^{\rm ILC} - 2\sum_i w^i_{\ell}f^i_{\ell} C_{\ell}^{\mathrm{ILC}, g^i} + \sum_{i,j} w^i_{\ell} w^j_{\ell} f^i_{\ell} f^j_{\ell} C_{\ell}^{g^i g^j} \, ,
\end{equation}
where $C_\ell^{\rm ILC}$ is the power spectrum of the usual standard ILC map with no tracer subtraction.

The derivation of these results is nearly identical to that in Refs.~\cite{Yu_2017, Lizancos:2023}. The key difference is that those results sought a linear combination of tracers with maximal correlation to the CMB lensing field, therefore requiring only one set of coefficients $c_{a,\ell}$ and one optimal filter $f_\ell$ at each $\ell$. In contrast, here we seek a linear combination of tracers with maximal correlation to the (CIB+tSZ) field. Since the CIB has a nontrivial spectral dependence, to optimally clean out the CIB, we must find the maximally correlated linear combination of tracers at \textit{each frequency}, giving us frequency-dependent coefficients $c^i_{a,\ell}$ and frequency-dependent filters $f^i_\ell$.

For this method, we must assume some specific theoretical model for the correlation of the CIB and tSZ fields and tracer maps.\footnote{The tracer-CIB cross-correlation can be directly measured at high frequencies (see Fig.~\ref{fig:cibXg_data} below), but the de-(CIB+tSZ) method also requires knowledge of this cross-correlation at lower frequencies used in the ILC construction, where it is much more difficult to measure directly; thus some level of theoretical modeling is likely always necessary.} This correlation is used to determine the coefficients $c^i_{a,\ell}$ for the linear combination of tracer samples maximally correlated with the (CIB+tSZ) field, and also to determine the fraction of tracer maps $f^i_\ell$ to remove at each frequency. Nevertheless, a slight model misspecification would only affect the optimality of the method by some small amount, as all the tracer samples are correlated with the CIB and tSZ fields. We test this assumption later, in \S \ref{sec:Discussion}.

\subsection{Modification of Constrained ILC: de-CIB Applied to tSZ-deprojected ILC Map}
\label{sec:de_CIB_only}

With this method, the idea is to start with a tSZ-deprojected ILC map, and then subtract off whatever portion of the tracer fields remains in the ILC map. Specifically, we obtain the cleaned map via 
\begin{equation}
    T^{\rm clean}_{\ell m} = T^{\rm ILC}_{\ell m} - f_\ell g^{\rm OPT}_{\ell m},
\end{equation}
where, in this case, 
\begin{equation}
    f_\ell = \frac{C_\ell^{T^{\rm ILC},g^{\rm OPT}}}{C_\ell^{g^{\rm OPT}g^{\rm OPT}}} \,.
\end{equation}
To find the optimal combination of tracers in this case, we have that 
\begin{equation}
    \label{eq.gILC}
    g^{\rm OPT}_{\ell m} \equiv \sum_a c_{a,\ell} \, g^a_{\ell m} \,,
\end{equation}
where $c_{a,\ell}$ is defined as in Eq.~\eqref{eq.decib_coeffs} but where $\mathrm{(CIB+tSZ)}_i$ is now replaced with ``OPT" and the coefficients are thus no longer frequency-dependent. The superscript ``OPT" denotes that we are finding the combination of tracers that has maximal correlation with the ILC map. Since the tSZ signal has been deprojected in the ILC, the only remaining contaminant that is correlated with the tracers in the ILC map is the CIB. Thus, in cleaning out the remaining tracers, we are really cleaning out the CIB, motivating the name \textit{de-CIB} for this method.

In this method, we first deproject the tSZ signal in the usual frequency-dependent way, and not using the external LSS data. When deprojecting the tSZ signal, we cannot find separate combinations of tracers to clean out the contaminants at each frequency. To see why, consider two possibilities: 
\begin{enumerate}
    \item First, we can attempt to perform a similar procedure as with de-(CIB+tSZ), except here, we let $f^i_{\ell} = \frac{C_{\ell}^{\mathrm{CIB}_i, g^i}}{C_{\ell}^{g^i,g^i}}$ with $g^i_{\ell m}=c^i_{a,\ell}g^a_{\ell m}$ and $c^i_{a,\ell} = \sum_b (\rho^{\ell})^{-1}_{ab} \rho^{\ell}_{b,\mathrm{CIB}_i} \sqrt{\frac{C_{\ell}^{\mathrm{CIB}_i\mathrm{CIB}_i}}{C_{\ell}^{g_a g_a}}}$. We then try to find ILC weights that minimize the variance of a linear combination of the modified frequency maps, as in the de-(CIB+tSZ) procedure, but subject to the constraint that the tSZ signal has zero response in the ILC map. The problem here is that the effective tSZ response in each of the modified frequency maps is no longer well-defined (i.e., no longer equal to the usual tSZ SED), because the tSZ field is correlated with the tracers that we subtract.  Thus, the tSZ field is modified in a frequency-dependent manner.
    
    \item Alternatively, we could switch the order of tracer subtraction and ILC weight determination. First we could find ILC weights from the usual constrained ILC where the tSZ signal is deprojected. We could then modify the frequency maps as in the previous scenario by subtracting the portion of the tracer fields contained in the CIB at each frequency. We then apply the weights to the modified frequency maps. The problem with this method is that we are double-subtracting the tSZ signal. First, we implicitly remove tSZ signal when subtracting the fraction of tracers from each frequency map, since the tSZ field is correlated with the tracers. We then subtract tSZ signal once again by applying the weights computed from the usual tSZ-deprojected ILC.
\end{enumerate}

Due to these challenges, we thus use the procedure of subtracting the tracer fields from the tSZ-deprojected ILC map instead of subtracting them from each frequency.  However, we note that the resulting cleaned map will not formally have minimum variance.

\section{LSS Tracer Maps as Additional ``Frequency" Maps in Harmonic ILC}
\label{sec.method2}

For the second method to clean the CIB and tSZ signals from blackbody CMB+kSZ maps, we note that the CMB and kSZ signals are not correlated at the two-point level with the CIB or tSZ signals and are also not correlated with the LSS tracer maps.\footnote{This statement is violated at low $\ell$ by the integrated Sachs-Wolfe (ISW) effect and by a small (but for our purposes, negligible) amount at high $\ell$ by the Rees-Sciama effect.  Thus our method should not be used at $\ell \lesssim 100$, where the ISW signal is large.} Therefore, to create an ILC map that preserves the CMB+kSZ blackbody signal and removes the CIB and tSZ contaminants, we use the fact that the CMB and kSZ fields have zero ``response'' in the tracer maps.  We can then include the tracer maps as ``frequency" maps in the set of maps used in the ILC, with the CMB+kSZ signal of interest having zero response at these channels:
\begin{equation}
\hat{T}^{\rm ILC}_{\ell m} = \sum_{i=1}^{N} w^i_{\ell} T^i_{\ell m} + \sum_{i=1}^{N_g} w^{N+i}_{\ell} g_{i, \ell m},
\end{equation}
where $g_i$ are the tracer samples and $N_g$ is the number of tracer maps used.  Our new spectral response vector for the CMB+kSZ signal of interest, $\textbf{\textit{a}}$, is then
\begin{equation}
    \textbf{\textit{a}} \leftarrow \textbf{\textit{a}} + \underbrace{[0,...,0]}_{N_g} \,,
\end{equation}
and the new weights vector $w^i_{\ell}$ now has length $N+N_g$.  Note that $a^i$ is now also a vector of length $N+N_g$, consisting of $N$ ones followed by $N_g$ zeros. The signal-preservation constraint of the ILC then remains unchanged, allowing the signal of interest to propagate in an unbiased fashion to the ILC map, as usual: 
\begin{equation}
    \sum_{i=1}^{N+N_g} w_{\ell}^{i} a_i = \sum_{i=1}^{N} w_\ell^i a_i = 1 \,.
\end{equation}

This formulation is equivalent to the standard ILC (see \S \ref{sec.ILC}), but with a modified covariance matrix $\left(\hat{R}^{\ell}_{ij}\right)^\prime$. The modified covariance matrix now includes $N_g$ extra rows and $N_g$ extra columns to account for the cross-correlations of the tracer maps with each frequency channel $T_i$, cross-correlations of the tracer maps with each other, and the auto-correlation of each tracer map. Schematically, $\left(\hat{R}^{\ell}\right)^\prime$ can be written as
\begin{equation}
    \left(\hat{R}^{\ell}\right)^\prime =
    \begin{bmatrix}
    \hat{R}^{\ell} & X_{\ell} \\
    X^T_{\ell} & G_{\ell}  
    \end{bmatrix}
    \label{eq:R_ij^prime}
\end{equation}
where $\hat{R}$ is the original covariance matrix of the data of size $N \times N$, and $X$ is a matrix of size $N \times N_g$ defined as
\begin{equation}
    X^{ia}_{\ell} = \sum_{\ell' = \ell-\Delta \ell /2}^{\ell+\Delta \ell /2} \frac{2\ell'+1}{4\pi} C_{\ell'}^{ig_a} \,,
\end{equation}
where $i \in \{1,...,N\}$ indexes the temperature maps at different frequency channels and $a \in \{1,...,N_g\}$ indexes the different tracer maps.  Finally, $G$ is a matrix of size $N_g \times N_g$ that accounts for the auto- and cross-correlations of the tracer maps with each other:
\begin{equation}
    G^{ab}_{\ell} = \sum_{\ell' = \ell-\Delta \ell /2}^{\ell+\Delta \ell /2} \frac{2\ell'+1}{4\pi} C_{\ell'}^{g_ag_b} \,.
\end{equation}

The ILC weights are then given by 
\begin{equation}
    w^i_{\ell} = \frac{  \left( \hat{R}_{\ell}^{-1} \right)^\prime_{ij} a_j }{ \left( \hat{R}_{\ell}^{-1} \right)^\prime_{km} a_k a_m } \,,
\end{equation}
where here $i,j,k,m \in \{1,...,N+N_g\}$.

We can now calculate the power spectrum of the ILC map:
\begin{equation}
    C_{\ell}^{\hat{T}\hat{T}}  = \sum_{i,j=1}^{N} w^i_{\ell} w^j_{\ell} C_{\ell}^{ij} + 2\sum_{i=1}^{N}\sum_{a=1}^{N_g} w^i_{\ell} w^{N+a}_{\ell} C_{\ell} ^{ig_a} + \sum_{a,b=1}^{N_g}  w^{N+a}_{\ell} w^{N+b}_{\ell} C_{\ell}^{g_ag_b} .
\end{equation}

A drawback of this method is that deprojecting the tSZ effect via a constrained ILC becomes non-trivial. To do so, one would need to determine the additional entries in $b_i$ from Eq.~\eqref{eq.ilc_deproj_constraint} corresponding to the tSZ field's response in the tracer maps (corresponding to $i \in \{N+1,...,N+N_g\}$ in our formalism above). Unlike with the CMB, these correlations do not vanish because the tSZ signal is correlated with the LSS tracer density fields.  Thus, determining the values of $b_i$ to use would rely on detailed modeling of these correlations (\textit{e.g.},~\cite{Tanimura_2019, Koukoufilippas:2019ilu, Pandey_2019_yg, Alonso_2018, schaan2020act}), which is contrary to our overall goal of deprojecting contaminants in a (mostly) model-independent way. Nevertheless, this method can simultaneously clean both the CIB and tSZ signals without necessitating any theoretical models of these signals or their cross-correlations with one another and the tracer fields; it is fully data-driven. This is possible because the CMB+kSZ signal of interest has zero response in \textit{every} tracer map, and thus, we do not have to find any optimal linear combination of these tracers as we did in \S \ref{sec:deCIB} for de-CIBing and de-(CIB+tSZ)ing.

\section{Constraint Requiring Zero Cross-Correlation with Tracer Maps in Harmonic ILC}
\label{sec:method1}

Our final method uses the LSS tracers to clean the CIB and tSZ contaminants via a novel extension of the constrained ILC technique.  In this approach, we impose the explicit requirement that the cross-correlation of the ILC map with an LSS tracer map vanishes.  The method can then be extended to require that an arbitrary number of such cross-correlations vanish (cf. \S \ref{subsubsec.mcilc}).  The premise relies on the fact that the LSS tracer maps contain only contaminants, and no contributions from the signal of interest, as in the method presented in \S \ref{sec.method2}.

\subsection{One Deprojected Component}
\label{subsec:method1_withoutdeprojtsz}
In this method, we start with a standard harmonic ILC, with a preserved CMB+kSZ component, as described in \S \ref{sec.ILC}.  The twist here is that we add an additional explicit constraint to the ILC, in which we require that the final ILC map has zero cross-correlation with the LSS tracer density map:
\begin{equation}
    \label{eq.g_constraint}
    \langle T_{\ell m}^{\rm ILC} g_{\ell m} \rangle = 0 \Longleftrightarrow \sum_i w^i_{\ell} \langle T^i_{\ell m} g_{\ell m} \rangle = 0 \,.
\end{equation}
We define $c_i^{\ell} \equiv \langle T^i_{\ell m} g_{\ell m} \rangle = C_{\ell}^{ig}$, so that the final constraint becomes 
\begin{equation}
    \sum_i w^i_{\ell} c_i^{\ell} = 0 \,.
\end{equation}

This problem is then equivalent to a constrained ILC with one deprojected component, which can be solved as usual with Lagrange multipliers to obtain the weights.  The resulting weights are identical to those given in Eq.~\eqref{eq.cilc} with the replacement $b_i \rightarrow c_i^\ell$ (crucially, the constraint now is different at each $\ell$, whereas previously $b_i$ was $\ell$-independent).
Then the power spectrum of this ILC map is given by 
\begin{align}
   C_{\ell}^{\hat{T}\hat{T}} &=  w^i_{\ell} w^j_{\ell} C_{\ell}^{ij} \notag \nonumber
   \\ &= \frac{ \left[ c_k^{\ell} (\hat{R}^{-1}_\ell)_{kl} c_l^{\ell} a_p(\hat{R}^{-1}_\ell)_{pi}-a_k (\hat{R}^{-1}_\ell)_{kl} c_l^{\ell} c_p^{\ell} (\hat{R}^{-1}_\ell)_{pi} \right] \left[ c_k^{\ell} (\hat{R}^{-1}_\ell)_{kl} c_l^{\ell} a_p(\hat{R}^{-1}_\ell)_{pj}-a_k (\hat{R}^{-1}_\ell)_{kl} c_l^{\ell} c_p^{\ell}(\hat{R}^{-1}_\ell)_{pj} \right]}{\left[(a_k (\hat{R}^{-1}_\ell)_{kl} a_l)(c_m^{\ell} (\hat{R}^{-1}_\ell)_{mn} c_n^{\ell})-(a_k (\hat{R}^{-1}_\ell)_{kl} c_l^{\ell})^2 \right]^2} C_{\ell}^{ij} \,.
\end{align} 

With a single tracer map, this method is performed as described above. However, if one has multiple tracer maps, assuming the goal is to remove just the CIB, one must find the optimal linear combination of tracer maps correlated with the CIB. Since our constraint involves the cross-correlation of the final ILC map (which is a combination of temperature maps from different frequencies) with some combination of tracers, we cannot simply find the combination of tracers with highest correlation with the CIB at any given frequency. Instead, we require some frequency-independent version of the CIB, similar to what the Compton-$y$ field is for the tSZ effect. In reality, the CIB decorrelates across frequencies (\textit{e.g.}, \cite{Planck:2013cib,Mak2017,Lenz_2019}), so such a simplification is not entirely possible; however, it is a good approximation for finding some combination of tracers that are highly correlated with the CIB, and any inexactness in the modeling would only affect the optimality of this linear combination of tracers by a small amount.

Thus, we simply use Eq.~\eqref{eq.decib_gi} and \eqref{eq.decib_coeffs} from \S \ref{sec:deCIB} but with all the quantities becoming frequency-independent, \textit{i.e.}, $g^i_{\ell m} \rightarrow g^{\mathrm{CIB}}_{\ell m}$ in Eq.~\eqref{eq.decib_gi}, $c^i_{a,\ell} \rightarrow c_{a,\ell}$ in Eq.~\eqref{eq.decib_coeffs}, and $\mathrm{CIB}_i \rightarrow a^{\mathrm{CIB}}(\hat{n})$ in Eq.~\eqref{eq.decib_coeffs}. The $a^{\mathrm{CIB}}(\hat{n})$ field is the analog of the Compton-$y$ field for the CIB, described in detail in Appendix \ref{app:mbb}, and $g^{\mathrm{CIB}}$ is the linear combination of tracer maps with maximal correlation to $a^{\mathrm{CIB}}(\hat{n})$. Note that this $g^{\mathrm{CIB}}$ is independent of frequency, as required for the zero-tracer-correlation constraint in the ILC. Then this problem is the same as described above, but with $g_{\ell m} \rightarrow g^{\mathrm{CIB}}_{\ell m}$ in Eq.~\eqref{eq.g_constraint} so that $c^\ell_i \equiv \langle T^i_{\ell m} g^{\mathrm{CIB}}_{\ell m} \rangle$.

\subsection{Two Deprojected Components}
\label{subsec:method1_withdeprojtsz}
Suppose that we now want to additionally deproject the tSZ signal using its known frequency dependence.  This problem is equivalent to solving an ILC with two deprojected components, which gives the weights in Eq.~\eqref{eq.ILC_weights_two_deproj_comps}, but with $c_i$ replaced with the $\ell$-dependent quantity $c_i^{\ell}$ (where $c^\ell_i \equiv \langle T^i_{\ell m} g^{\mathrm{CIB}}_{\ell m} \rangle$, as before). Then the power spectrum of this ILC map is given by
\begin{align}
    \label{eq.ILC_constraint_g_two_deproj_power_spec}
   C_{\ell}^{\hat{T}\hat{T}} &= w^i_{\ell} w^j_{\ell} C_{\ell}^{ij} \notag \nonumber
   \\ &= \left( \hat{R}_{\ell}^{-1} \right)_{ik} \left( \hat{R}_{\ell}^{-1} \right)_{jm} \frac{1}{Q_{\ell}^2} \left( (B_{\ell}C_{\ell}-F_{\ell}^2)a_k + (E_{\ell}F_{\ell}-C_{\ell}D_{\ell})b_k + (D_{\ell}F_{\ell}-B_{\ell}E_{\ell})c_k^{\ell} \right) \notag \nonumber
   \\ &\qquad \times \left( (B_{\ell}C_{\ell}-F_{\ell}^2)a_m + (E_{\ell}F_{\ell}-C_{\ell}D_{\ell})b_m + (D_{\ell}F_{\ell}-B_{\ell}E_{\ell})c_m^{\ell} \right) C_{\ell}^{ij} .
\end{align} 

An alternative to removing both the CIB and tSZ signals is to do the above, but instead of deprojecting the tSZ signal directly, we deproject $g^{\mathrm{tSZ}}$, where $g^{\mathrm{tSZ}}$ is the linear combination of tracer maps with maximal correlation to the Compton-$y$ field. Thus, the weights are still given by Eq.~\eqref{eq.ILC_weights_two_deproj_comps}, but both $b_i$ and $c_i$ are replaced with $\ell$-dependent quantities $b^\ell_i$ and $c^\ell_i$, respectively. Here $b^\ell_i \equiv \langle T^i_{\ell m} g^{\mathrm{tSZ}}_{\ell m} \rangle$ and $c^\ell_i \equiv \langle T^i_{\ell m} g^{\mathrm{CIB}}_{\ell m} \rangle$. The power spectrum of this ILC map is then given by Eq.~\eqref{eq.ILC_constraint_g_two_deproj_power_spec} but with $b_i \rightarrow b^\ell_i$.

\subsection{Discussion}
We note that usually an ILC only uses information about the SEDs of various components, where these SEDs are fully deterministic quantities. With the new method presented in this section, we have a constraint in the ILC that depends on a realization of a random field. Another interesting feature of this method is that, when using only a single tracer map, the galaxy shot noise does not explicitly enter into any of the calculations since the auto-spectrum of the tracer sample, \textit{i.e.}, $C_\ell^{gg}$, never appears in any of the expressions. While the formalism for this method is self-consistent, this points to the suboptimality of the method for high shot noise samples. As the shot noise increases (as the number of tracers approaches zero), the constraint requiring zero cross-correlation of the ILC map with tracers is no longer effective in cleaning the CIB or tSZ signals, as the tracer maps provide no information about these signals (the cross-correlation of the CIB and tSZ signals with the tracers will asymptotically go to zero). Thus, there is an implicit dependence on the shot noise in this method, and this implicit dependence of cross-correlations on the shot noise would play a role in our other new methods as well. 

When we have multiple tracer maps, the shot noise \textit{does} explicitly enter our calculations for finding the optimal linear combination of tracer maps, and will thus affect the optimality of these coefficients. Finding this optimal linear combination of tracer maps is the only step in this method that requires theoretical modeling of the correlations between the CIB and tSZ fields and tracer fields.\footnote{As before, the tracer-CIB cross-correlations can be directly measured at high frequencies, but some level of theoretical modeling is likely necessary at low frequencies.} Just as in \S \ref{sec:deCIB} for the de-CIB and de-(CIB+tSZ) methods, small modeling misspecifications would only affect the optimality of these coefficients, and thus of our results, by some small amount.

We note that this method is a spatial deprojection of the tracer maps. Comparing the two variations described in \S \ref{subsec:method1_withdeprojtsz}, the method of deprojecting the tSZ signal directly using its known frequency dependence is a spectral deprojection that is dependent upon using several frequency channels, whereas the method of deprojecting both $g^{\mathrm{CIB}}$ and $g^{\mathrm{tSZ}}$ involves only spatial deprojection.  However, the latter still requires multiple frequency channels, as one must have at least as many frequencies as total constraints in the ILC in order for the Lagrange multiplier problem to yield a solution.  Interestingly, because the CIB and tSZ signals have non-zero correlation, $g^{\mathrm{CIB}}$ and $g^{\mathrm{tSZ}}$ will likewise have non-zero correlation. Since we are performing spatial deprojections, in the limit that the CIB and tSZ fields are perfectly correlated, $g^{\mathrm{CIB}}$ and $g^{\mathrm{tSZ}}$ are perfectly correlated, and we are able to deproject two components for the price of one in terms of the noise penalty resulting from the constrained ILC procedure. However, if these fields are not significantly correlated, it is unclear \textit{a priori} which method (spectral deprojection of the tSZ field or spatial deprojection of $g^{\mathrm{tSZ}}$) will have a higher noise penalty. We investigate such matters further in \S \ref{sec:results} and \S \ref{sec:Discussion}.

We note that we cannot optimally deproject both the CIB and tSZ signals using a single constraint on the tracer maps. This is because the CIB and tSZ signals have different effective SEDs, and the form of Eq.~\eqref{eq.g_constraint} requires a frequency-independent linear combination of tracers. For the de-(CIB+tSZ) method, we are able to simultaneously remove both signals by finding the optimal combination of tracer maps for correlation with the total CIB+tSZ signal at \textit{each frequency}, but this cannot be done here.


\section{Modeling choices}
\label{sec:modeling}

In this section, we describe the modeling choices used in this work, first for the \emph{unWISE} galaxy catalogs used as LSS tracers in our new methods, and then for the other mm-wave sky components considered in our analysis (primary CMB, tSZ, CIB, kSZ, and radio sources). More details on the theory and modeling can be found in Appendices~\ref{sec:spectra} and \ref{sec:models}, respectively.

\subsection{unWISE Galaxy Catalog }
\label{subsec:unwise}

As a concrete example of our new methods, we consider using the \emph{unWISE} galaxy catalog to remove CIB and tSZ contamination in CMB+kSZ power spectrum measurements.  In this section, we discuss the \emph{unWISE} galaxy catalog; for more details regarding \emph{unWISE}, we refer the reader to \cite{Schlafly18, krolewski_2020, krolewski2021cosmological, Kusiak_2022}. 

The \emph{unWISE} galaxy catalog \cite{Schlafly_2019, krolewski_2020, krolewski2021cosmological} is constructed from the \emph{Wide-Field Infrared Survey Explorer} (\emph{WISE}) satellite mission including the post-hibernation, non-cryogenic \emph{NEOWISE} data. The original \emph{WISE} mission mapped the sky in four bands, at  3.4, 4.6, 12, and 22 $\mu$m (W1, W2, W3, and W4) \cite{Wright_2010}. Based on color cuts on the magnitude in the W1 and W2 bands (see Table~1 in Ref.~\cite{Kusiak_2022}), the \emph{unWISE} catalog was constructed, resulting in over 500 million galaxies over the full sky, divided into three subsamples (blue, green, and red) of mean redshifts $\bar{z} = 0.6$, 1.1, and 1.5, respectively.

\begin{figure}[t!]
    \centering
    \includegraphics[scale=0.5]{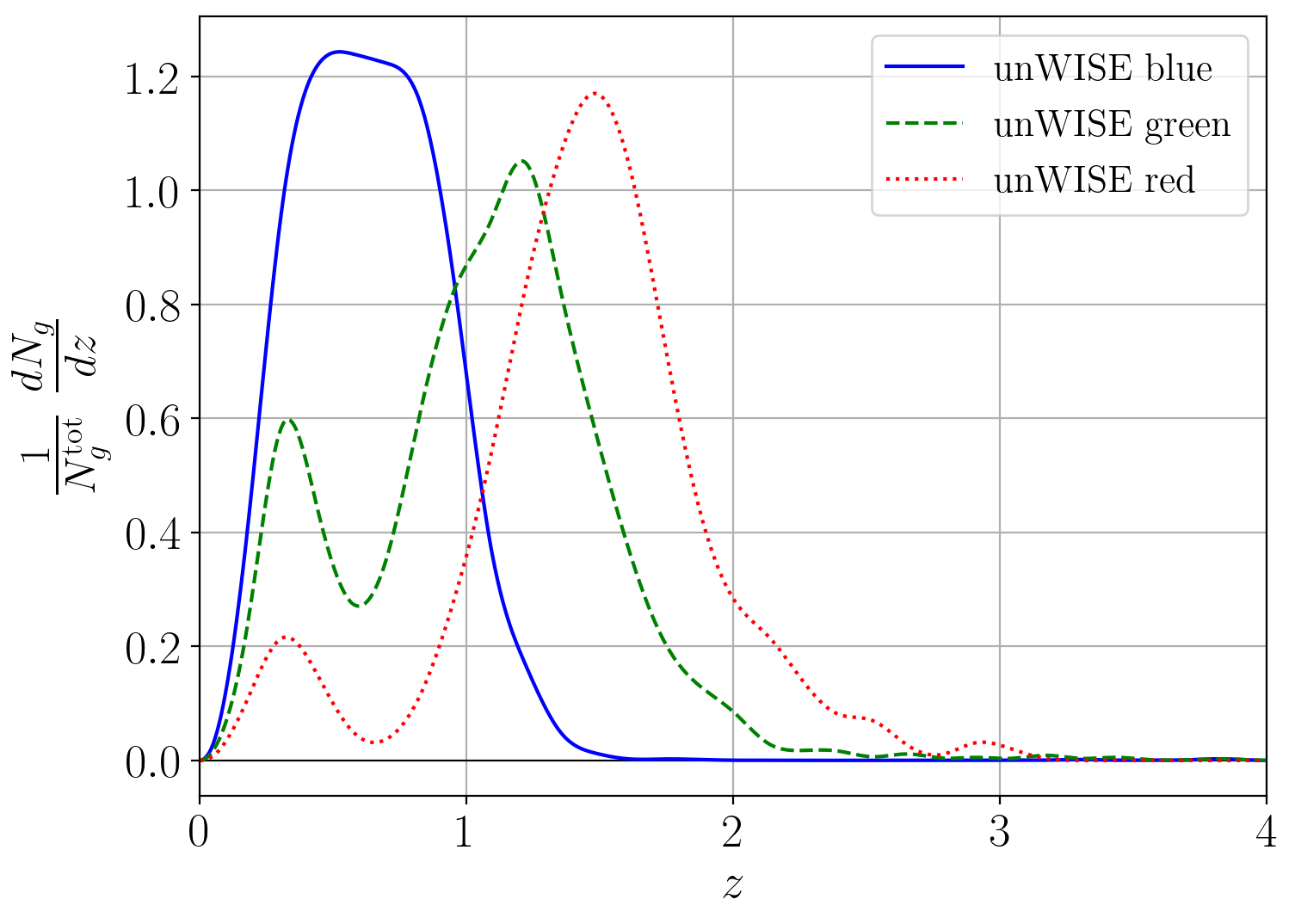}
    \caption{Normalized redshift distributions $\frac{1}{N_g^{\rm{tot}}}\mathrm{d}N_g/\mathrm{d}z$ for each of the \emph{unWISE} galaxy samples: blue (solid), green (dashed), and red (dotted), obtained by cross-matching the \emph{unWISE} objects with the \emph{COSMOS} catalog. See Table~\ref{table:unwise} for other important characteristics of the \emph{unWISE} samples, \textit{e.g.}, redshift statistics, or the number density of galaxies.
    \label{fig:dndz_COSMOS}}
\end{figure}

In Fig.~\ref{fig:dndz_COSMOS} we present the redshift distributions of \emph{unWISE} obtained by direct cross-matching of the samples with the \emph{COSMOS} photometric galaxies in \cite{krolewski_2020}, used in this analysis. Table~\ref{table:unwise} shows other characteristics of each of the \emph{unWISE} subsamples: mean redshift and approximate redshift width (also obtained from cross-matching with the \emph{COSMOS} galaxies), as well as the number density of galaxies and the faint-end logarithmic slope of the luminosity function $s$, necessary to compute the lensing magnification terms (see Appendix~\ref{subsec:gal_hm}). 

    \begin{table}[t!]
    \setlength{\tabcolsep}{10pt}
    \renewcommand{\arraystretch}{1.7}
    \begin{tabular}{| c| c |c|c|c| } 
    \hline
    \emph{unWISE} & $\bar{z}$ & $\delta_{z}$ & $\bar{n}_g$ [deg$^{-2}$] & $s$ \\ 
    \hline\hline
    blue & 0.6 & 0.3 & 3409 & 0.455\\
    green & 1.1& 0.4 & 1846 & 0.648\\ 
    red &  1.5 & 0.4 & 144 & 0.842\\ 
    \hline
    \end{tabular}
    \caption{Important properties of each \emph{unWISE} sample: $\bar{z}$, mean redshift; $\delta_z$, approximate width of the redshift distribution, both obtained from $\mathrm{d}N_g/\mathrm{d}z$ as measured by matching to objects with precise photometric redshifts in the \emph{COSMOS} field \citep{Laigle_2016} (see \S \ref{subsec:unwise}); $\bar{n}$, the number density per deg$^2$; and $s$, the faint-end logarithmic slope of the luminosity function $s = \mathrm{d} \log_{10} N_g/ \mathrm{d}m$. See \cite{Schlafly19, krolewski_2020, krolewski2021cosmological} for further details.}
    \label{table:unwise}
    \end{table}

As qualitatively assessed in \cite{Kusiak_2022}, the emission from galaxies in the \emph{unWISE} samples is approximately 70-90\% stellar-dominated emission and 10-30\% a mixture of stellar and thermal dust emission, with a contribution from the $3.3 \, \mu {\rm m}$ PAH emission for the blue sample; 50-70\% stellar-dominated and 30-50\% mixture, with a small contribution from the $3.3 \, \mu {\rm m}$ PAH emission for the green sample; and the red sample is stellar-dominated. The average halo masses of \emph{unWISE} were constrained to be $\approx 10^{13} M_{\odot}/h$ \cite{Kusiak_2022}.

The \emph{unWISE} catalog has been used in multiple analyses, \textit{e.g.}, to constrain the ionized gas density with the projected-field kSZ estimator in \cite{Kusiak_2021} or to measure the late-time cosmological parameters $\sigma_8$, the amplitude of low-redshift density fluctuations, and $\Omega_m$, the matter density fraction in \cite{krolewski2021cosmological,Farren2023}. 

The halo occupation distribution (HOD) (see Appendix~\ref{sec:spectra} for the discussion of the HOD \cite{Zheng_2007, zehavi2005_sdss} within the larger halo model \cite{Cooray:2002dia, seljak2000, peacock2000}) of the \emph{unWISE} galaxies was also already constrained in \cite{Kusiak_2022} for each of the \emph{unWISE} samples by fitting their auto-power spectra and cross-spectra with \emph{Planck} CMB lensing into the standard HOD model \cite{Zehavi_11, Zheng_2007}. However, that analysis was only performed on relatively large scales, up to $\ell_{\rm{max}}=1000$. In our work, we want to consider correlations involving these galaxy samples out to much smaller scales. Therefore we re-fit the model considered in \cite{Kusiak_2021} to the \emph{unWISE} galaxy auto-correlation data only, as they have the most constraining power. This choice is also motivated by the treatment of the shot noise in \cite{Kusiak_2021}, where the authors allowed it to be negative due to its effective indistinguishability from the galaxy -- galaxy one-halo term on large scales and complications related to the mask treatment.  Physically, we expect that the shot noise should be near the value given by the inverse of the number density of galaxies for each sample (Table~\ref{table:unwise}).  When re-fitting the HOD model out to small scales, we thus enforce this expectation.  We describe the procedure of re-fitting the \emph{unWISE} HOD below, and the HOD constraints in Appendix \ref{subsubsec:unwise_hod}. 

\subsection{Other Fields in the mm-wave Sky}

In this work, we model the sky comprising the primary CMB, tSZ, kSZ, and CIB fields, as well as the radio point source contribution.  All fields are modeled analytically, predominantly via the halo model.  We provide details of our theoretical predictions in Appendix~\ref{sec:spectra}, and give more details on specific choices of the modeling in Appendix~\ref{sec:models}, which we also summarize below. 

For the CIB, we consider two models, fitted to the standard Shang \textit{et al.}~(2012) \cite{Shang_2012} CIB halo model, described in Appendix~\ref{subsec:cib_hm}. The first model, which we refer to as the H13 model, was determined by fitting the \emph{Herschel} Multi-tiered Extragalactic Survey (\emph{HerMES}) \cite{Oliver_2012} data from the SPIRE instrument aboard the \emph{Herschel Space Observatory} \cite{Pilbratt_2010}, and was described in \cite{Viero_2013_hermes}. The second CIB model \cite{Planck:2013cib}, which we refer to as P14 fitted the \emph{Planck} nominal mission CIB power spectrum results. Details of these two CIB models are presented in Appendices~\ref{subsec:cib_hermes} and \ref{subsec:cib_planck}. For our purposes, an important distinction is that these two CIB models predict different correlation coefficients between the CIB and the \emph{unWISE} galaxies (see Fig.~\ref{fig:corr_cib_g}).  We choose to consider both models because they encompass our measurements of the CIB -- \emph{unWISE} cross-power spectra, as described in Appendix~\ref{subsec:lenz_cibg} and shown in Fig.~\ref{fig:cibXg_data}. The physical reason why these two models behave differently is because they have different values of the underlying halo model parameters, e.g., the population of CIB-sourcing halos or the evolution of the dust properties.

For the tSZ field, we use the standard halo model approach, with the pressure profile from Battaglia {\it et al.}~(2012)~\cite{Battaglia_2012} (the ``AGN feedback model at $\Delta=200$'' from their Table~1), as described in detail in Appendix~\ref{subsec:tsz_hm} and Appendix~\ref{subsec:battaglia_y}. For the kSZ power spectrum, we use the sum of simulated kSZ power spectra from the simulations of \cite{Battaglia_2010_kSZ} and \cite{Battaglia_2013_kszpatchy}, accounting for the late-time and patchy kSZ contributions, respectively (see Appendix~\ref{subsec:ksz}). We also consider the contribution from radio point sources, for which we assume a simple analytical power-law model, described in Appendix~\ref{subsec:radio}. 

We compute not only the auto-correlations of the various mm-wave sky components at the frequencies considered in this work, but also cross-correlations between these fields (except for the radio point sources), as well as with the \emph{unWISE} galaxies, if required by the new methods described above.


\section{Results}
\label{sec:results}

\begin{figure}[htb]
    \centering
    \includegraphics[scale=0.4]{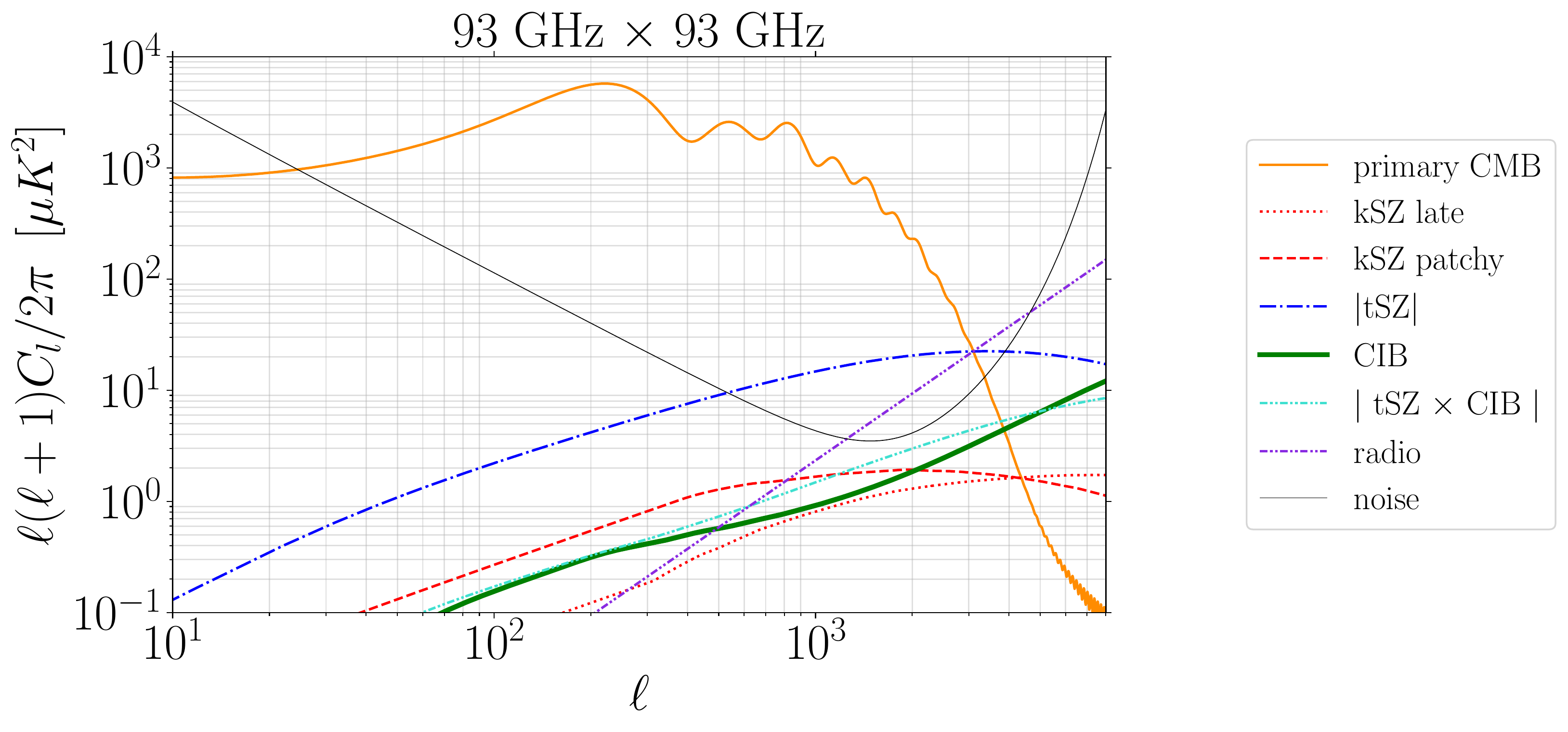}
    \includegraphics[scale=0.4]{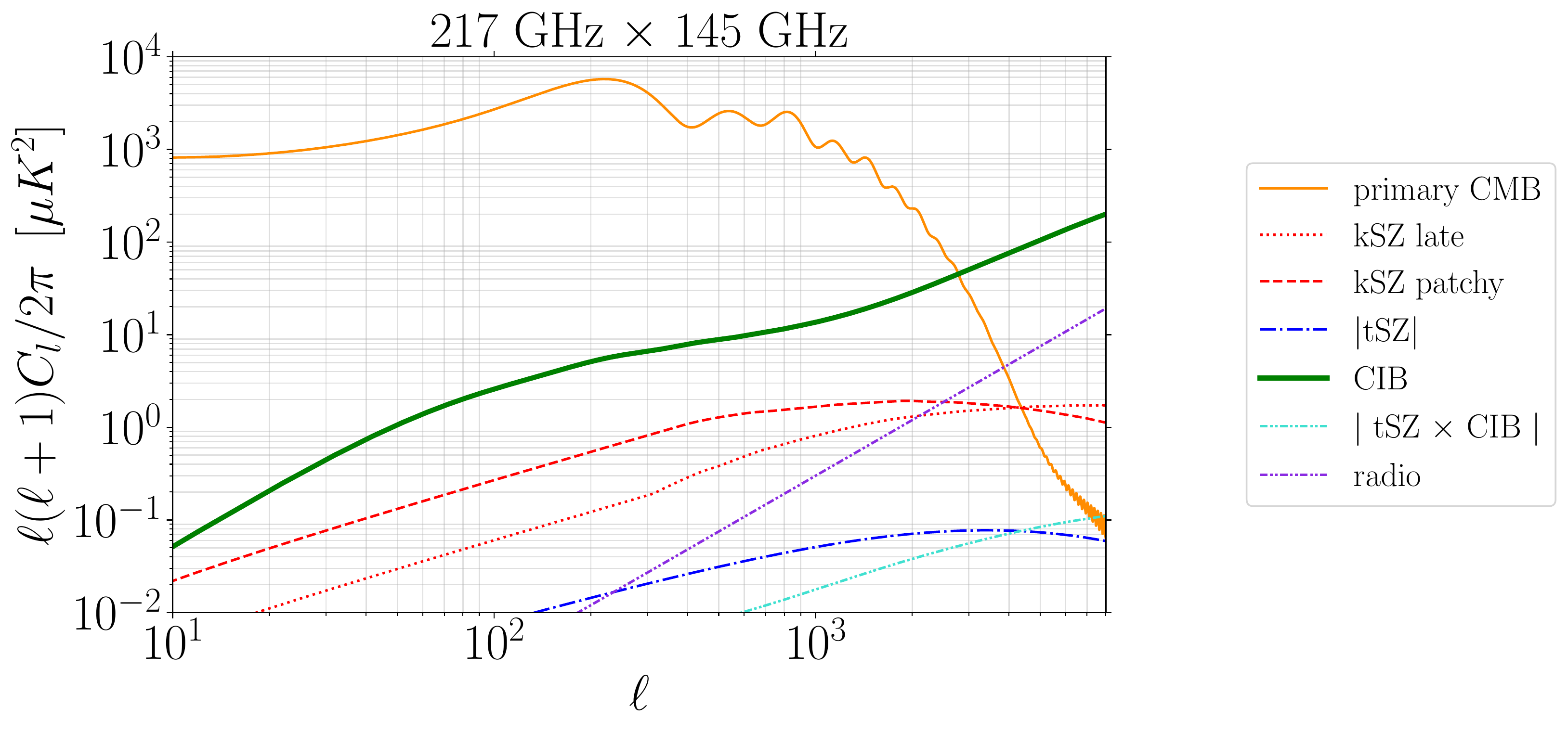}
    \includegraphics[scale=0.4]{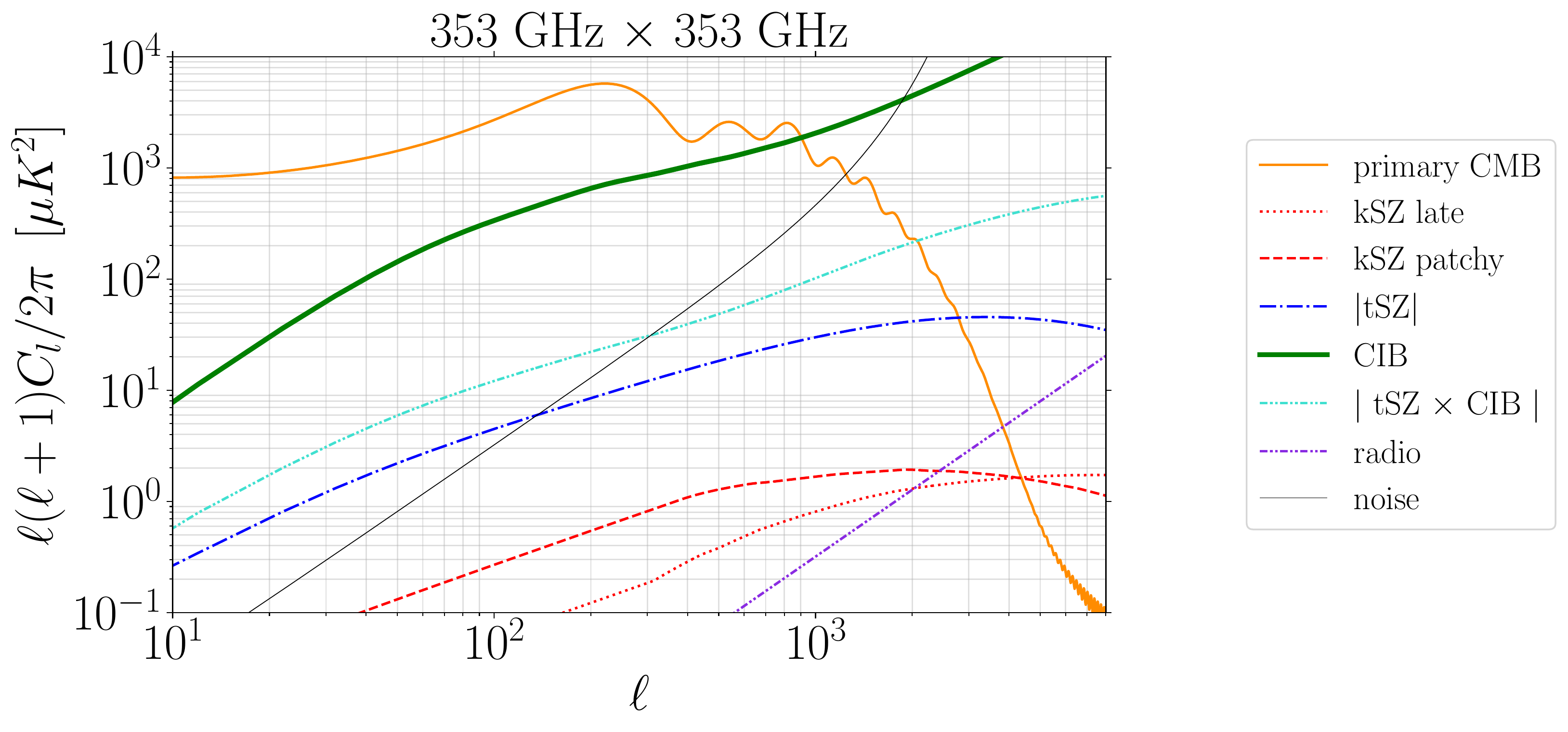}
    \caption{Microwave sky components considered in this analysis (lensed primary CMB, kSZ effect, tSZ effect, CIB (H13 CIB model), radio, and tSZ -- CIB correlation), along with the instrumental noise at 93 GHz (SO), 217 GHz $\times$ 145 GHz (\emph{Planck} $\times$ SO), and 353 GHz (\emph{Planck}), from top to bottom. The frequency-independent components are the primary CMB (solid orange curves) and the kSZ effect, broken down into its ``patchy'' \cite{Battaglia_2013_kszpatchy} (dashed red) and post-reionization, or ``late'' \cite{Battaglia_2010_kSZ} (dotted red) contributions. The frequency-dependent components, including the tSZ (blue dash-dotted; we show its absolute value), CIB (thick solid green), and tSZ $\times$ CIB (densely dash-dot-dotted cyan; we show its absolute value) are calculated with {\tt class\_sz}, and the radio (dash-dot-dotted purple) contribution is computed following Eq.~\eqref{eq:radio}. The noise curves (thin solid grey) are calculated according to specifications discussed in Appendix~\ref{subsec:noise} for the \emph{Planck} and SO frequencies. }
    \label{fig:all_comp}
\end{figure}

We model a sky containing the lensed primary CMB, the tSZ effect, the kSZ effect, the CIB (for both the H13 and P14 CIB models), radio sources, and detector and atmospheric noise, with power spectra described in Appendix~\ref{sec:spectra} and modeling choices discussed in Appendix~\ref{sec:models}, at eight frequencies: 93, 145, 225, and 280 GHz (SO) and 100, 143, 217, and 353 GHz (\emph{Planck}). The noise curves are shown in Fig.~\ref{fig:noise_curves}, and the noise properties (white noise and beam) are provided in Table \ref{table:noise}. The sky component power spectra are generated using \verb|class_sz| from $\ell_{\mathrm{min}}=30$ to  $\ell_{\mathrm{max}}=10^{4}$ with an $\ell$-space binning of $\Delta \ell = 10$ and then interpolated with a cubic spline. We use an $\ell$-space binning of $\Delta \ell=20$ for covariance matrix calculations in our ILC methods (see Eq.~\eqref{eq:Rij}, for example).  We convert the units for all considered sky component spectra to $\mu \mathrm{K}^2$ (see Appendix~\ref{app:mbb} for details). We show all sky components of our model in Fig.~\ref{fig:all_comp} for selected frequencies. We compute the auto- and cross-correlations of the three \emph{unWISE} galaxy samples, as well as their cross-correlations with the tSZ and CIB fields, following the same prescription. 

\begin{table}[t]
    \centering
    \setlength{\tabcolsep}{8pt}
    \renewcommand{\arraystretch}{1.2}
    \begin{tabular}{ |c|c|c|c|c|c|c|c|c|c| } 
    \hline
      & CMB &  kSZ  &  tSZ  & CIB & Radio & $g$ & Noise & Constraints & Method\\  
      & & & & & & & & in ILC& Type\\
    \hline\hline
    Standard ILC &\checkmark &\checkmark &\checkmark &\checkmark &\checkmark &- &\checkmark & 1 &Baseline \\
    \hline
    Standard ILC &\checkmark &\checkmark &- &- &\checkmark &- &\checkmark & 1 &Idealized\\
    (no CIB or tSZ)*&&&&&&&&& \\
    \hline
    ILC with $g$ freq maps &\checkmark &\checkmark &\checkmark &\checkmark &\checkmark &\checkmark &\checkmark &1 &New\\
    \hline
    de-(CIB+tSZ) &\checkmark &\checkmark &\checkmark &\checkmark &\checkmark &\checkmark &\checkmark &1 &New\\
    \hline
    ILC with constraint on $g^{\mathrm{CIB}}$ &\checkmark &\checkmark &\checkmark &\checkmark &\checkmark &$\nearrow$ &\checkmark &2 &New\\
    \hline
    ILC (deproj tSZ) &\checkmark &\checkmark &$\nearrow$ &\checkmark &\checkmark &- &\checkmark &2 &Baseline\\
    \hline
    ILC &\checkmark &\checkmark &$\nearrow$ &- &\checkmark &- &\checkmark &2 &Idealized\\
    (deproj tSZ, no CIB)*&&&&&&&&& \\
    \hline
    de-CIB (deproj tSZ) &\checkmark &\checkmark &$\nearrow$ &\checkmark &\checkmark &\checkmark &\checkmark &2 &New\\
    \hline
    ILC &\checkmark &\checkmark &$\nearrow$ &$\nearrow$ &\checkmark &- &\checkmark &3 &Baseline\\
    (deproj tSZ and CIB)&&&&&&&&& \\
    \hline
    ILC with constraint on $g^{\mathrm{CIB}}$  &\checkmark &\checkmark &$\nearrow$ &\checkmark &\checkmark &$\nearrow$ &\checkmark &3 &New\\
    (deproj tSZ) & & & & & & & & &\\
    \hline
    ILC with constraints on $g^{\mathrm{CIB}}$  &\checkmark &\checkmark &\checkmark &\checkmark &\checkmark &$\nearrow$ &\checkmark &3 &New\\
    and $g^{\mathrm{tSZ}}$ & & & & & & & & &\\
    \hline
    \end{tabular}
    \caption{Summary of the methods considered in this work: standard ILC (\S \ref{subsec.standard_ILC}); standard ILC with no CIB or tSZ included in sky model (for idealized comparison only); ILC with the LSS tracer fields $g^a$ as additional ``frequency" maps (\S \ref{sec.method2}); de-(CIB+tSZ) applied to a standard ILC map (end of \S \ref{sec:deCIB}); ILC with constraint requiring zero cross-correlation with $g^{\mathrm{CIB}}$ (\S \ref{subsec:method1_withoutdeprojtsz}); constrained ILC with tSZ deprojected (\S \ref{subsubsec.one_deproj_comp}); constrained ILC with tSZ deprojected and no CIB in the sky model (for idealized comparison only); de-CIB applied to a tSZ-deprojected ILC map (\S \ref{sec:deCIB}); constrained ILC with both tSZ and CIB deprojected (\S \ref{subsubsec.two_deproj_comps}); tSZ-deprojected ILC with an additional constraint requiring zero cross-correlation with $g^{\mathrm{CIB}}$ (\S \ref{subsec:method1_withdeprojtsz}); and ILC with two additional constraints requiring zero cross-correlation with $g^{\mathrm{CIB}}$ and $g^{\mathrm{tSZ}}$ (\S \ref{subsec:method1_withdeprojtsz}). The components we consider are the lensed primary CMB, kSZ signal, tSZ signal, CIB, radio contribution, and noise, as well as  ``$g$'' (the galaxy overdensity maps or some linear combination thereof). Here $\checkmark$ indicates that the component is included in the sky model, - indicates that the component is not included in the sky model, and $\nearrow$ indicates that the component is included in the sky model but deprojected in the ILC (or in the ILC portion of the method if the method has multiple parts). For $g$, $\nearrow$ indicates that $g$ is explicitly (spatially) deprojected in the ILC, $\checkmark$ indicates that the galaxy overdensities are incorporated in the cleaning method in some other way, and - indicates that the galaxy overdensities are not incorporated in the cleaning method. The penultimate column gives the number of constraints used in the ILC (or ILC portion of the method), including the signal preservation constraint. The final column gives the type of method: ``baseline" (standard method from the literature that uses only the mm-wave sky maps [no LSS tracer maps] and that uses only frequency-space information [no spatial information]), ``idealized" (method in which the CIB and/or tSZ contaminants are not included in the model from the outset), and ``new" (new methods developed in this work). The asterisk (*) indicates an idealized method that cannot be applied to actual data and is included in this work only for comparison purposes. } 
    \label{table:methods_components}
\end{table}

\begin{figure}[t]
    \centering
    \includegraphics[trim={0.23cm 0 0 0}, scale=0.74]{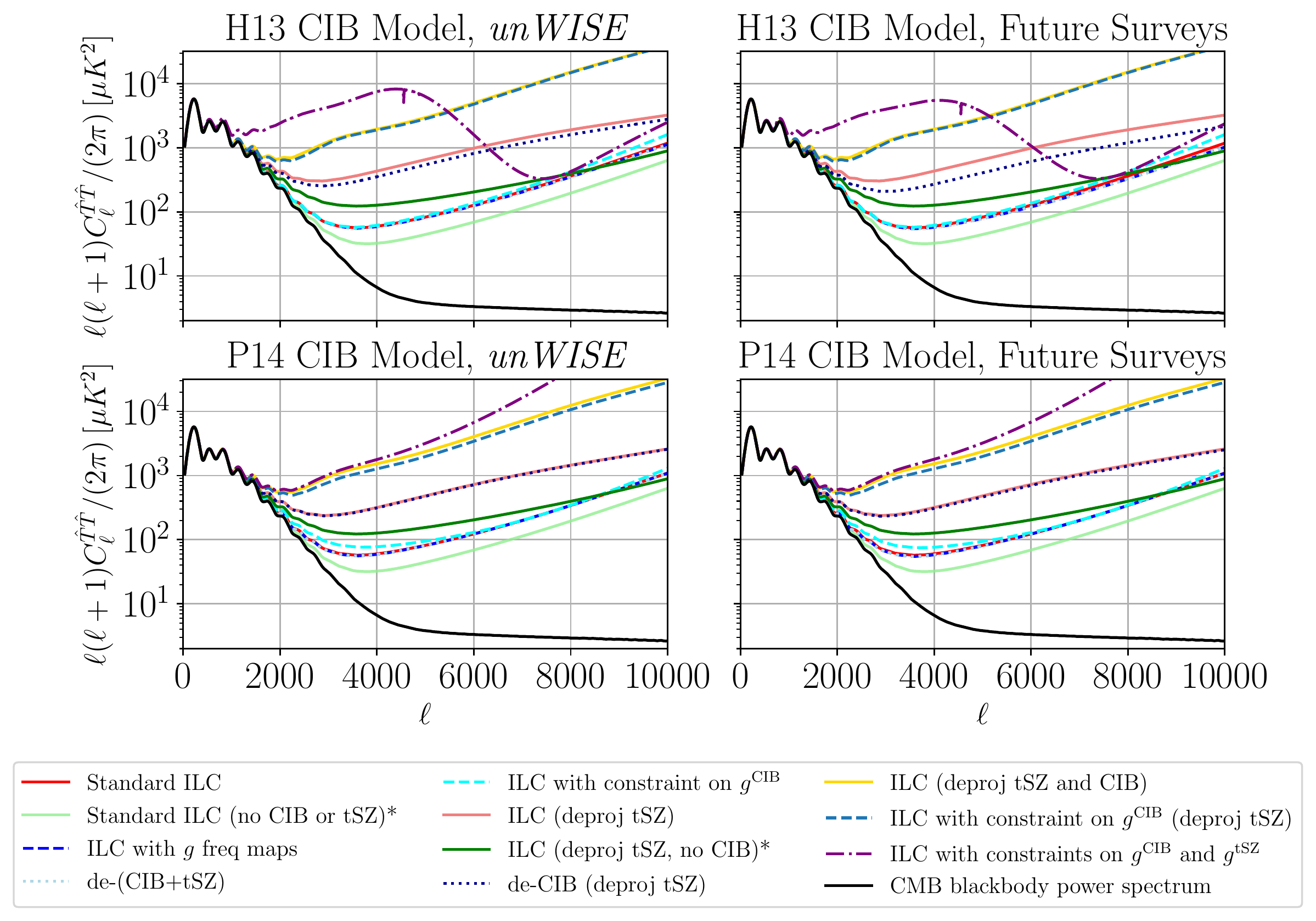}
    \caption{Auto-power spectra of final cleaned maps for both the H13 (top) and P14 (bottom) CIB models, for various cleaning methods (see Table~\ref{table:methods_components}), plotted as $D_{\ell}^{\hat{T}\hat{T}}=\ell(\ell+1)C_{\ell}^{\hat{T}\hat{T}}/(2\pi)$ where $\hat{T}$ is the cleaned map. ``Baseline" methods are shown as solid curves in shades of red and yellow. ``Idealized" results are shown as solid curves in shades of green. Our new methods are shown as dashed or dotted curves in shades of blue and purple. The asterisk (*) indicates an idealized method that cannot be applied to actual data and is included in this work only for comparison purposes. Our new methods (dashed and dotted curves in shades of blue and purple) decrease the auto-spectra of the cleaned maps from those of the baseline methods (solid curves in shades of red and yellow) toward those of the idealized results (solid curves in shades of green), with more significant impacts using the H13 CIB model since the CIB and galaxies have higher correlation in the H13 CIB model than in the P14 CIB model. In particular, one can compare the methods as follows: de-CIB (deproj tSZ) [new] and ILC with constraint on $g^{\rm CIB}$ [new] with ILC (deproj tSZ) [baseline] and ILC (deproj tSZ, no CIB) [idealized]; ILC with $g$ freq maps [new] and de-(CIB+tSZ) [new] with standard ILC [baseline] and standard ILC (no CIB or tSZ) [idealized]; and ILC with constraint on $g^{\rm CIB}$ (deproj tSZ) [new] and ILC with constraints on $g^{\rm CIB}$ and $g^{\rm tSZ}$ [new] with ILC (deproj tSZ and CIB) [baseline]. The total CMB blackbody temperature power spectrum (lensed primary CMB and patchy + late-time kSZ signal) is also shown for comparison (solid black). The left panels show results using \emph{unWISE}, while the right panels show forecasts for a future LSS survey, such as \emph{Euclid}, that contains $\approx$ 1.5 billion galaxies, assuming the same redshift distribution and HOD as for \emph{unWISE}. This figure illustrates the impact of the galaxy shot noise values on our results and the associated improvements with future data.}
    \label{fig:Tclean_auto_spectra}
\end{figure}

\begin{figure}[t]
    \centering
    \includegraphics[scale=0.73]{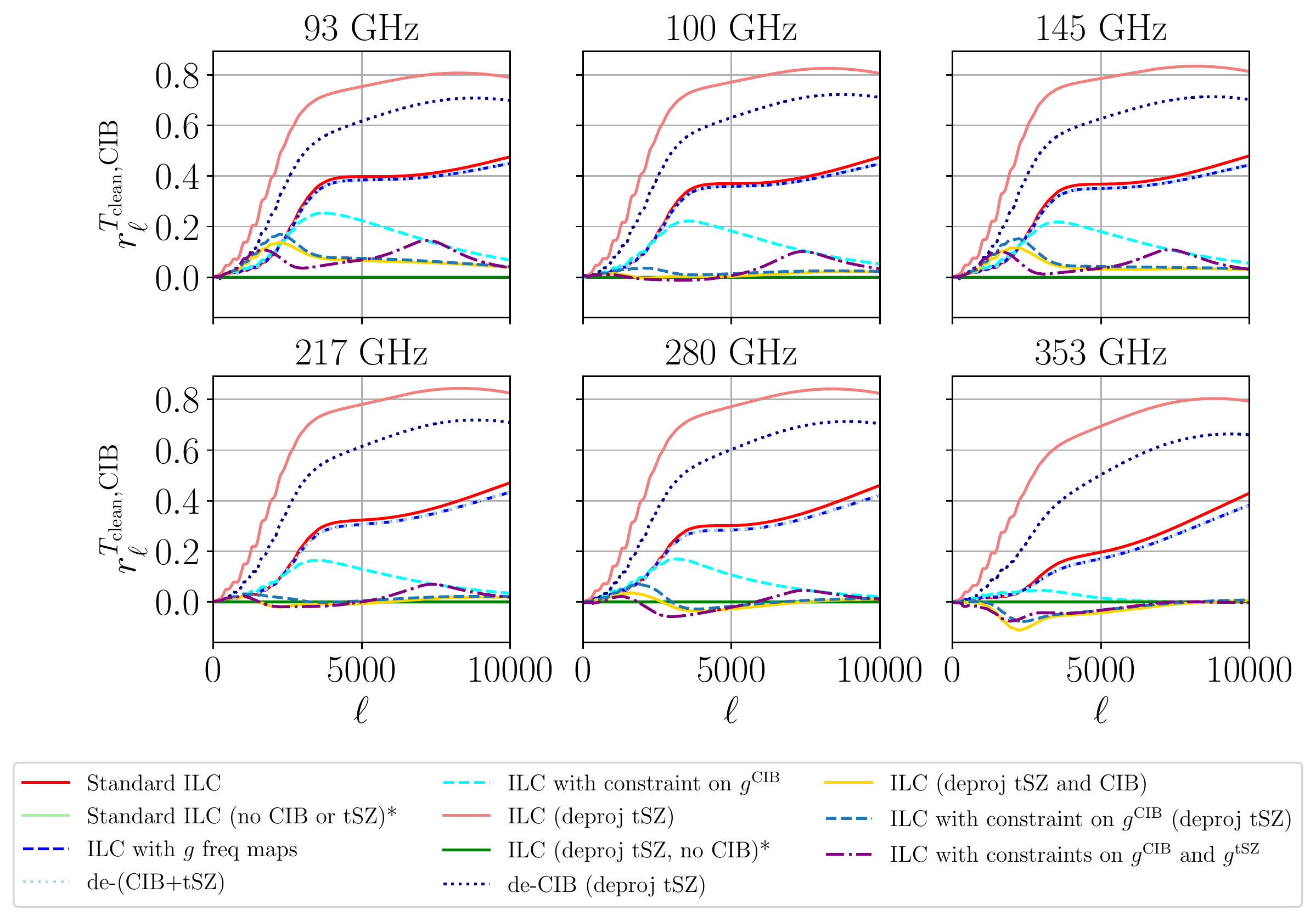}
    \caption{Correlation coefficients of the cleaned maps with the CIB at six frequencies, for each method. The asterisk (*) indicates an idealized method that cannot be applied to actual data and is included in this work only for comparison purposes. Our new methods (dashed and dotted curves in shades of blue and purple) decrease the correlation coefficient of the cleaned map with the CIB from those of the baseline methods (solid curves in shades of red and yellow) toward those of the idealized methods (solid curves in shades of green). In particular, one can compare the methods as follows: de-CIB (deproj tSZ) [new] and ILC with constraint on $g^{\rm CIB}$ [new] with ILC (deproj tSZ) [baseline] and ILC (deproj tSZ, no CIB) [idealized]; ILC with $g$ freq maps [new] and de-(CIB+tSZ) [new] with standard ILC [baseline] and standard ILC (no CIB or tSZ) [idealized]; and ILC with constraint on $g^{\rm CIB}$ (deproj tSZ) [new] and ILC with constraints on $g^{\rm CIB}$ and $g^{\rm tSZ}$ [new] with ILC (deproj tSZ and CIB) [baseline]. Results for 143 and 225 GHz are omitted for concision.}
    \label{fig:TcleanxCIB_corr}
\end{figure}

\begin{figure}[t]
    \centering
    \includegraphics[scale=0.72]{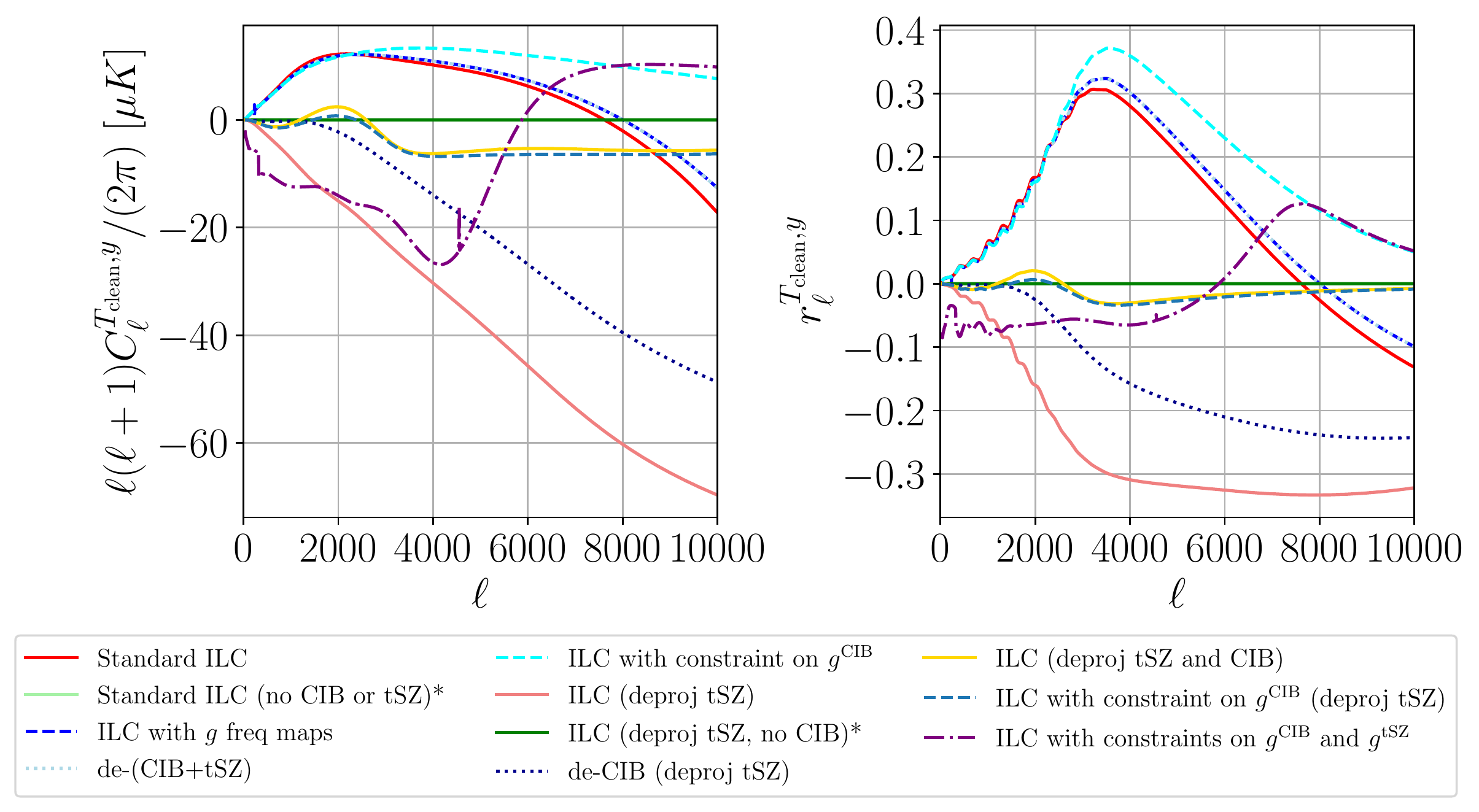}
    \caption{\textbf{Left}: Cross-power spectra of the cleaned maps with the tSZ field (in dimensionless Compton-$y$ units), plotted as $\ell(\ell+1)C_{\ell}^{T_{\rm clean},y}/(2\pi)$ (in $\mu \mathrm{K}$). \textbf{Right}: Correlation coefficients of the cleaned maps with the tSZ field. The asterisk (*) indicates an idealized method that cannot be applied to actual data and is included in this work only for comparison purposes. One can compare the methods as follows: de-CIB (deproj tSZ) [new] and ILC with constraint on $g^{\rm CIB}$ [new] with ILC (deproj tSZ) [baseline] and ILC (deproj tSZ, no CIB) [idealized]; ILC with $g$ freq maps [new] and de-(CIB+tSZ) [new] with standard ILC [baseline] and standard ILC (no CIB or tSZ) [idealized]; and ILC with constraint on $g^{\rm CIB}$ (deproj tSZ) [new] and ILC with constraints on $g^{\rm CIB}$ and $g^{\rm tSZ}$ [new] with ILC (deproj tSZ and CIB) [baseline]. }
    \label{fig:TcleanxtSZ}
\end{figure}

We compare the results of the methods proposed in \S \ref{sec:deCIB}-\S \ref{sec:method1} in terms of their ability to recover the CMB blackbody temperature power spectrum, including the primary CMB and the kSZ signal. We group the methods according to the number of constraints in the ILC, including the signal preservation constraint; the number of deprojected components is one less than the total number of constraints.  Within each group, we include our new methods as well as a ``baseline" method and an ``idealized" method for comparison. The baseline methods are methods that use only the mm-wave sky maps (no LSS tracer maps) and that use only frequency-space information (no spatial information), \textit{e.g.}, standard ILC. Such methods have been used elsewhere previously in the literature. The idealized methods are methods in which the CIB and/or tSZ contaminants are not included in the sky model from the outset; such methods cannot be applied to actual data and are included here only for comparison purposes. 

We note that some of our baseline and new methods that involve constraints in the ILC (either spectral or spatial) are analogous to bias-hardening in other contexts in which one tries to orthogonalize the reconstructed signal with respect to some contaminant, regardless of the noise penalty incurred (similar to how the term is used in descriptions of bias-hardened CMB lensing reconstruction, \textit{e.g.}, Ref.~\cite{Toshiya2013}). Such methods are partially bias-hardened due to the constraints.

The methods involving one constraint in the ILC --- the signal preservation constraint --- are:
\begin{itemize}
    \item standard ILC [baseline] (see \S \ref{subsec.standard_ILC});
    \item standard ILC with no CIB or tSZ effect included in the sky model [idealized]*\footnote{\label{footnote:asterisk}The ``idealized" methods considered in this work cannot be applied to actual data and are included here only for comparison purposes.} (see \S \ref{subsec.standard_ILC});
    \item de-(CIB+tSZ) applied to a standard ILC map [new method] (see \S \ref{sec:de_CIBplustSZ} and note that we do not consider de-CIB applied to a standard ILC map on its own, as one would naturally clean both CIB and tSZ signals when applying this method to a standard ILC map);
    \item ILC with $g$ maps as additional ``frequency" maps [new method] (see \S \ref{sec.method2}).
\end{itemize}
The methods involving two constraints in the ILC (one deprojected component) are:
\begin{itemize}
    \item ILC with deprojected tSZ component [baseline] (see \S \ref{subsubsec.one_deproj_comp});
    \item ILC with deprojected tSZ component and no CIB included in the sky model [idealized]*\footnote{See footnote \ref{footnote:asterisk}.} (see \S \ref{subsubsec.one_deproj_comp});
    \item de-CIB applied to a tSZ-deprojected ILC map [new method] (see \S \ref{sec:de_CIB_only});
    \item ILC with a zero-tracer-correlation constraint for $g^{\mathrm{CIB}}$ [new method] (see \S \ref{subsec:method1_withoutdeprojtsz} and note that $g^{\mathrm{CIB}}$ is the linear combination of tracer maps with maximal correlation to $a^{\mathrm{CIB}}(\hat{n})$, which is described in Appendix \ref{app:mbb}).\footnote{As explained in \S \ref{sec:method1}, for the de-(CIB+tSZ) method, we are able to simultaneously clean the CIB and tSZ signals by finding the combination of tracers that has maximal correlation with the combined CIB+tSZ signal at each frequency. For ILC methods involving a single zero-tracer-correlation constraint, we cannot simultaneously clean the CIB and tSZ signals (which have different effective SEDs) by using this single additional constraint because our linear combination of tracers must be frequency-independent to fit the form of Eq.~\eqref{eq.g_constraint}. }
\end{itemize}
The deprojection of a component in the above methods will incur some noise penalty in the resulting cleaned map. This penalty will be even more pronounced in the following methods, which involve three constraints (two deprojected components):
\begin{itemize}
    \item ILC with deprojected CIB and tSZ components [baseline] (see \S \ref{subsubsec.two_deproj_comps});
    \item ILC with deprojected tSZ component and a zero-tracer-correlation constraint for $g^{\mathrm{CIB}}$ [new method] (see \S \ref{subsec:method1_withdeprojtsz});
    \item ILC with zero-tracer-correlation constraints for both $g^{\mathrm{CIB}}$ and $g^{\mathrm{tSZ}}$ [new method] (see \S \ref{subsec:method1_withdeprojtsz} and note that $g^{\mathrm{tSZ}}$ is the linear combination of tracer maps with maximal correlation to the Compton-$y$ field).
\end{itemize}
For direct deprojection of the CIB component in an ILC, we model the CIB SED as a modified blackbody with an effective dust temperature of 24.0 K and spectral index of 1.2 (see Ref.~\cite{Madhavacheril_2020} and Table 9 of Ref.~\cite{Planck:2013cib}) for the H13 CIB model. For the P14 model, we model the CIB SED as a modified blackbody with an effective dust temperature of 20.0 K and spectral index of 1.45, as determined in Appendix \ref{app:mbb}. Importantly, we note that, in reality, the CIB decorrelates across frequencies, so such ``effective CIB SEDs" will not be exact, preventing the CIB field in a real data analysis from being fully deprojected in a constrained ILC approach (see Fig.~\ref{fig:cib_corr} for CIB correlation coefficients in the H13 and P14 models).

Summaries of each of the methods, including the sky components included in each method, are detailed in Table \ref{table:methods_components}. For a review of the standard and constrained ILC procedures, see \S \ref{sec.ILC}. For further details on the new methodologies, see \S \ref{sec:deCIB} for de-CIB and de-(CIB+tSZ), \S \ref{sec.method2} for the ILC with the tracers $g$ included as additional ``frequency'' maps, and \S \ref{sec:method1} for the ILC with the additional constraint(s) of requiring zero cross-correlation of the cleaned map with the $g^{\mathrm{CIB}}$ and/or $g^{\mathrm{tSZ}}$ tracer fields.

All evaluation metrics are derived analytically. As described in Appendix~\ref{sec:models}, the CIB -- galaxy cross-correlation measurements are encompassed by the H13 and P14 CIB models. Thus, the true forecasts of our methods likely lie somewhere in between the forecasts using these two models. For most of the remainder of the main text of the paper, we provide results for the H13 CIB model, with the results for the P14 CIB model given in Appendix \ref{app:planck_results}.

Part of the de-CIB method involves finding the linear combination of the \emph{unWISE} blue, green, and red samples (i.e., $g^{\rm OPT}$ in Eq.~\eqref{eq.gILC}) that is maximally correlated with the final ILC map, which contains residual CIB from each of the input frequency maps. We note that the blue and green samples exhibit higher correlation with the CIB, resulting in higher values of the coefficients for these samples. This is because, although the redshift distribution of CIB sources most aligns with that of the red sample~(compare the \emph{unWISE} redshift distributions in Fig.~\ref{fig:dndz_COSMOS} and the CIB redshift kernel in, \textit{e.g.},~\cite{Planck:2013cib,Sabyr_2022}), its number density of galaxies is significantly lower than that of the blue or green samples. 

We note that a prediction of our models is that at low $\ell$ both the \emph{unWISE} blue and green samples have high correlation with the CIB, but relatively low correlation with each other.  When optimally combined, this may artificially produce a correlation coefficient for $g^{\mathrm{OPT}}$ with the ILC map that is greater than unity. To mitigate this effect, for multipoles where the correlation coefficient of the ILC map with $g^{\mathrm{OPT}}$ is greater than 0.95, we replace the linear combination of samples with the single sample with the highest correlation with the ILC map. In reality, such effects would not be a problem since the correlation of two physical fields cannot be greater than unity (and moreover, even in our theoretical models, such issues only appear at very low $\ell$ where our methods are not useful in practice, as CMB measurements there are already cosmic-variance-limited, and the kSZ power spectrum cannot be measured due to the large primary CMB sample variance).

We repeat a similar procedure with the de-(CIB+tSZ) method. The coefficients $c^i_{a,\ell}$ for the  \emph{unWISE} blue, green, and red samples with maximal CIB+tSZ correlation at each frequency are determined, and we follow a similar replacement strategy at low $\ell$, where if the correlation of the (CIB+tSZ) field with the optimal combination at some frequency is greater than 0.95, we replace the linear combination of samples with the single sample with the highest (CIB+tSZ) correlation at each frequency.

We assess how well the methods remove contaminant signals for both the H13 and P14 CIB models. Fig.~\ref{fig:Tclean_auto_spectra} shows the auto-power spectra of the cleaned maps from each of our methods using \emph{unWISE} in the left panels. The trade-off between component deprojection and noise in the resulting cleaned map is evident from this plot; as more constraints are added to the ILC, the noise in the cleaned map increases significantly (although often less so for our new zero-tracer-cross-correlation constraint than for the traditional frequency-space constraints).  Of note is the auto-spectrum of a map produced by ILC with constraints on both $g^{\mathrm{CIB}}$ and $g^{\mathrm{tSZ}}$, which displays an irregular shape for the H13 CIB model; this result will be discussed further in \S \ref{sec:Discussion}. Fig.~\ref{fig:Tclean_auto_spectra_ratios} in Appendix \ref{app:additional_plots} shows a few of the key ratios of cleaned map auto-spectra from the methods we consider. We note a significant reduction in the cleaned map auto-spectrum when applying the de-CIB method to a tSZ-deprojected ILC map. We also note a similar (though less pronounced) effect when adding the galaxy maps as additional ``frequency" maps to a standard ILC or when using the de-(CIB+tSZ) method. Moreover, we compare the ratios of some of our new methods to each other, finding that the ILC with galaxy maps as additional frequency maps and de-(CIB+tSZ) have the lowest cleaned map auto-spectra of the new methods. In fact, those two methods give almost exactly the same results in our calculations.

A crude estimate of the signal-to-noise ratio (SNR) for the kSZ power spectrum can be defined as
\begin{equation}
    \label{eq.snr}
    \rm SNR = \sqrt{\sum_{\ell}\frac{(C_{\ell}^{\mathrm{kSZ}})^2}{\sigma^2(C_{\ell}^{\mathrm{tot}})}} \qquad \text{with} \qquad \sigma^2(C_{\ell}^{\mathrm{tot}}) = \frac{2}{(2\ell+1)f_{\mathrm{sky}}} (C_{\ell}^{\mathrm{tot}})^2 \, ,
\end{equation}
where $C_\ell^{\rm kSZ}$ is the kSZ power spectrum, $C_\ell^{\rm tot}$ is the total cleaned map power spectrum, $f_{\rm sky}$ is the unmasked fraction of the sky, and we sum over multipoles $2 \leq \ell \leq 10^4$. Using our models and various cleaning methods, we obtain the kSZ power spectrum SNR results given in Table \ref{table:snr}, assuming $f_{\rm sky} \simeq 0.4$, similar to that from SO \cite{SO2019}. The relevant columns here are those labeled ``using \emph{unWISE}," which provide the SNRs given current data, \textit{i.e.}, the \emph{unWISE} galaxies. The latter two columns give projections of the methods for future LSS surveys, which will be discussed further in \S \ref{sec:forecasts}. We note that these are not forecasts of actual SNRs that would realistically be obtained (as marginalization over additional parameters in the sky model is necessary); rather, these numbers are simply useful for comparing the \textit{relative} SNR from the various methods.

We also estimate the SNR for the total blackbody (CMB+kSZ) power spectrum, found by replacing $C_\ell^{\mathrm{kSZ}}$ in Eq.~\eqref{eq.snr} with $C_\ell^{\mathrm{CMB}} + C_\ell^{\mathrm{kSZ}}$ and summing over multipoles $100 \leq \ell \leq 10^4$ to avoid scales where the ISW signal is large and would lead to biases in our methods.  These SNRs are provided in Table \ref{table:snr_blackbody}. We note that these approximations of SNR for the total blackbody power spectrum are likely fairly accurate, as the only major approximation in this case is our neglect of Galactic foregrounds, which are relevant at $\ell \lesssim 1000$.

To assess how well each method performs in removing CIB contamination, we compute the correlation coefficients of the cleaned maps with the CIB field at each frequency (Fig.~\ref{fig:TcleanxCIB_corr}), as well as their cross-power spectra (Fig.~\ref{fig:TcleanxCIB_spectra} of Appendix \ref{app:additional_plots}). (For concision, in this figure and the figures that follow, we omit separate plots for 143 GHz and 225 GHz since they are very similar to the 145 GHz and 217 GHz plots, respectively.) As expected, the idealized results (green curves) have near-zero correlation with the CIB, whereas the baseline methods of standard ILC and tSZ-deprojected ILC (red curves) display significant correlation with the CIB. Our new methods (dashed and dotted blue and purple curves) lower the correlation of the cleaned maps with the CIB from the baseline method results, with the constrained ILC methods requiring zero $g^{\mathrm{CIB}}$ and/or $g^{\mathrm{tSZ}}$ tracer cross-correlation performing particularly well. We note that the ILC with both CIB and tSZ signals deprojected does not have a correlation coefficient of exactly zero simply because a CIB SED has to be assumed for the deprojection (and this effective SED is not the exact same as the actual SED of our modeled CIB power spectra). This is a realistic representation of what may happen when using this method on actual data, as the effective CIB SED is not known exactly and the CIB decorrelates across frequencies \cite{Madhavacheril_2020}. Moreover, we note that since the CIB and tSZ fields have non-zero correlation, for methods that do not explicitly deproject the tSZ field, some of the residual correlation between the cleaned map and the CIB may actually be due to the residual tSZ signal.

\begin{figure}[t]
    \centering
    \includegraphics[scale=0.72]{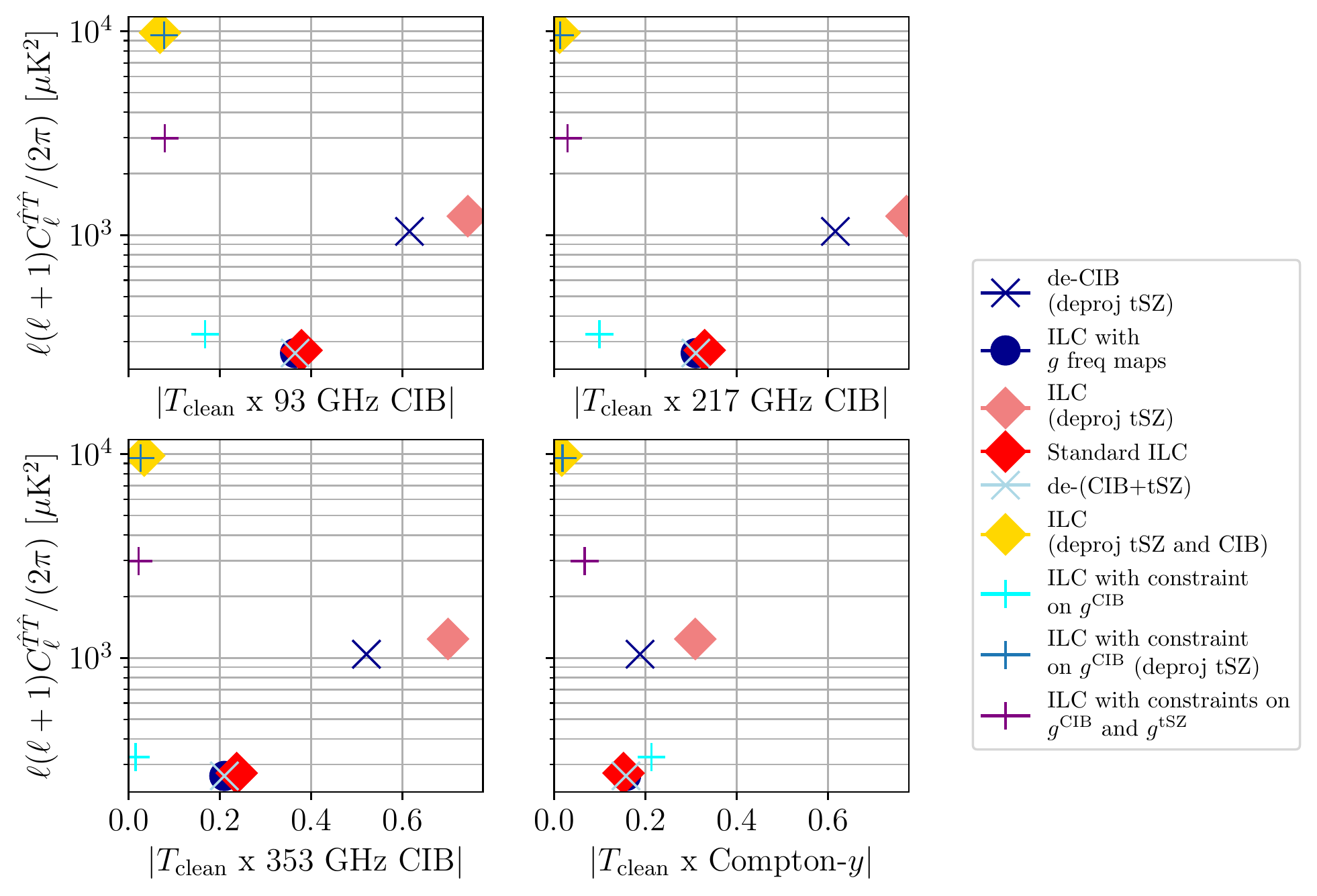}
    \caption{Bias-variance trade-off for different methods, assessed in terms of trade-offs among the noise penalty, CIB correlation, and tSZ correlation. The mean $D_\ell^{\hat{T}\hat{T}}$ over multipoles $2000 \leq \ell \leq 10000$ is plotted against the mean absolute $T_{\rm clean} \times \mathrm{CIB}$ or $T_{\rm clean} \times \mathrm{Compton}-y$ correlation coefficient, also computed over multipoles $2000 \leq \ell \leq 10000$. Correlation coefficients are shown for three select CIB frequencies as well as the Compton-$y$ field. Note that this figure is specific to our simulated joint SO + \textit{Planck} noise configuration. One can compare the methods as follows: de-CIB (deproj tSZ) [new] and ILC with constraint on $g^{\rm CIB}$ [new] with ILC (deproj tSZ) [baseline] and ILC (deproj tSZ, no CIB) [idealized]; ILC with $g$ freq maps [new] and de-(CIB+tSZ) [new] with standard ILC [baseline] and standard ILC (no CIB or tSZ) [idealized]; and ILC with constraint on $g^{\rm CIB}$ (deproj tSZ) [new] and ILC with constraints on $g^{\rm CIB}$ and $g^{\rm tSZ}$ [new] with ILC (deproj tSZ and CIB) [baseline]. } 
    \label{fig:tradeoff_hermes}
\end{figure}

We also compute the cross-power spectra and correlation coefficients of the cleaned maps with the tSZ field (Fig.~\ref{fig:TcleanxtSZ}). Since the tSZ field is correlated with the CIB, any residual CIB in the maps leads to nonzero correlation with the tSZ signal, even if the tSZ signal is deprojected, as shown in Fig.~\ref{fig:TcleanxtSZ}. From these results, we note a trade-off: our new methods (that do not explicitly deproject the tSZ signal) must choose between removing more of the CIB signal or more of the tSZ signal; those that result in lower correlation of the cleaned map with the CIB result in higher correlation of the cleaned map with the tSZ signal, and vice versa.  This is particularly notable for the constrained ILC requiring zero $g^{\mathrm{CIB}}$ tracer cross-correlation, which does well for CIB removal and much less well for tSZ removal, presumably due to the \emph{unWISE} galaxies' lower correlation with the Compton-$y$ field (and also the fact that in that method we are finding the linear combination of samples to specifically optimize for CIB cleaning). It will therefore be crucial to select a method that achieves a balance of cleaning both the CIB and tSZ contaminants (or optimizing to clean whichever is more relevant to a given analysis), while also not significantly adding noise to the cleaned map. Fig.~\ref{fig:tradeoff_hermes} evaluates how well each method achieves such a balance, showing the mean $D_\ell^{\hat{T}\hat{T}}$ over multipoles $2000 \leq \ell \leq 10000$ plotted against the mean absolute correlation coefficients of $T_{\rm clean}$ with the CIB and Compton-$y$ fields computed over the same multipoles. We note that this figure is specific to our simulated joint SO + \textit{Planck} noise configuration. Assessing the bias-variance trade-off for other instrument configurations is beyond the scope of this work.

\section{Forecasts Using Future LSS Surveys}
\label{sec:forecasts}

Future LSS surveys with wide-field survey instruments, such as \emph{Euclid} \cite{Euclid2011, Euclid:2021}, \emph{Roman} \cite{Spergel2013wfirst, Akeson2019, Yung:2022}, and Rubin LSST \cite{LSST}, will measure properties of over a billion galaxies. Since the shot noise of the samples scales inversely with the number of galaxies, the use of galaxies from these future surveys will significantly decrease the shot noise on the galaxy map power spectra, thereby increasing the correlation of the CIB with the galaxy samples (assuming the redshift distributions are unchanged for the sake of this argument). This increased correlation will, in turn, increase the magnitude of the improvements using our new methods, particularly at high $\ell$. In this section, we provide approximate forecasts for these improvements, given the specifications of \emph{Euclid}.

\emph{Euclid} is expected to image 1.5 billion galaxies out to high redshifts $z>2$ \cite{Euclid2011}, whereas the \emph{unWISE} catalog contains 500 million galaxies. As an illustrative exercise, we assume the same HOD as for \emph{unWISE} and a redshift distribution of galaxies into blue, green, and red that is identical to that in Table \ref{table:unwise}. Therefore, we divide each of the shot noise values in Table \ref{table:gg_shotnoise} by a factor of three to obtain the results in the right panels of Fig.~\ref{fig:Tclean_auto_spectra}. We see significant improvements to the de-CIB method when the galaxy shot noise is lower.

Estimates of the SNRs for the kSZ power spectrum for such a future survey are shown in Table \ref{table:snr} in the latter two columns, for both the H13 and P14 CIB models. SNR estimates for the total CMB blackbody temperature power spectrum for such a future survey are also shown, in Table \ref{table:snr_blackbody}. Most of our methods involving the galaxy maps exhibit higher SNRs with these future surveys than with current data, as expected since the number of galaxies will significantly increase. Of note is that for the methods involving ILC with zero-tracer-cross-correlation constraints on $g^{\mathrm{CIB}}$ and/or $g^{\mathrm{tSZ}}$, the SNR forecasts sometimes decrease in spite of the increase in the number of galaxies. This is because the galaxy shot noise only enters into the calculations for these methods for the determination of $g^{\mathrm{CIB}}$ and/or $g^{\mathrm{tSZ}}$, as explained in \S \ref{sec:method1}; slight changes in the shot noise could affect the relative contributions of the different tracer maps to the linear combination with optimal correlation with the CIB and/or tSZ signals. Importantly, we note that such a decrease in the SNR would not occur in reality. This is because it is unphysical to change the shot noise without changing the cross-correlation of the CIB and tSZ fields with the tracers as well. Our forecasts do not account for changes in these cross-correlations, and therefore, likely underestimate the improvements we would obtain using the new methods with larger samples of galaxies. In particular, although the ILC methods with zero-tracer-cross-correlation constraints are sometimes projected to have lower SNR in Tables \ref{table:snr} and \ref{table:snr_blackbody} with larger galaxy samples, these SNRs would likely increase in reality due to higher correlation of the tracers with the CIB and tSZ fields when there are more galaxies included. As described in \S \ref{sec:method1}, aside from the determination of coefficients for the different tracer maps, these methods have an implicit dependence on the shot noise, but one would have to recompute the theoretical prediction of cross-correlations with the tracers in a proper, self-consistent way as the sample changes (which we are not currently doing, for simplicity). Recomputing these theoretical cross-correlations would likely result in higher SNRs than those in Tables \ref{table:snr} and \ref{table:snr_blackbody} for the other methods as well using these future LSS surveys.

\begin{table}[t]
    \centering
    \setlength{\tabcolsep}{8pt}
    \renewcommand{\arraystretch}{1.2}
    \begin{tabular}{ |c|c|c|c|c| } 
    \hline
      & H13 CIB Model &  P14 CIB Model & H13 CIB Model & P14 CIB Model \\  
      & (using \emph{unWISE})&  (using \emph{unWISE}) & (future LSS survey) & (future LSS survey) \\ 
    \hline\hline
    Standard ILC &115.32 &115.25 &115.32 &115.25  \\
    \hline
    Standard ILC (no CIB or tSZ)* &202.74 &202.74 &202.74 &202.74 \\
    \hline
    ILC with $g$ freq maps &116.80 &116.75 &120.04 & 118.37\\
    \hline
    de-(CIB+tSZ) &116.80 &116.75 &120.04 &118.37\\
    \hline
    ILC with constraint on $g^{\mathrm{CIB}}$ &110.67 &95.43 &110.58 &96.65\\
    \hline
    ILC (deproj tSZ) &19.43 &25.17 &19.43 &25.17\\
    \hline
    ILC (deproj tSZ, no CIB)* &62.21 &62.21 &62.21 &62.21\\
    \hline
    de-CIB (deproj tSZ) &23.22 &25.52 &29.25 &25.93\\
    \hline
    ILC (deproj tSZ and CIB) &5.94 &7.33 &5.94 &7.33\\
    \hline
    ILC with constraint on $g^{\mathrm{CIB}}$ &6.36 &8.47 &6.34 &8.43\\
    (deproj tSZ) & & & &\\
    \hline
    ILC with constraints on $g^{\mathrm{CIB}}$ &18.74 &6.66 &19.72 &6.60\\
    and $g^{\mathrm{tSZ}}$ & & & &\\
    \hline
    \end{tabular}
    \caption{Crude estimate of the signal-to-noise ratio for the kSZ power spectrum (Eq.~\eqref{eq.snr}) in each of the methods considered in this work, for both the H13 and P14 CIB models. The asterisk (*) indicates an idealized method that cannot be applied to actual data and is included in this work only for comparison purposes.  We show results using current \emph{unWISE} data and also results for a future LSS survey containing three times more galaxies than \emph{unWISE}, such as \emph{Euclid} (we assume the same HODs and redshift distributions as for \emph{unWISE}). We note that the overall magnitudes of these numbers are rough estimates; we present these numbers solely to compare the \textit{relative} magnitudes of the SNR for the various methods, CIB models, and LSS surveys. Of note is that the de-CIB method applied to a tSZ-deprojected ILC map has a significantly higher SNR than simply a tSZ-deprojected ILC map, with improvements in SNR up to 20\% using \emph{unWISE} and projected improvements up to 50\% for future LSS surveys.}
    \label{table:snr}
\end{table}

\begin{table}[t]
    \centering
    \setlength{\tabcolsep}{8pt}
    \renewcommand{\arraystretch}{1.2}
    \begin{tabular}{ |c|c|c|c|c| } 
    \hline
      & H13 CIB Model &  P14 CIB Model & H13 CIB Model & P14 CIB Model \\  
      & (using \emph{unWISE})&  (using \emph{unWISE}) & (future LSS survey) & (future LSS survey) \\ 
    \hline\hline
    Standard ILC &1168.84 &1156.47 &1168.84 &1156.47  \\
    \hline
    Standard ILC (no CIB or tSZ)* &1326.73 &1326.73 &1326.73 &1326.73 \\
    \hline
    ILC with $g$ freq maps &1170.30 &1168.32 &1172.69 & 1178.59\\
    \hline
    de-(CIB+tSZ) &1170.30 &1168.32 &1172.62 &1178.59\\
    \hline
    ILC with constraint on $g^{\mathrm{CIB}}$ &1166.39 &1046.06 &1166.25 &1053.80\\
    \hline
    ILC (deproj tSZ) &798.37 &834.22 &798.37 &834.22\\
    \hline
    ILC (deproj tSZ, no CIB)* &946.23 &946.23 &946.23 &946.23\\
    \hline
    de-CIB (deproj tSZ) &828.71 &836.47 &857.11 &838.48\\
    \hline
    ILC (deproj tSZ and CIB) &686.67 &720.92 &686.67 &720.92\\
    \hline
    ILC with constraint on $g^{\mathrm{CIB}}$ &700.39 &737.77 &699.57 &736.97\\
    (deproj tSZ) & & & &\\
    \hline
    ILC with constraints on $g^{\mathrm{CIB}}$ &492.59 &710.77 &503.45 &709.39\\
    and $g^{\mathrm{tSZ}}$ & & & &\\
    \hline
    \end{tabular}
    \caption{Estimate of the signal-to-noise ratio for the total blackbody (CMB+kSZ) temperature power spectrum in each of the methods considered in this work, for both the H13 and P14 CIB models. The asterisk (*) indicates an idealized method that cannot be applied to actual data and is included in this work only for comparison purposes.  We show results using current \emph{unWISE} data and also results for a future LSS survey containing three times more galaxies than \emph{unWISE}, such as \emph{Euclid} (we assume the same HODs and redshift distributions as for \emph{unWISE}). Of note is that the de-CIB method applied to a tSZ-deprojected ILC map has a higher SNR than simply a tSZ-deprojected ILC map, with improvements in SNR up to 4\% using \emph{unWISE} and projected improvements up to 7\% for future LSS surveys.}
    \label{table:snr_blackbody}
\end{table}


\section{Discussion}
\label{sec:Discussion}

In this work, we have developed new tSZ- and CIB-removal methods using LSS tracers to enhance detection of the total CMB blackbody temperature power spectrum, including the kSZ signal (which dominates on small angular scales). We have utilized the crucial fact that the cross-correlation of the LSS tracers with the primary CMB vanishes, as it also does with the kSZ signal due to the equal probability of the line-of-sight electron velocity being positive and negative. However, these LSS tracers are highly correlated with the CIB and tSZ signals. Specifically, we used the \emph{unWISE} galaxy samples as the LSS tracers. To forecast the performance of our methods, we have analytically modeled the microwave sky comprised of the lensed primary CMB, kSZ effect, tSZ effect, CIB (considering two possibilities for the CIB model, H13 and P14), radio sources, and detector + atmospheric noise at eight frequencies from 93 GHz to 353 GHz, for a combined SO and \emph{Planck}-like experiment. The specifications of each of the six new methods presented in this work (de-CIB applied to an ILC with deprojected tSZ component; de-(CIB+tSZ) applied to standard ILC; ILC with the tracers $g$ as additional ``frequency" maps; ILC with a zero-tracer-correlation constraint using $g^{\mathrm{CIB}}$; tSZ-deprojected ILC with a zero-tracer-correlation constraint using $g^{\mathrm{CIB}}$; and ILC with zero-tracer-correlation constraints using both $g^{\mathrm{CIB}}$ and $g^{\mathrm{tSZ}}$) are discussed in \S \ref{sec:deCIB}, \S \ref{sec.method2}, and \S \ref{sec:method1} and summarized in Table~\ref{table:methods_components}. The final cleaned map auto-spectra are shown in Fig.~\ref{fig:Tclean_auto_spectra} for the two CIB models considered, and for both \emph{unWISE} and future surveys, and the estimates of the relative kSZ power spectrum and total CMB blackbody temperature power spectrum SNRs are presented in Table~\ref{table:snr} and Table~\ref{table:snr_blackbody}, respectively.

The SNRs and Fig.~\ref{fig:Tclean_auto_spectra} show a clear consequence of explicit deprojection of components in the ILC.  The spectra cluster into three groups: those with no deprojected sky components, those with only the tSZ component deprojected, and those with both the CIB and tSZ components deprojected. When both the CIB and tSZ components are deprojected, the resulting ILC map auto-spectrum blows up at high $\ell$ due to the significant noise penalty incurred, making this an unfavorable method, although it is robust against residual contamination from these fields (see Figs.~\ref{fig:TcleanxCIB_spectra}--\ref{fig:TcleanxtSZ}).  In addition to the noise penalty, one further disadvantage of this method is that we have to assume some specific CIB SED, which is not known from first principles.

For the first two groups of methods (no deprojection and tSZ deprojection), we assess how well our new methods do in moving away from the baseline method in that group toward the idealized method in that group. The de-CIB method exhibits a larger move toward the idealized method in its group (tSZ deprojection) than do the de-(CIB+tSZ) method and the ILC with $g$ as ``frequency'' maps method in their group (no deprojection); in fact, applying de-CIBing to a tSZ-deprojected ILC map results in a SNR improvement of 20\% with \emph{unWISE} data and a projected 50\% improvement for future LSS surveys, when compared to a tSZ-deprojected ILC map with no special CIB-removal procedure.  This is a non-negligible improvement that can be obtained with the \emph{unWISE} data that are already on hand.  Moreover, the methods involving an ILC with an explicit additional zero-tracer-correlation constraint using $g^{\mathrm{CIB}}$ fall somewhere between these groups, depending on the CIB model. This spatial deprojection of $g^{\mathrm{CIB}}$ results in less additional noise in the cleaned map than does the spectral deprojection of the tSZ or CIB component.

The approach of using external galaxy maps in a standard minimum-variance CMB ILC could also be extended using other foreground maps, \textit{e.g.}, neutral atomic hydrogen (HI) maps~\cite{HI4PI_2016,Peek2018}.
However, in the method of requiring zero cross-correlation of our ILC map with the $g$ maps, we are effectively putting in prior knowledge that the $g$ maps are indeed purely tracers of a contaminant field, and contain no contribution from the signal of interest, which is a valid assumption for the CMB and kSZ signals at $\ell \gtrsim 100$.  We are thus effectively performing a ``spatial deprojection'' of any component traced by the $g$ field from the final ILC map.  For this procedure to work, it is crucial that we have multiple frequencies (as for any ILC method) and that the SED of the contaminants (the CIB and tSZ fields, in the case of LSS tracer $g$ maps) is different from that of the preserved component (the CMB/kSZ signal here).  It is also important that the $g$ maps are not correlated with the signal of interest in our ILC (which is violated by the ISW signal at low multipoles in this case, as discussed earlier).

Based on our forecasts, the methods of using ILC with a zero-tracer-cross-correlation constraint on $g^{\mathrm{CIB}}$ (with no tSZ deprojection) and using ILC with galaxy maps as additional ``frequency'' maps (also with no tSZ deprojection) appear to have the best balance of removing contamination while maintaining the kSZ and CMB blackbody power spectra SNRs. However, the idea of applying delensing-like techniques to this problem is still useful, as the de-CIB method can be used in conjunction with an ILC that deprojects the tSZ component, without significantly decreasing the kSZ and CMB blackbody power spectrum SNRs. We further note that, if the goal is simply to produce a cleaned map with minimal correlation to the CIB and tSZ contaminants (without the concern of added noise or decreased SNR), our method of tSZ-deprojected ILC with the additional zero-tracer-cross-correlation constraint on $g^{\mathrm{CIB}}$ performs just as well as the method of ILC with the explicit deprojection of both the CIB and tSZ components.  Moreover, the former results in a cleaned map with less noise than the latter, and also allows one to not have to model a specific CIB SED. 

Of note is the ILC with additional zero-tracer-correlation constraints using both $g^{\mathrm{CIB}}$ and $g^{\mathrm{tSZ}}$ (purple dash-dot curve). From Fig.~\ref{fig:Tclean_auto_spectra}, the auto-spectrum of the resulting ILC map from this method looks vastly different for the H13 and P14 CIB models. We note that in Figs.~\ref{fig:Tclean_auto_spectra}, \ref{fig:TcleanxCIB_corr}, and \ref{fig:TcleanxtSZ}, the results from this method match fairly well with those from the other ILC methods with two deprojected components at low $\ell$, but at high $\ell$, the results from this method begin to converge with those for the ILC with the zero-tracer-correlation constraint using only $g^{\mathrm{CIB}}$ (dashed cyan curve). This is because the CIB and tSZ signals are highly correlated at high $\ell$ in the H13 CIB model, but not in the P14 CIB model, as shown in Fig.~\ref{fig:corr_tsz_g_cib}. When the CIB and tSZ signals are highly correlated, so too are $g^{\mathrm{CIB}}$ and $g^{\mathrm{tSZ}}$. As explained in \S \ref{sec:method1}, spatially deprojecting two highly correlated combinations of tracers allows both tracer combinations to be deprojected for the price of one in terms of the noise penalty resulting from the ILC procedure. 

An important note for all of the methods is that we can use these approaches on actual data without theoretical models of the different sky components, galaxy maps, and correlations among them. The theoretical models presented in this work are simply needed for forecasting the impacts of the different methods on kSZ and CMB blackbody power spectrum detection.  A slight exception to this is the determination of coefficients for the optimal linear combination of tracer maps, used in the de-CIB (or de-(CIB+tSZ)) and ILC with zero-tracer-correlation constraint methods, where correlations between the galaxies and CIB and tSZ signals are modeled to provide optimal results. Nevertheless, as discussed in \S \ref{sec:deCIB} and \S \ref{sec:method1}, even in these cases, slight model misspecifications would only affect the optimality of the methods by some small amount. To validate this claim, we experiment with running the de-CIB procedure on a sky model containing the H13 CIB field (along with all other mm-wave fields), but using the P14 CIB model to determine the optimal tracer combination.  We find only a $\approx 10\%$ difference in the auto-spectrum of the cleaned map, averaged over multipoles $2000 \leq \ell \leq 10000$, with even less discrepancy at $\ell < 2000$, when using the P14 CIB model to determine the optimal tracer combination, versus using the H13 CIB model for this purpose. We note that this is a ``worst-case scenario" example, since the H13 and P14 models are extremely different, and the two models encompass the true data that would be used to determine the optimal tracer combination in an actual analysis.

We note that our new methods would bias the cleaned maps on scales where the ISW effect or Rees-Sciama effect is significant, since those signals are correlated with galaxies, CIB, and the tSZ effect (\textit{e.g.}, \cite{Ferraro:2022, Ferraro:2014, Shajib:2016, Krolewski:2021isw, Planck:2015isw}).  Fortunately, the Rees-Sciama effect is always very small (well below the kSZ signal) \cite{Seljak:1995}, and the ISW effect is limited to very large scales \cite{Sachs:1967}. Thus, our methods work well at $200 \lesssim \ell \lesssim 10000$, ideal for constructing a CIB- and tSZ-cleaned map for kSZ analyses. We also note that these methods may be useful in the context of CMB temperature bispectrum estimation and associated constraints on primordial non-Gaussianity, which are susceptible to tSZ- and CIB-related foreground biases~\cite{Hill_2018,Coulton_2023}; however, we leave an investigation of applications to higher-order statistics to future work.

As previously mentioned, the H13 and P14 CIB models encompass the observational data, so the true forecasts of our new methods likely lie somewhere between those using the two CIB models. As evident from Fig.~\ref{fig:Tclean_auto_spectra}, our methods appear to be more effective for the H13 CIB model, for which the majority of results in \S \ref{sec:results} are shown (and for which the CIB is also more strongly correlated with the \emph{unWISE} galaxies than in the P14 CIB model, as can be seen from the correlation coefficients shown in Fig.~\ref{fig:corr_cib_g}). Nevertheless, the impacts of these methods on enhancing kSZ and CMB blackbody power spectrum measurements will likely be \textit{larger} than what is shown. As discussed in \S \ref{sec:forecasts}, this is because the galaxy shot noise is a significant factor in our results. Decreasing the shot noise increases the correlation of the CIB with the galaxy samples, thereby increasing the magnitude of the improvements with our new methods, particularly at high $\ell$. Since the shot noise scales inversely with the number of galaxies, our results will become even more significant with larger LSS surveys, such as \emph{Euclid}, \emph{Roman}, and Rubin LSST, in the future.

There are several areas where the current analysis could be improved. First, precisely constraining the exact CIB model and the \emph{unWISE} galaxy HOD would improve the forecasts presented in this work, which, as we noted, lie in between the P14 and H13 CIB models. Second, the CIB and tSZ contaminant-removal methods presented in this work would benefit from including models for additional sky components, such as the cross-correlations with the radio sources, as well as the recently studied extragalactic CO emission lines \cite{maniyar_2023}, whose cross-correlations with the CIB at 150 and 220 GHz tend to be on the order of the CIB -- tSZ correlations.

Next steps for this work would include validating the new methods using map-level simulations and further applying them to data.  When applying these methods to real data, it would be advantageous to perform ILC in the needlet basis \cite{Delabrouille2009} to obtain weights that vary as a function of both position and scale, thus providing a more optimal weighting scheme for non-Gaussian foregrounds such as the tSZ field.  Thus, future work would also include generalizing our new methods to the needlet basis.  While this would be trivial for some methods, such as an ILC with $g$ as additional ``frequency'' maps, other approaches would require more careful construction. For the data application, the \emph{unWISE} catalog, with its large number density and high redshift overlap with the CIB, seems particularly well-suited for the methods presented here, and, as noted before, it is not necessary to constrain its galaxy clustering model (\textit{e.g.}, HOD) for most of the methods, which is a particularly difficult task at high $\ell$.  

We note that gains with our methods could be even larger even using only current data, as we could use arbitrarily many different galaxy catalogs to increase correlation of the tracers with the CIB and tSZ contaminants, \emph{e.g.}, the Dark Energy Survey~\cite{2016MNRAS.460.1270D}, 2-Micron All-Sky Survey~\cite{2006AJ....131.1163S}, and the Baryon Oscillation Spectroscopic Survey~\cite{2013AJ....145...10D}, and even other LSS tracers like galaxy lensing or CMB lensing maps. Future surveys will significantly expand upon these samples, and thus yield even larger improvements when using our techniques.


\section{Acknowledgements}
\label{sec:acknow}

We thank Boris Bolliet for help with modifying \verb|class_sz| and Oliver Philcox for useful discussions. We thank Marcelo Alvarez and Fiona McCarthy for help with CIB modeling and Alex Krolewski for help with \emph{unWISE} modeling. We also thank Simone Ferraro, Emmanuel Schaan, and Bernardita Ried Guachalla for comments on the manuscript. Some of the results in this paper have been derived using the \verb|healpy| and \verb|HEALPix| packages \cite{healpy_paper1, healpy_paper2}. This research used resources of the National Energy Research Scientific Computing Center (NERSC), a U.S.~Department of Energy Office of Science User Facility located at Lawrence Berkeley National Laboratory. This research also used computing resources from Columbia University's Shared Research Computing Facility project, which is supported by NIH Research Facility Improvement Grant 1G20RR030893-01, and associated funds from the New York State Empire State Development, Division of Science Technology and Innovation (NYSTAR) Contract C090171, both awarded April 15, 2010. AK and JCH acknowledge support from NSF grant AST-2108536. This material is based upon work supported by the National Science Foundation Graduate Research Fellowship Program under Grant No.~DGE 2036197 (KMS). JCH acknowledges additional support from NASA grants 21-ATP21-0129 and 22-ADAP22-0145, DOE grant DE-SC00233966, the Sloan Foundation, and the Simons Foundation. 
\begin{appendices}

\section{Theoretical Models of the Component Auto- and Cross-Spectra}
\label{sec:spectra}

In this appendix, we present the theoretical models used for the sky components considered in this work: the tSZ, kSZ, and CIB fields; the galaxy halo occupation distribution (HOD); and the radio source contribution. For the tSZ, CIB, and galaxy fields we give analytical halo model expressions. For the kSZ power spectrum, we describe the simulations from which it was obtained, and for the radio contribution, we describe a simple analytical model assumed in this work.

\subsection{Angular Power Spectra Predictions in the Halo Model}
\label{subsec:hm_predictions}
In this subsection, we give the predictions for the component auto- and cross-power spectra in the halo model used in our analysis, \textit{i.e.}, the tSZ effect, CIB, and galaxy overdensity fields. These components are computed with the \verb|class_sz| code version 1.01 \cite{Bolliet:2017lha,Bolliet_2022}, an extension of \verb|class| \cite{CLASS} version 2.9.4, which enables halo model computations of various cosmological observables.  We describe the exact choice of selected parameters in Appendix~\ref{sec:models}.

The halo model is an analytical framework that statistically describes the matter density field and other cosmological observables (\textit{e.g.}, galaxy density or CIB). The halo model assumes that all matter exists in the form of dark matter ``halos''. For more details about this model we refer the reader to Refs.~\cite{Cooray:2002dia}, \cite{seljak2000}, and \cite{peacock2000}. In the halo model, the power spectrum is the sum of one-halo and two-halo terms, which account respectively for the correlations of dark matter particles (or some other field) within one halo and between two distinct halos.  Thus, we can write the angular power spectrum as
\begin{equation}
    C_\ell^{ij}=C_\ell^{ij,\mathrm{1h}} +  C_\ell^{ij,\mathrm{2h}},
    \label{eq:cl_ij}
\end{equation}
where $ C_\ell^{ij,\mathrm{1h}}$ ($C_\ell^{ij,\mathrm{2h}}$) is the one-halo (two-halo) term of the correlation between tracers $i$ and $j$ ($i$ and $j$ can be the same tracer). We use $i$ and $j$ to refer to specific tracers here, not frequency channels as in the earlier sections.

We compute the one-halo term by integrating the \emph{multipole-space kernels} of tracers $i$ and $j$, $u_\ell^i(M,z)$ and $u_\ell^j(M,z)$, over halo mass $M$ and redshift $z$:
\begin{equation}
    C_\ell^{ij,\mathrm{1h}}=\int_{z_\mathrm{1}}^{z_\mathrm{2}}  \mathrm{d} z \frac{\mathrm{d}^2 V}{\mathrm{d} z \mathrm{d} \Omega} \int_{M_\mathrm{1}}^{M_\mathrm{2}}  \mathrm{d}M \frac{\mathrm{d}n}{\mathrm{d}M} u_\ell^i(M,z) u_\ell^j(M,z) \,,
    \label{eq:cl1h_ij}
\end{equation}
where $\mathrm{d} V $ is the cosmological volume element, defined in terms of the comoving distance $\chi(z)$ to redshift $z$ as $\mathrm{d}V= \chi^2\mathrm{d}\chi=\frac{c\chi^2}{H(z)}(1+z)\mathrm{d}\ln(1+z)$, where $H(z)$ is the Hubble parameter and $c$ is the speed of light. Note that $\mathrm{d} \Omega $ is the solid angle of this volume element and $\mathrm{d}n/\mathrm{d}M$ is the differential number of halos per unit mass and volume, defined by the halo mass function (HMF), where in our analysis we use the Tinker \textit{et al.}~(2008) analytical fitting function~\cite{Tinker_2008}. 
 
The two-halo term of the power spectrum of tracers $i$ and $j$ is given by
\begin{equation}
C_\ell^{ij, \mathrm{2h}}=\int_{z_\mathrm{1}}^{z_\mathrm{2}}  \mathrm{d} z \frac{\mathrm{d}^2 V}{\mathrm{d} z \mathrm{d} \Omega}  \left| \int_{M_\mathrm{1}}^{M_\mathrm{2}}  \mathrm{d}M_i \frac{\mathrm{d}n}{\mathrm{d}M_i} b(M_i, z) u_\ell^i(M_i, z)  \right| \left| \int_{M_\mathrm{1}}^{M_\mathrm{2}} \mathrm{d}M_j \frac{\mathrm{d}n}{\mathrm{d}M_j}  b(M_j, z) u_\ell^j(M_j, z)  \right| P_{\mathrm{lin}}\left(\frac{\ell+\tfrac{1}{2}}{\chi},z\right),
\label{eq:cl2h_ij}
\end{equation}
where $P_{\mathrm{lin}}(k,z)$ is the linear matter power spectrum (computed with \verb|CLASS| within  \verb|class_sz|) and $b(M, z)$ is the linear bias describing the clustering of the two tracers (\textit{e.g.},~\cite{gatti2021crosscorrelation, pandey2021crosscorrelation}). We model the linear halo bias using the Tinker \textit{et al.} (2010) \cite{Tinker_2010} fitting function. In this work, we use $z_\mathrm{1}=0.005$
and  $z_\mathrm{2}=12$ for the redshift range.  

Note that in principle the mass limits in the integrals in Eqs.~\eqref{eq:cl1h_ij} and \eqref{eq:cl2h_ij}, $M_\mathrm{1}$ and $M_\mathrm{2}$, can be different for different tracers, and can also depend on redshift, \textit{i.e.}, we can introduce $M_\mathrm{1}^i (z)$ and $M_\mathrm{1}^j (z)$ (and similarly for $M_\mathrm{2} (z)$). In that case, in the one- and two-halo terms of the auto-correlations of tracer $i$, one will use those mass limits $M_\mathrm{1}^i (z)$ and $M_\mathrm{2}^i (z)$ instead. For the two-halo term of the cross-correlations between tracer $i$ and tracer $j$, each mass integral in Eq.~\eqref{eq:cl2h_ij} will be integrated over its own mass range, while for the one-halo term, the lower (upper) mass limit will be chosen as the maximum (minimum) of the  two values for tracer $i$ and tracer $j$, \textit{i.e.}, $M_\mathrm{1} (x) = {\rm max} \left( M_\mathrm{1}^i (z), M_\mathrm{1}^j (z) \right)$ and $M_\mathrm{2} (z) = {\rm min} \left( M_\mathrm{2}^i (z), M_\mathrm{2}^j (z) \right)$.  We introduce this possibility because one of the models of the CIB emission we consider in this work uses a redshift-dependent $M_\mathrm{1}^\mathrm{CIB} (z)$ (see \S~\ref{subsec:cib_hermes} for more details). However, in general, unless stated otherwise, we assume $M_\mathrm{1} = 7\times10^8 \, M_\odot/h$ and ${M_\mathrm{2}} = 3.5\times10^{15} \, M_\odot/h$, motivated by the mass range considered in \cite{Kusiak_2022} for the \emph{unWISE} galaxies.

In the following subsections, we present the expressions for the multipole-space kernels of the tracers of interest for us, \textit{i.e.}, the tSZ effect, CIB, and galaxy overdensity which constitute the building blocks of the one-halo and two-halo terms of the auto- and cross-correlations, according to Eqs.~\eqref{eq:cl1h_ij} and \eqref{eq:cl2h_ij} (note that we include the kSZ effect in our analysis, but the cross-correlation between kSZ and other tracers of LSS vanishes at first order due to the oscillating line-of-sight velocity, and therefore we do not consider any cross-correlations involving this field).

\subsubsection{HOD}
\label{subsec:hm_hod}

The halo occupation distribution (HOD) is a model within the larger halo model framework that describes the distribution of galaxies within dark matter halos. In the HOD model, there exist two types of galaxies: centrals, located at the center of each halo, and satellite galaxies, distributed within the host halo according to some specified prescription. In this work, we follow the HOD model of the \emph{unWISE} galaxies from \cite{Kusiak_2022}, with some small changes detailed in \S~\ref{subsubsec:unwise_hod}.

The HOD model used in \cite{Kusiak_2022} is a modified version of the widely used Zheng \textit{et al.} (2007) HOD model \cite{Zheng_2007, Zehavi_11}, with some adjustments following the DES-Y3 HOD analysis \cite{Zacharegkas2021}. In this model, the expectation value for the number of central galaxies $N_c$ in a halo of mass $M$ is given by
\begin{equation}
    N_c (M) =\frac{1}{2}\left[1+\mathrm{erf}\left(\frac{\log M - \log M_\mathrm{min}^\mathrm{HOD}}{\sigma_{\log M}}\right)\right] \,,
    \label{eq:N_c}
\end{equation}
where $M_\mathrm{min}^\mathrm{HOD}$ is the characteristic minimum mass of halos that can host a central galaxy,  $\sigma_{\mathrm{log} M}$, is the width of the cutoff profile \cite{Zheng_2007}, and erf is the error function. 

The expectation value for the number of satellite galaxies $N_s$ in a halo is given by a power law and coupled to $N_c$ in the following way:
\begin{equation}
    N_s(M) = N_c(M) \left[\frac{M-M_0}{M_1^\prime}\right]^{\alpha_s},
    \label{eq:N_s}
\end{equation}
where $\alpha_s$ is the index of the power law of the satellite profile, $M_0$ is the mass scale above which the number of satellites grows, and $M_1^\prime$ sets the amplitude.  

This standard HOD prescription consists of five free parameters: two for the central galaxies ($M_\mathrm{min}^\mathrm{HOD}$,  $\sigma_{\mathrm{log} M}$), and three for the satellite profile ($\alpha_s$, $M_0$, and $M_1^{\prime}$). The authors of \cite{Kusiak_2022} followed the DES-Y3 HOD model \cite{Zacharegkas2021}, and only chose to constrain four HOD parameters: $\sigma_{\rm{log}M}$, $\alpha_s$,  $M_\mathrm{min}^\mathrm{HOD}$, and $M_1^{\prime}$, setting $M_0=0$.

\subsubsection{Galaxy Overdensity}
\label{subsec:gal_hm}

The galaxy overdensity multipole-space kernel $u_\ell^g(M,z)$ is defined as
\begin{equation}
u_\ell^g(M,z)=W_g(z)\bar{n}_g^{-1}\left[N_c+ N_s u_\ell^m(M,z)\right],
\label{eq:ulg}
\end{equation}
where $N_c$ and $N_s$ are the HOD expectation value for the number of central and satellite galaxies (Eqs.~\eqref{eq:N_c} and \eqref{eq:N_s}), $\bar{n}_g$ is the mean number density of galaxies given by 
\begin{equation}
    \bar{n}_g(z) =\int_{M_\mathrm{min}}^{M_\mathrm{max}}\mathrm{d}M \frac{\mathrm{d}n}{\mathrm{d}M}\left(N_c+N_s\right),
    \label{eq:n_g}
\end{equation}
and $W_g(z)$ is the galaxy window function defined as 
\begin{equation}
    W_g(z)= \frac{H(z)}{c \chi^2(z)} \frac{1}{N_g^\mathrm{tot}}\frac{\mathrm{d}N_g}{\mathrm{d}z},\quad \mathrm{with} \quad N_g^\mathrm{tot}=\int \mathrm{d}z\frac{\mathrm{d}N_g}{\mathrm{d}z},
\label{eq:wgz}
\end{equation}
where $\frac{1}{N_g^\mathrm{tot}}\frac{\mathrm{d}N_g}{\mathrm{d}z}$ is the normalized galaxy distribution of the given galaxy sample, and $u_{\ell}^m$ is the Fourier transform of the dark matter density profile. We model $u_{\ell}^m$ using the standard Navarro, Frenk, and White (NFW) dark matter profile \cite{Navarro_1997}, with truncation at $r_{\mathrm{out}} = \lambda^{\rm{NFW}} r_{200c}$, described by an analytic formula \cite{Scoccimarro:2000gm} 
\begin{align}
    u_\ell^m (M,z) &= \left(\cos(q)[\mathrm{Ci}((1+\lambda^{\rm{NFW}} c_{200c})q)-\mathrm{Ci}(q)]+\sin(q) [\mathrm{Si}((1+ \lambda^{\rm{NFW}} c_{200c})q)-\mathrm{Si}(q)]- \frac{\sin(\lambda^{\rm{NFW}} c_{200c}q)}{(1+ \lambda^{\rm{NFW}} c_{200c})q)} \right) \nonumber
    \\& \qquad \times \frac{M}{\rho_{m,0}} f_{_\mathrm{NFW}}(\lambda^{\rm{NFW}} c_{200c})
    \label{eq:ulm}
\end{align}
where $\rho_{m,0}$ is the mean matter density at $z=0$, $\mathrm{Ci}(x)=\int_x^{\infty}\mathrm{d}t\cos(t)/t$ and $\mathrm{Si}(x)=\int_0^x\mathrm{d}t\sin (t)/t$ are the cosine and sine integrals, with the argument $q$ defined as $q=k \frac{r_{200c}}{c_{200c}} $, where $k=(\ell +1/2)/\chi$ is the wavenumber, $r_{200c}$ is the mass-dependent radius that encloses mass $M_{200c}$ (the mass enclosed within the spherical region whose density is 200 times the critical density of the universe), and $c_{200c}$ is the concentration parameter computed with the concentration-mass relation defined in Ref.~\cite{Bhattacharya_2013}. 
Finally, the NFW function $f_{_\mathrm{NFW}}$  is given by
\begin{equation}
f_{_\mathrm{NFW}}(x)=[\ln(1+x)-x/(1+x)]^{-1}.
\end{equation} 

The galaxy multipole-space kernel $u_\ell^g(M,z)$ defined in Eq.~\eqref{eq:ulg} will enter all cross-correlations involving galaxies, as well as the two-halo term of the auto-correlation $C_\ell^{gg}$, according to Eqs.~\eqref{eq:cl1h_ij} and \eqref{eq:cl2h_ij}; however, the one-halo term of $C_\ell^{gg}$ needs to be modified. Following Section~2.2 in Ref.~\cite{vandenBosch2013}, we use the second moment of the galaxy multipole-space kernel, $u_\ell^g(M,z)$, not simply the square of Eq.~\eqref{eq:ulg}, which is given by (see Eqs.~15 and 16 in Ref.~\cite{Koukoufilippas:2019ilu})
\begin{equation}
\langle |u_\ell^g(M,z)|^2 \rangle=W_g(z)\bar{n}_g^{-2}\left[N_s^2 u_\ell^m(M,z)^2 + 2N_s u_\ell^m(M,z)\right],
\label{eq:ulg^2}
\end{equation}
where $N_s$ is the expectation value for the number of satellites, given in Eq.~\eqref{eq:N_s}, and $\bar{n}_g$ is the mean number density of galaxies (Eq.~\eqref{eq:n_g}). Thus, the one-halo term for the galaxy auto-correlation is given by
\begin{equation}
    C_\ell^{gg,\mathrm{1h}}=\int_{z_\mathrm{1}}^{z_\mathrm{2}}  \mathrm{d} z \frac{\mathrm{d}^2 V}{\mathrm{d} z \mathrm{d} \Omega} \int_{M_\mathrm{1}}^{M_\mathrm{2}}  \mathrm{d}M \frac{\mathrm{d}n}{\mathrm{d}M} \langle |u_\ell^g(M,z)|^2 \rangle , 
\label{eq:clgg1h}
\end{equation}
while the two-halo term is 
\begin{equation}
    C_\ell^{gg,\mathrm{2h}}=\int_{z_\mathrm{1}}^{z_\mathrm{2}}  \mathrm{d} z \frac{\mathrm{d}^2 V}{\mathrm{d} z \mathrm{d} \Omega} \left|\int_{M_\mathrm{1}}^{M_\mathrm{2}}  \mathrm{d}M \frac{\mathrm{d}n}{\mathrm{d}M} b(M,z) u_\ell^g(M,z) \right|^2 P_{\mathrm{lin}}\left(\frac{\ell+\tfrac{1}{2}}{\chi},z\right). \label{eq:clgg2h}
\end{equation}

For all correlations involving galaxy overdensity, we must also take into account the galaxy lensing magnification contribution, which arises from the fact that the luminosity function of galaxies is steep at the faint end, near the threshold for detection. The lensing magnification is a real, not observational, effect; therefore it must be included in the analytical halo model. It is quantified by the logarithmic slope of the galaxy number counts as a function of apparent magnitude $m$ near the magnitude limit of the survey defined as $s = \frac{\mathrm{d log_{10}} N}{\mathrm{d}m}$ (we will discuss relevant values for the \emph{unWISE} galaxies later in the text).

The observed galaxy overdensity field $g^{\mathrm{obs}}$ is the sum of the true overdensity field $g$ and the galaxy lensing magnification field $\mu_{g}$.
As an example, the observed cross-correlation of galaxy overdensity and the Compton-$y$ parameter includes the lensing magnification term $C_{\ell}^{y \mu_g}$:
\begin{equation}
    C_{\ell}^{y g, \rm{ obs}} = C_{\ell}^{ y g} +  C_{\ell}^{y \mu_g}.
\end{equation}
For the auto-correlation of a galaxy sample, the lensing magnification will appear both in the cross-correlation with galaxy overdensity, as well as in its own auto-correlation:
\begin{equation}
    C_{\ell}^{gg, \rm{ obs}} = C_{\ell}^{ gg} +  2 C_{\ell}^{ g\mu_g} +
    C_{\ell}^{ \mu_g \mu_g}. 
\end{equation} 

Each correlation involving the lensing magnification can be similarly written in the halo model as a sum of the one-halo and two-halo terms, according to Eqs.~\eqref{eq:cl1h_ij} and \eqref{eq:cl2h_ij}.  The multipole-space kernel for galaxy lensing magnification $u_\ell^{\mu_g}$ can be written as 
\begin{equation}
    u_\ell^{\mu_g}(M,z) = (5s-2) W_{\mu_g}(z) u^m_\ell(M,z),
    \label{eq:ulmu}
\end{equation}
where $u_{\ell}^m$ is the dark matter profile defined in Eq.~\eqref{eq:ulm} and the lensing magnification bias window function $W_\mathrm{\mu_{_{g}}}$ is
\begin{equation}
 W_\mathrm{\mu_{_{g}}}(z)=\frac{3}{2}\frac{\Omega_\mathrm{m}(H_0/c)^2}{\chi^2(z)}(1+z) \chi(z) I_g(z)\quad\mathrm{with}\quad I_g(z)=\int_z^{z_\mathrm{max}}\mathrm{d}z_g \frac{1}{N_g^\mathrm{tot}}\frac{\mathrm{d}N_g}{\mathrm{d}z}\frac{\chi(z_g)-\chi(z)}{\chi(z_g)},
 \label{eq:wmu}
\end{equation}
where $\chi(z_g)$ is the comoving distance to galaxies at redshift $z_g$ and $\frac{1}{N_g^\mathrm{tot}}\frac{\mathrm{d}N_g}{\mathrm{d}z}$ is the normalized galaxy distribution of the given galaxy sample from Eq.~\eqref{eq:wgz}. 

 For all components considered in this work that involve galaxy overdensity, we assume that their auto- and cross-correlations include their respective lensing magnification contributions, without explicitly saying so.

In this work, we compute not only galaxy auto-correlations but also cross-correlations of different galaxy samples. The one-halo and two-halo terms of the cross-correlation between two galaxy samples, ``$g_1$'' and ``$g_2$'' is straightforwardly computed following Eqs.~\eqref{eq:cl1h_ij} and \eqref{eq:cl2h_ij}. The galaxy multipole-space kernels $u_\ell^{g_1}(M,z)$ $u_\ell^{g_2}(M,z)$ are given by Eq.~\eqref{eq:ulg}, where the sample-specific parameters like the redshift distribution or the redshift and mass range can be adjusted for a given catalog. 

\subsubsection{Cosmic Infrared Background}
\label{subsec:cib_hm}

In this subsection, we give the halo model description of the CIB emission, which is based on the model presented in Shang \textit{et al.} \cite{Shang_2012}, which was further used in many other analyses, including \cite{Viero_2013_hermes, websky_Stein_2020, McCarthy_2021, Sabyr_2022}. Following \cite{Shang_2012}, we can define the CIB in the halo model analogously to the galaxy HOD, but with additional prescriptions that describe the infrared emission of each galaxy.

First, we can write the specific intensity of the CIB $I_{\nu}$ at frequency $\nu$ as 
\begin{equation}
    I_{\nu} = \int \frac{\mathrm{d} \chi}{\mathrm{d}z} a(z) \bar{j}_\nu(z) \mathrm{d}z \,, 
\end{equation}
where $\bar{j}_\nu(z)$ is the average emissivity~(\textit{e.g.},~\cite{McCarthy_2021}): 
\begin{equation}
    \bar{j}_\nu(z) = \int_{}^{} \mathrm{d}M \frac{\mathrm{d}n}{\mathrm{d}M} \frac{L_{(1+z)\nu}(M,z)}{4\pi} \,, 
    \label{eq:j_nu}
\end{equation}
where $L_{(1+z)\nu}(M,z)$ is the infrared luminosity of a halo of mass $M$ at redshift $z$ and the factor of $(1+z)$ in the frequency accounts for redshifting of the emitted radiation. The expression for $\bar{j}_\nu(z)$ should be compared to the analogous equations for galaxy number density within the HOD in Eq.~\eqref{eq:n_g}.

The luminosity $L_{(1+z)\nu}(M,z)$ is the sum of the contributions from the central and satellite galaxies, defined by $L_{(1+z)\nu}^{c}(M,z)$ and $L_{(1+z)\nu}^{s}(M,z)$, respectively. In this work, we follow the assumption made in \cite{McCarthy_2021} that both the central and satellite luminosity depend on the same galaxy luminosity model $L^{\rm{gal}}_{(1+z)\nu} (M,z)$ (which is only dependent on the mass of the host halo $M$ and redshift $z$), weighted by the number of centrals and satellites, respectively. Thus, the central luminosity, $L_{(1+z)\nu}^{c}(M,z)$ can be defined as
 \begin{equation}
     L_{(1+z)\nu}^{c}(M,z) = N_{c}^{\rm{CIB}} (M,z) L^{\rm{gal}}_{(1+z)\nu} (M,z) \,,
     \label{eq:L_cent}
 \end{equation}
where the number of CIB central galaxies in a halo, $N_{c}^{\rm{CIB}}$, similarly to the galaxy HOD, is either zero or one, depending on whether the halo mass $M$ is smaller or larger than the parameter $M_{\rm{min}}^{\rm{CIB}}$, the minimum mass to host a central galaxy that sources CIB emission. This requirement can be  written as
\begin{equation}
    N_{c}^{\rm{CIB}} (M,z) = 
    \begin{cases}
      0, & \text{if}\ M < M_{\rm{min}}^{\rm{CIB}} \\
      1, &  \text{if}\ M \geq M_{\rm{min}}^{\rm{CIB}} \,.
    \end{cases}
\end{equation}
The satellite luminosity, $L_{(1+z)\nu}^{s}(M,z)$, that is, the luminosity of a halo due to its satellite galaxies, is given by
\begin{equation}
    L_{(1+z)\nu}^{s}(M,z) = \int \mathrm{d} M_s \frac{\mathrm{d} N}{\mathrm{d} M_s}  L^{\rm{gal}}_{(1+z)\nu} (M,z) \,,
    \label{eq:L_sat}
\end{equation}
where $\frac{\mathrm{d} N}{\mathrm{d} M_s}$ is the subhalo mass function.

The luminosity of galaxies $L^{\rm{gal}}_{(1+z)\nu}$ is governed by the \emph{luminosity-mass relation} (or \emph{L--M} relation), that can be written as 
\begin{equation}
    L^{\rm{gal}}_{(1+z)\nu} = L_0 \Phi(z) \Sigma(M) \Theta((1+z)\nu) \,,
    \label{eq:LM_relation}
\end{equation}
where $L_0$ is a normalization factor, $\Phi(z)$ describes the redshift evolution, $\Sigma(M)$ the mass dependence, and $\Theta$ is the SED of the infrared emission. We describe parametrized functions for the \emph{L--M} relation below, and discuss the specific values of the model parameters in Appendix~\ref{sec:models}.

The redshift evolution of the \emph{L--M} relation is parametrized by a power law index $\delta^{\rm{CIB}}$ in the form of
\begin{equation}
    \Phi(z) = (1+z)^{\delta^{\rm{CIB}}} \,.
    \label{eq:Phi_cib}
\end{equation}
It is well-motivated by observations~\cite{Stark_2009, Gonzalez_2011}, however, to extend the redshift evolution of the \emph{L--M} relation, by including the so-called plateau redshift $z_p$, where $\delta^{\rm{CIB}} = 0$ at $z \geq z_p$:
\begin{equation}
    \Phi(z) = 
    \begin{cases}
      (1+z)^{\delta^{\rm{CIB}}}, & \text{if}\ z < z_p \,, \\
      1, &  \text{if}\ z \geq z_p \,.
    \end{cases}
    \label{eq:Phi_cib_zp}
\end{equation} 
This approach was taken in \cite{websky_Stein_2020} and \cite{Viero_2013_hermes}, where the authors assumed $z_p = 2$, motivated by observations \cite{Stark_2009, Gonzalez_2011}. We also follow this prescription and include $z_p$ as a parameter in our model, to be specified below.

The mass dependence of the \emph{L--M} relation is written as 
\begin{equation}
\Sigma (M) = \frac{M}{\sqrt{2\pi \sigma^2_{L-M}}} e^{-(\mathrm{log}_{10} M - \mathrm{log}_{10} M_{\rm{eff}}^{\rm{CIB}})/2 \sigma^2_{L-M}} \,, 
\end{equation}
with two free parameters, $M_{\rm{eff}}^{\rm{CIB}}$ the peak of the specific IR emissivity, and $\sigma^2_{L-M}$ which controls
the range of halo masses that source the CIB emission. 

The CIB SED is the standard modified blackbody combined with a power law decline at high frequencies 
\begin{equation}
    \Theta = 
    \begin{cases}
       \nu^{\beta^{\rm{CIB}}}B_{\nu}(T_d(z))& \text{if}\ \nu < \nu_0, \\
      \nu^{—\gamma^{\rm{CIB}}} &  \text{if}\  \nu \geq \nu_0,
    \end{cases}
    \label{eq:theta_cib}
\end{equation}
where $B_{\nu}(T)$ is the Planck function at temperature $T$, $\nu_0$ is the break frequency that has to satisfy the continuous derivative requirement
\begin{equation} 
   \frac{\mathrm{d} \ln \Theta(\nu,z) }{\mathrm{d} \ln \nu} = -\gamma^{\rm{CIB}} \,, 
\end{equation}
 and $T_d(z)$ is the dust temperature at redshift $z$ that we parameterize as 
 \begin{equation}
      T_d(z) = T_0 (1+z)^{\alpha^{\rm{CIB}}}, 
 \end{equation}
 with $T_0$ and $\alpha^{\rm{CIB}}$ being free parameters.

To sum up, the CIB model considered in this work \cite{Shang_2012} has ten free parameters \{$L_0$, $\alpha^{\rm{CIB}}$, $\beta^{\mathrm{CIB}}$, $\gamma^{\mathrm{CIB}}$, $T_0$, $M_{\rm{eff}}$, $\sigma^2_{L-M}$, $\delta^{\rm{CIB}}$,  $M_{\rm{min}}^{\rm{CIB}}$, ($z_p$)\}. Their exact values are presented in Appendix~\ref{sec:models} for the two CIB models considered in this work. 

In CIB modeling and analysis \cite{Shang_2012, Planck:2013cib, McCarthy_2021}, one usually implements a flux cut above which bright sources are detected and can be removed (thus suppressing the Poisson power associated with these objects). In our modeling, we follow this prescription and remove all halos whose total luminosity is larger than the luminosity corresponding to a given flux cut value,  $ S_{\nu}$ defined as 
\begin{equation}
    S_{\nu} = \frac{L_{(1+z)\nu}(M,z)}{4\pi(1+z)\chi^2} \,.
    \label{eq:flux_cut}
\end{equation}
We give the flux cut values for each CIB frequency that we use in this work in Table~\ref{table:fluxcut}; for the \emph{Planck} frequencies we use the values given in Ref.~\cite{Planck:2013cib}, and for SO frequencies those in the SO forecast paper~\cite{Ade_2019_so_forecast}. 

\begin{table}[t]
\setlength{\tabcolsep}{10pt}
\renewcommand{\arraystretch}{1.1}
    \begin{tabular}{|c|c|}
        \hline
         Frequency & Flux cut   \\
         $\left[ \mathrm{GHz} \right]$ & $\left[ \mathrm{mJy} \right]$ \\
         \hline 
         \hline 
         93 &   7  \\
         100 & 400 \\
         143 & 350 \\
         145 &  15 \\
          217 & 225  \\
          225 & 20 \\
          280 &  25 \\
          353 & 315  \\
          545 & 350 \\
          \hline
          
    \end{tabular}
    \caption{Point source flux cut values (in mJy) for CIB frequencies (in GHz) considered in this work. The \emph{Planck} frequency (100, 217, 353, 545 GHz) flux cut values come from Table~1 in Ref.~\cite{Planck:2013cib}, while for the SO frequencies (93, 145, 280 GHz), we use the flux cut values from the SO forecast paper \cite{Ade_2019_so_forecast}. The flux cut is implemented according to Eq.~\eqref{eq:flux_cut} for each frequency (in both auto- and cross-correlations involving the CIB). The 545 GHz channel is not considered in our forecasting calculations, but since we use this frequency in comparison to data, we include it here as well for completeness.}
    \label{table:fluxcut}
\end{table}

Finally, putting all of the pieces together, the CIB multipole-space kernel $u_\ell^{\nu} (M,z)$ at frequency $\nu$ can be written (analogously to $u_\ell^{g} (M,z)$ in Eq.~\eqref{eq:ulg}) as
\begin{equation}
    u_\ell^{\nu} (M,z) = W_{I_{\nu}}(z) \bar{j}_\nu^{-1} \frac{L_{(1+z)\nu}^{c} + L_{(1+z)\nu}^{s} u^{m}_{\ell}(M,z)}{4\pi} \,,
\end{equation}
where the CIB window function $W_{I_{\nu}} (z)$ is defined as 
\begin{equation}
       W_{I_{\nu}} (z) = a (z) \bar{j}_\nu(z) \,.
\end{equation}

The one- and two-halo terms of the CIB are then computed according to Eqs.~\eqref{eq:cl1h_ij} and \eqref{eq:cl2h_ij} using the above formula for $u_\ell^{\nu}$. Note that in this work, we consider not only the auto-correlations of the CIB emission at the same frequency $\nu$,  $C_\ell^{\nu \nu}$, but also cross-correlations of the CIB emission at different frequencies $\nu$ and $\nu^\prime$, \textit{i.e.}, $C_\ell^{\nu \nu^\prime}$. These can be computed analogously according to the prescription presented in this subsection.

\subsubsection{Thermal Sunyaev-Zel'dovich Effect}
\label{subsec:tsz_hm}

The electron pressure, \textit{i.e.}, Compton-$y$, multipole-space kernel $ u_\ell^y (M,z)$ is given by~(\textit{e.g.},~\cite{KomatsuSeljak2002,HillPajer2013})
\begin{equation}
    u_\ell^y(M,z) = \frac{\sigma_\mathrm{T}}{m_\mathrm{e} c^2} \frac{4\pi r_s}{\ell_s^2} \int_{x_{\rm{min}}}^{x_\mathrm{max}} \mathrm{d}x \, x^2 \, \mathrm{sinc}(w_\ell x)  P_e(x,M,z),
    \quad \mathrm{with}\quad w_\ell =\frac{\ell+\tfrac{1}{2}}{\ell_s} \,, 
\label{eq:uy}
\end{equation}
where $\sigma_\mathrm{T}$ is the Thomson cross-section, $m_e$ is the electron mass, and the mass-dependent $r_s$ and $\ell_s$ are the characteristic radius and the characteristic multipole of the pressure profile, related via:
\begin{equation}
    \ell_s=\frac{d_A}{r_s}=\frac{1}{(1+z)}\frac{\chi}{r_s} \,,
\end{equation}
where $d_A = \chi /(1+z)$ is the angular diameter distance to redshift $z$. The integration variable $x=r/r_s$ is the ratio between the distance from the center of the halo $r$ and its characteristic radius $r_s$. The pressure profile $P_e(x,M,z)$ is a quantity that parameterizes the radial pressure, and there exist various choices for $P_e$ in the literature. We set $x_{\rm{min}} = 10^{-5}$ and $x_{\rm{max}} = 4$. 

In this work, we consider the Battaglia \textit{et al.} (2012)~\cite{Battaglia_2012} (B12 hereafter) pressure profile $P_e(x,M,z)$, which is parameterized using the generalized NFW formula
\begin{equation}
     P_e(x,M,z) = P_{\Delta} P_0 \left(\frac{x}{x_c}\right)^{\gamma^y} \left[1+\left(\frac{x}{x_c}\right)^{\lambda^y}\right]^{-\beta^y} \,,
\end{equation}
where
\begin{equation}
     P_{\Delta} = \frac{G \Delta M_{\Delta}\rho_c(z) \Omega_b}{2 R_{\Delta} \Omega_m}
\end{equation}
for any spherical overdensity definition $\Delta$ relative to the critical density $\rho_c$.  As in \cite{Battaglia_2012} (and the literature after \cite{pandey2021crosscorrelation, pandey2020}), we set $\lambda^y = 1.0$ and $\gamma^y = -0.3$, and parameterize $P_0$, $x_c$, and $\beta^y$ according to a scaling relation, which for a parameter $X$ can be written as
\begin{equation}
    X (M_{\Delta})  = X_{\Delta} \left(\frac{M_{\Delta}}{10^{14} M_\odot}\right)^{\alpha^y} (1+z)^{\omega^y}, 
\end{equation}
where $X$ denotes any of the parameters $P_0$, $x_c$, and $\beta^y$ , $X_{\Delta}$ is the value of that parameter at $M_{\Delta} = 10^{14} M_\odot$ at $z = 0$, and $\alpha^{y}$ and $\omega^y$ are free parameters. We set those parameters to the standard B12 values \cite{Battaglia_2012} (the ``AGN feedback model at $\Delta=200$'' from their Table~1), which are also summarized in more detail in Appendix~\ref{subsec:battaglia_y}.

To obtain the auto- and cross-correlations of the tSZ field at some frequency, we multiply the auto- and cross-correlations of the Compton-$y$ field by the standard tSZ spectral function at each frequency $\nu$, \textit{i.e.}:
\begin{equation}
    g(\nu) = T_{\rm{CMB}} \left(x \, \rm{coth}\left(\frac{x}{2}\right) -4\right) , 
    \label{eq:g_nu_tsz} \,
\end{equation}
where $x = h_P \nu /(k_B T_{\rm{CMB}} )$, with $h_P$ the Planck constant, $k_B$ the Boltzmann constant, and $T_{\rm{CMB}}$ the CMB temperature today, for which we take $T_{\rm{CMB}}=2.726$ K.

\subsection{Kinematic Sunyaev-Zel'dovich Effect}
\label{subsec:ksz}
In this work, we do not compute the kSZ power spectrum within the halo model, but rather use simulated kSZ power spectra from \cite{Battaglia_2010_kSZ} and \cite{Battaglia_2013_kszpatchy} for the late-time and patchy kSZ contributions, respectively.  As noted before, we do not consider any cross-correlations between the kSZ effect and other fields, as they are zero to first order due to the electron velocity being as likely to be positive as to be negative. 

To compute the post-reionization (late-time) contributions to the kSZ effect, the authors of Ref.~\cite{Battaglia_2010_kSZ} used a suite of periodic-box, high-resolution hydrodynamical, TreePM-SPH simulations of the cosmic web with active galactic nuclei (AGN) feedback.  These simulations are the same as those from which the pressure profile model for the Compton-$y$ field described above is obtained.

For the patchy reionization contributions to the kSZ effect, the authors of Ref.~\cite{Battaglia_2013_kszpatchy} found the amplitude of the kSZ signal at $\ell=3000$ by using a method calibrated from radiation-hydrodynamic simulations at the epoch of reionization (the past light cone spanning redshifts $z$ = 5.5 to $z$ = 20). The predictions were made in relatively large volumes (compared to previous approaches) ($L \geq$ 2 Gpc $h^{−1}$). They employed a semi-analytic model for reionization based on Ref.~\cite{Battaglia2013_paper1} and assumed the \emph{Wilkinson Microwave Anisotropy Probe} (WMAP9) cosmology.

We obtain the total kSZ power spectrum by taking the sum of the late-time and patchy contributions.  The kSZ SED is identical to that of the primary CMB blackbody.

\subsection{Radio Source Contribution}
\label{subsec:radio}
In our modeled mm-wave sky, we also include a contribution from Poisson-distributed radio point sources. Following the Atacama Cosmology Telescope (ACT) analyses in \cite{Dunkley_2013, Choi_2020}, we assume that the radio source auto-power can be modeled as a Poissonian with a fiducial radio SED, which can be written as 
\begin{equation}
    D_{\ell}^{\rm{radio}} = A_s \left(\frac{\ell(\ell+1)}{\ell_0(\ell_0+1)} \right)^2 \left( \frac{\nu_i \nu_j}{\nu_0^2}\right)^{\beta_s} \,, 
    \label{eq:radio}
\end{equation}
where --- following \cite{Choi_2020} --- we set the Poissonian amplitude, $A_s$, referenced at $\ell_0 = 3000$ at $\nu_0 = 150$ GHz, to be $A_s=3.74 \, \mu {\rm K}^2$ (the mean Poisson value of the radio power of the deep ACT region, given in Table~7 of \cite{Choi_2020}) and $\beta_s=-0.5$. We convert the  predictions for the radio auto- and cross-frequency power spectra from intensity units to thermodynamic temperature units, as is done for the CIB (see Appendix~\ref{app:mbb}).

In this work, we do not consider any cross-correlations between the radio sources and other sky components, as they tend to be smaller than the other contributions and have only recently been modeled \cite{Li_2022}.  However, incorporating these additional components should be considered in future forecasting work. 

\section{Specific modeling choices and comparison to data}
\label{sec:models}

In this appendix, we present the specific modeling choices used for the forecasting calculations in our analysis, particularly for the \emph{unWISE} galaxy HOD, the two CIB models considered, and the tSZ field. For the detailed theoretical models and analytical expressions of these components within the halo model, as well as other components used in this work, we refer the reader to Appendix~\ref{sec:spectra}. We also describe the atmospheric and detector noise models, as well as detailed computational settings used to calculate our analytical theory predictions presented in Appendix~\ref{sec:spectra} with \verb|class_sz| code version 1.01 \cite{Bolliet:2017lha,Bolliet_2022}.\footnote{\url{https://github.com/borisbolliet/class_sz}}  To assess the level of validity of our assumptions, we compare the theory-predicted curves with measurements, when possible.

\subsection{\emph{unWISE} HOD}
\label{subsubsec:unwise_hod}

As mentioned above, the halo occupation distribution of the \emph{unWISE} galaxies was already obtained in \cite{Kusiak_2022}, but only fitting angular power spectra up to $\ell = 1000$. Ref.~\cite{Kusiak_2022} constrained the HOD for each of the \emph{unWISE} samples (blue, green, and red) by jointly fitting each sample's auto-spectrum and its cross-correlation with \emph{Planck} CMB lensing from \cite{krolewski_2020} (measured with \verb|NaMaster|) in the standard HOD model \cite{Zehavi_11, Zheng_2007}, with some modifications following the DES-Y3 HOD analysis \cite{Zacharegkas2021}. For our current analysis, since our goal is to clean CIB and tSZ contamination from CMB+kSZ maps using the \emph{unWISE} catalog out to small scales, we need a more faithful HOD model on scales much smaller than $\ell_{\rm{max}} = 1000$, where the kSZ signal dominates.

Furthermore, there is an additional caveat in the treatment of the shot noise in \cite{Kusiak_2022}. The authors concluded that on the scales considered in that work, the Poissonian shot noise contribution is effectively indistinguishable from the one-halo term of $C_\ell^{gg}$, and did not put any prior forcing the shot noise value to be close to its standard constant Poissionian value, $C_\ell^{\mathrm{shot-noise}}=1/\bar{n}_g$ (where $\bar{n}_g$ is the galaxy density in sr$^{-1}$; see Table~\ref{table:unwise} for the $\bar{n}_g$ values of the \emph{unWISE} subsamples), and allowed it to be negative. The best-fit model for the blue sample indeed yielded a negative value of the shot noise.  Since in this work, we want to use the \emph{unWISE} galaxy -- galaxy auto-correlations on much smaller scales where the shot noise of the survey is expected to dominate over the signal, we re-fit those measurements using exactly the same HOD model as in \cite{Kusiak_2022} (see Appendix~\ref{subsec:gal_hm} for the details regarding the HOD modeling), but putting a Gaussian prior on the log of shot noise, $ \mathrm{log}(A_\mathrm{SN})$, centered at its expected Poissonian value (corresponding to the log of $A_\mathrm{SN} = 1/\bar{n}_g= 8.93 \times 10^{-8}, 1.65 \times 10^{-7}, 2.11 \times 10^{-6} $ for the blue, green, and red sample, respectively, calculated for the values of $\bar{n}_g$ from Table~\ref{table:unwise}), with standard deviation of 0.2. We extend the analysis out to $\ell_{\rm{max}} = 4000$, modify the shot noise prior, and update the pixel window function treatment (described in Appendix~\ref{app:pwf}). These are the only changes in modeling the \emph{unWISE} HOD between this work and Ref.~\cite{Kusiak_2022}. 

The re-fitting procedure is performed as follows: for each \emph{unWISE} sample, we use the same $C_\ell^{gg}$-only data points as in \cite{Kusiak_2022}, which are already divided by the pixel window function, but we use the data up to $\ell_{\rm{max}} = 4000$ (after removing the lowest-$\ell$ data point as in \cite{Kusiak_2022}), giving 39 bandpowers of width $\Delta \ell = 100$. These measurements are fit to constrain four HOD parameters, described in Appendix~\ref{sec:spectra}: $\alpha_{\mathrm{s}}$, $\sigma_{\mathrm{log} M}$, $M_\mathrm{min}^\mathrm{HOD}$, $M_1^\prime$.  We also fit the parameter $\lambda_{\mathrm{NFW}}$, which quantifies the NFW truncation radius $r_{\mathrm{out}}$, and the shot noise amplitude, thus totaling six parameters overall. The theoretical prediction is computed with \verb|class_sz|, and all other parameters, as well as the likelihood, except the changes described in this section, are kept the same as in \cite{Kusiak_2022}, including a fixed \emph{Planck} 2018 cosmology \cite{Planck2018}. Since re-fitting this model with a full Markov Chain Monte Carlo (MCMC) as performed in \cite{Kusiak_2022} is very computationally expensive, we simply find the maximum-likelihood model using the \verb|Cobaya| \cite{cobaya2021} minimizer routine instead. This is performed separately for each of the three \emph{unWISE} samples. 

\begin{figure}[htb]
    \centering
    \includegraphics[scale=0.4]{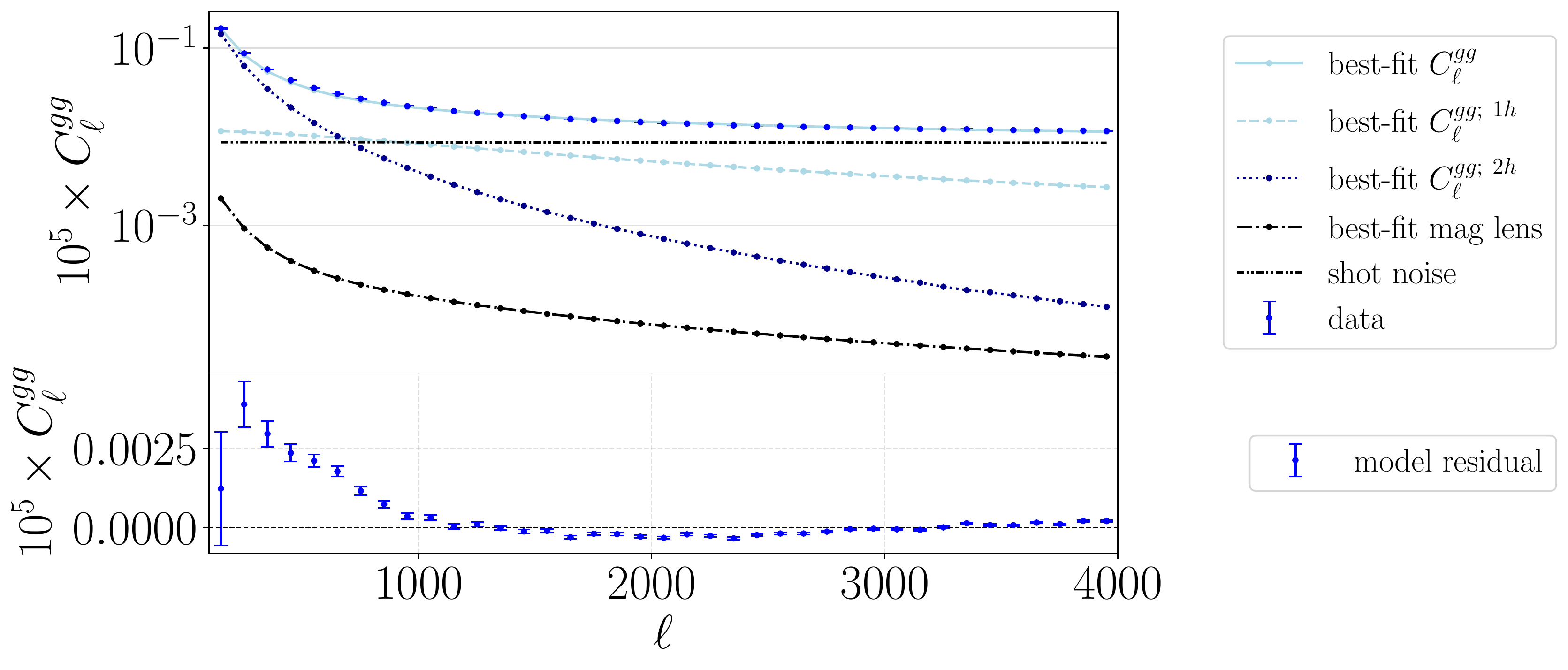}
    \includegraphics[scale=0.4]{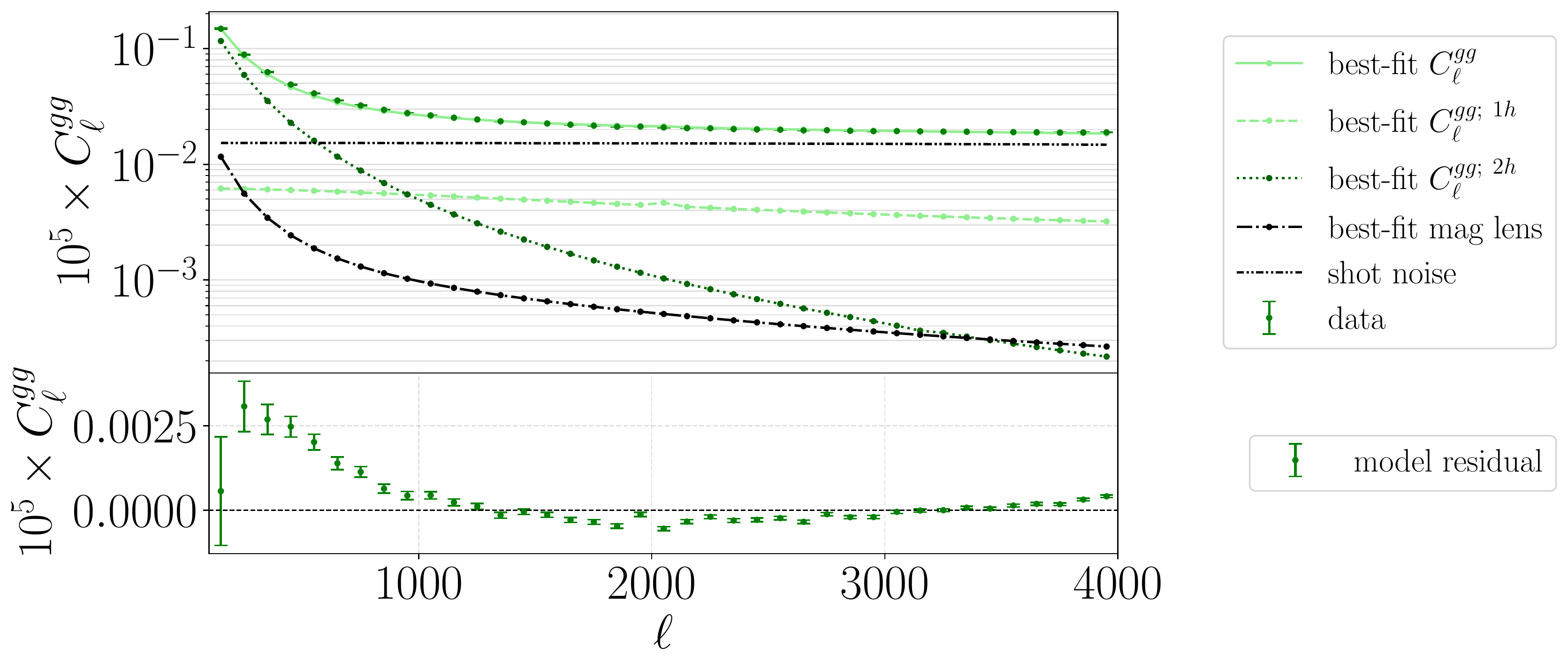}
    \includegraphics[scale=0.4]{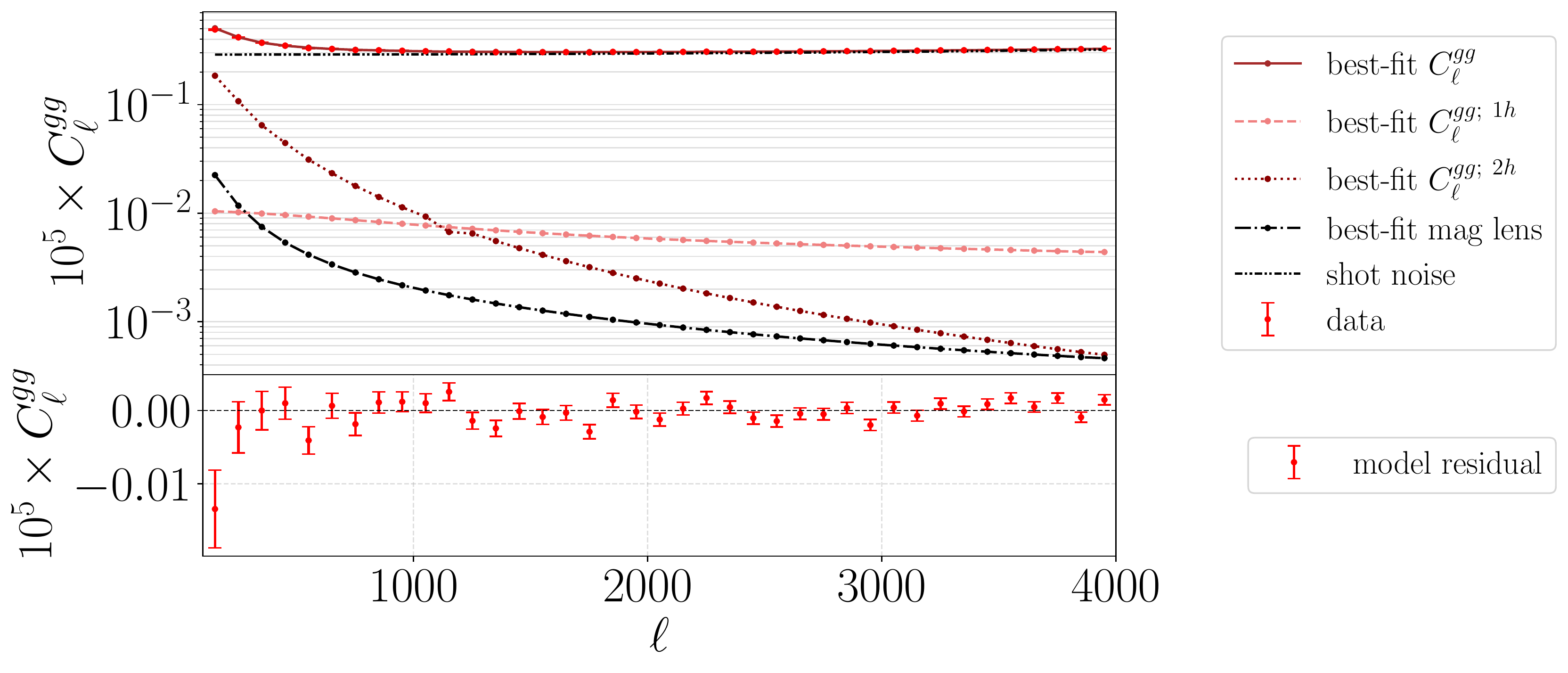}
    \caption{ Comparison of the \emph{unWISE} clustering measurements  $C_\ell^{gg}$ from \cite{krolewski_2020, Kusiak_2022} along with the theory predictions for these correlations computed with the \emph{unWISE} HOD constrained in this work (Appendix~\ref{subsubsec:unwise_hod}). Top panel: the solid curves are the best-fit total signal, the dotted curves show the best-fit one-halo contribution to $C_\ell^{gg}$, the dashed show the best-fit two-halo contribution to $C_\ell^{gg}$, the dash-dotted black show the total best-fit lensing magnification contribution, and the grey dash-dot-dotted show the best-fit shot noise contribution. Bottom panel: residuals of the model at each bin. From top to bottom: results for the \emph{unWISE} blue, green, and red samples (also color-coded). }
    \label{fig:unwise_gg}
\end{figure}

The results of this procedure are shown in Fig.~\ref{fig:unwise_gg} in comparison with the $C_\ell^{gg}$ measurements. The theory curves are computed with the values of the best-fit parameters from the minimization procedure, resulting in $\chi^2=$ 1188, 842, and 72.9, for the \emph{unWISE} blue, green, and red sample, respectively. The obtained best-fit values for the six model parameters $\{ \alpha_{\mathrm{s}}$, $\sigma_{\mathrm{log} M}$, $M_\mathrm{min}^\mathrm{HOD}$, $M^{\prime}_1$, $\lambda_{\mathrm{NFW}}$, $A_{\mathrm{SN}} \}$ for each of the \emph{unWISE} samples are presented in Table~\ref{table:hod_results}. Formally, the $\chi^2$ values of these fits are poor (with the red marginally better than the other two samples).  However, describing $C_\ell^{gg}$ accurately at this level of precision at small scales is a challenging task.  Looking at the model residuals in Fig.~\ref{fig:unwise_gg}, we consider the best-fit parameter values presented here to be a plausible model of \emph{unWISE} galaxy clustering.  However, a very detailed analysis is needed to build a model that adequately describes it at this level of precision.  Whether the HOD approach (which is not a first-principles, but widely-used, clustering model) is the best method to accurately do so is also left for future work. 
To validate the use of the constrained HOD presented in this section, we vary the HOD parameter values and study its impact on the CIB -- galaxy cross-correlations (which is the driving factor of the CIB-removal methods presented in this work). We find that they do not change the cross-correlation significantly, compared to variations in the CIB halo model (see Appendix~\ref{subsec:lenz_cibg} for a comparison of the CIB -- galaxy theoretical predictions to measurements).

    \begin{table}[t!]
    \setlength{\tabcolsep}{10pt}
    \renewcommand{\arraystretch}{1.4}
    \begin{tabular}{ |c|c|c|c|} 
    \hline
    \textbf{Parameter}  & \textbf{Blue}  & \textbf{Green} & \textbf{Red}  \\ 
    \hline\hline
    $\sigma_{\mathrm{log} M}$ & 0.02 & 0.03&   0.07\\ 
    \hline
    $\alpha_s$ & 1.06 &   1.14& 1.76\\ 
    \hline
    $\mathrm{log}_{10}(M_\mathrm{min}^\mathrm{HOD}/(M_\odot/h))$  & 11.69 &  12.23 & 12.56\\
    \hline
    $\mathrm{log}_{10}(M_{1}^\prime/(M_\odot/h))$ &  12.61 &  13.17 & 13.57\\ 
    \hline
    $\lambda_\mathrm{NFW}$ &   1.80 &  0.94 & 1.02 \\
    \hline
    $A_\mathrm{SN}$ & $0.87 \times 10^{-7}$ & $1.53 \times 10^{-7}$ & $28.8 \times 10^{-7}$ \\
    \hline
   \end{tabular}
    \caption{Summary of best-fit values for the six HOD model parameters $\{ \alpha_{\mathrm{s}}$, $\sigma_{\mathrm{log} M}$, $M_\mathrm{min}^\mathrm{HOD}$, $M^{\prime}_1$, $\lambda_{\mathrm{NFW}}$, $A_\mathrm{SN} \}$ obtained by fitting the measured \emph{unWISE} galaxy -- galaxy auto-correlations to the halo model predictions, separately for each of the three \emph{unWISE} galaxy samples.  See Appendix~\ref{subsubsec:unwise_hod} for more details on how these values are obtained. }
    \label{table:hod_results}
    \end{table}

The best-fit shot noise amplitudes for the galaxy auto-correlations determined in these fits are given in Table~\ref{table:hod_results}.  For cross-correlations between the \emph{unWISE} samples, we use the shot noise values from \cite{krolewski_2020}.  These values can generally be non-zero due to galaxies in different samples that live in the same host halo.  The shot noise for the red-blue cross-correlation is considered to vanish due to their wide difference in redshift; for blue-green it is $6.22 \times 10^{-9}$; and for green-red it is $4.67 \times 10^{-8}$ (note that there is a typo in the Appendix of Ref.~\cite{krolewski_2020}, where these values are given, confirmed with the authors). The shot noise values for auto- and cross-correlations of \emph{unWISE} subsamples are summarized in Table~\ref{table:gg_shotnoise}.

\begin{table}[t!]
\setlength{\tabcolsep}{10pt}
\renewcommand{\arraystretch}{1.2}
    \begin{tabular}{|c|c|c|c|}
        
        \hline
         Sample & \emph{unWISE} blue & \emph{unWISE} green &  \emph{unWISE} red \\
         \hline 
         \hline
         \emph{unWISE} blue &  $ 0.87 \times 10^{-7}$ &  $0.06 \times 10^{-7}$ & 0\\
         \hline
         \emph{unWISE} green & $0.06 \times 10^{-7}$ & $ 1.53 \times 10^{-7}$ & $0.47 \times 10^{-7}$\\
         \hline
         \emph{unWISE} red & 0 & $0.47 \times 10^{-7}$ & $ 28.8 \times 10^{-7}$\\
         \hline
    \end{tabular}
    \caption{ \emph{unWISE} galaxy shot noise values. The shot noise values for the auto-correlations are obtained from fitting the \emph{unWISE} galaxy -- galaxy data to the halo model prediction (see Appendix~\ref{subsubsec:unwise_hod} for the description of this procedure), while the values for the cross-correlations between the different \emph{unWISE} samples are taken from Ref.~\cite{krolewski_2020}.}
    \label{table:gg_shotnoise}
\end{table}

\subsection{H13 CIB model}
\label{subsec:cib_hermes}
In this subsection, we describe the first CIB model considered in this work, which we refer to as the H13 model. The H13 CIB model \cite{Viero_2013_hermes} used the \emph{Herschel} Multi-tiered Extragalactic Survey (\emph{HerMES}) \cite{Oliver_2012} data from the SPIRE instrument aboard the \emph{Herschel Space Observatory} \cite{Pilbratt_2010} to constrain the standard Shang \textit{et al.} halo model of the CIB emission \cite{Shang_2012} presented in Appendix~\ref{subsec:cib_hm}. H13 was further used in the literature, \textit{e.g.}, in the \emph{WebSky} simulation suite \cite{websky_Stein_2020}.  We refer to this CIB model as H13; we validate our predictions against measurements from the \emph{WebSky} simulations (which also use this CIB model) since we do not have access to the data or theory curves from H13.

We present the parameter values of the H13 model in Table~\ref{table:cib_params_websky}, which approximately correspond to those in Table~V of Ref.~\cite{Viero_2013_hermes} and Section~3.2.2 of Ref.~\cite{websky_Stein_2020} (note that there is a typo on page 13 of Ref.~\cite{websky_Stein_2020}, confirmed with the authors, where the $\delta_{\rm{CIB}}$ parameter should have a value of $\delta_{\rm{CIB}} = 0.3$, not 2.4). The \emph{WebSky} \cite{websky_Stein_2020} CIB model uses a redshift-dependent lower mass integral limit in Eqs.~\eqref{eq:cl1h_ij} and~\eqref{eq:cl2h_ij}.\footnote{M. Alvarez, priv.~comm.}  We use this redshift-dependent lower mass limit, $M_1^\mathrm{CIB}(z)$ for all calculations involving the H13 CIB model, as described in Appendix~\ref{subsec:hm_predictions}. Furthermore, the \emph{WebSky} simulations assume that all CIB-sourcing halos host a central galaxy, therefore  $M_{\rm{min}}^{\rm{CIB}} = 0$, which we also implement in our modeling. There are, however, some differences between our (\emph{unWISE}-motivated) assumptions and the \emph{WebSky} implementation that cannot be easily resolved: \emph{WebSky} uses a different mass definition ($\Delta = 200 \bar{\rho}_m$) than we assume ($\Delta = 200 \rho_c$), as well as a different concentration-mass relation from Duffy \emph{et al.}~(2008) \cite{Duffy_2008}. 

To account for these changes, we re-fit the $L_0$ parameter, the overall free normalization of the infrared luminosity -- mass relation (\emph{L--M} relation), to mitigate differences coming from a different choice of the HMF or the mass definition than the original \emph{WebSky} implementation \cite{websky_Stein_2020}.  
We obtain $L_0 =$ 5.06 $ \times 10^{-7}$ Jy Mpc$^2$/$M_\odot$/Hz, which comes from benchmarking the \verb|class_sz| prediction to match the power spectra of the \emph{WebSky} \cite{websky_Stein_2020} CIB maps using parameters in Table~\ref{table:cib_params_websky}, and other specifications described in this section. We use the CIB flux cut values given in Table~\ref{table:fluxcut}.

\begin{table}[t!]
\setlength{\tabcolsep}{3pt}
\renewcommand{\arraystretch}{1.5}
\begin{tabular}{ |c|c|c|c|} 
\hline
\textbf{Parameter}  & \textbf{Parameter description}  & \textbf{Value H13} & \textbf{Value P14}  \\ 
\hline
\hline
$L_0$ [Jy Mpc$^2$/$M_\odot$/Hz] & Normalization of \emph{L--M} relation  & 5.06 $\times 10^{-7}$  &  {7.0 $\times 10^{-8}$} \\
\hline
$\alpha^{\rm{CIB}}$ & Redshift evolution of dust temperature & 0.2 & {0.36} \\
\hline
$T_0$  & Dust temperature at $z = 0$ & 20.7 K & {24.4 K }\\
\hline
$\beta^{\rm{CIB}}$  & Emissivity index of SED &   1.6  & {1.75}\\
\hline
$\gamma^{\rm{CIB}}$  & Power law index of SED at high frequency &  1.7  & {1.7}\\ 
\hline
$\mathrm{log_{10}} (M_\mathrm{eff}^\mathrm{CIB} / M_\odot ) $ & Most efficient halo mass &  12.3 & {12.6}\\
\hline
$M_{\rm{min}}^{\rm{CIB}} / M_\odot $ & Minimum halo mass to host a galaxy  & 0 & {10} \\
\hline
$\sigma^2_{L-M}$ & Distribution of halo masses sourcing CIB emission & 0.3  & {0.5}\\
\hline
$\delta_{\rm{CIB}}$ & Redshift evolution of \emph{L--M} relation   & 1.28 & {3.6}\\ 
\hline
$z_p$ & Plateau redshift of \emph{L--M} relation & 2 &  {no $z_p$}\\
\hline
\end{tabular}
\caption{CIB parameters of the H13 CIB model \cite{Viero_2013_hermes} (also used in the \emph{WebSky} simulations \cite{websky_Stein_2020}), and P14 CIB model \cite{Planck:2013cib} considered in this work. The P14 values are identical to those in Table~9 in \cite{Planck:2013cib} (as well as Table~I in \cite{McCarthy_2021}), besides $L_0$, the normalization of \emph{L--M} relation (Eq.~\eqref{eq:LM_relation}) --  two CIB models considered in this work (see Appendix~\ref{subsec:cib_hermes} for more details and parameter definitions and Appendix~\ref{subsec:cib_hm} for the theoretical description of the CIB emission in the halo model).}
\label{table:cib_params_websky}
\end{table}

For the CIB auto- and cross-correlations, we use the shot noise values from the \emph{Planck} CIB model (given in Table~6 in Ref.~\cite{Planck:2013cib}). For the SO frequencies (93, 145, 225, 280 GHz), the shot noise values are obtained via interpolation in log-space (after converting shot noise values to $\mu \mathrm{K}^2$) in one dimension at each \emph{Planck} frequency to obtain shot noise values for the \emph{Planck} -- SO cross-frequency power spectra, followed by another one-dimensional log-space interpolation for each SO frequency to obtain shot noise values for the SO -- SO auto- and cross-frequency power spectra. These values are summarized in Table~\ref{table:cib_shotnoise}.

\begin{table}[t]
\setlength{\tabcolsep}{10pt}
\renewcommand{\arraystretch}{1.2}
    \begin{tabular}{|c|c|c|c|c|c|c|c|c|c|}
        \hline
         Frequency $\left[ \mathrm{GHz} \right]$ & 93 &100 & 143 &145 & 217 & 225 &280 & 353 & 545 \\
         \hline 
         \hline
         93 & 0.10 &  0.12 & 0.34 & 0.36 & 1.22 &1.35 & 2.46 & 4.41 & 9.72\\
         \hline
         100 & 0.12& 0.15 & 0.42 & 0.44& 1.50 &1.66 & 3.02& 5.40& 12.00\\
         \hline
         143 & 0.34 & 0.42 & 1.20& 1.25& 4.30 &4.75 & 8.54& 15.00& 35.00\\
         \hline
          145 &0.36 & 0.44 & 1.25& 1.31& 4.49 &4.96 & 8.92& 15.69& 36.58\\
         \hline
          217 &1.22 & 1.50 & 4.30 & 4.49 & 16.00 &17.75 & 32.59& 59.00& 135.00\\
         \hline
          225 &1.35 &1.66 &4.75 &4.96 &17.75 &19.69 &36.18 &65.57 &150.50\\
         \hline
          280 &2.46 &3.02 &8.54 &8.92 &32.59 &36.18 &66.85 &122.05 & 286.25\\
          \hline
          353 &4.41 & 5.40 & 15.00 & 15.69& 59.00 &65.57  & 122.05& 225.00& 543.00\\
          \hline
          545 &9.72 &12.00 &35.00 &36.58 & 135.00 &150.50 & 286.25& 543.00& 1454.00\\
          \hline
          
    \end{tabular}
    \caption{CIB shot noise values for auto- and cross-frequency (in GHz) power spectra in Jy$^2$/sr. The \emph{Planck} frequency (100, 143, 217, 353, 545 GHz) shot noise values are taken from Table~6 in Ref.~\cite{Planck:2013cib}, and the SO frequency (93, 145, 225, 280 GHz) shot noise values are obtained via interpolation in log space (after converting shot noise values to $\mu \mathrm{K}^2$) --- see text for details.
    We note that we do not use the 545 GHz frequency in our final analysis; however, we include it here to increase the accuracy of interpolation.}
    \label{table:cib_shotnoise}
\end{table}

\begin{figure}[t]
    \centering
    \includegraphics[scale=0.45]{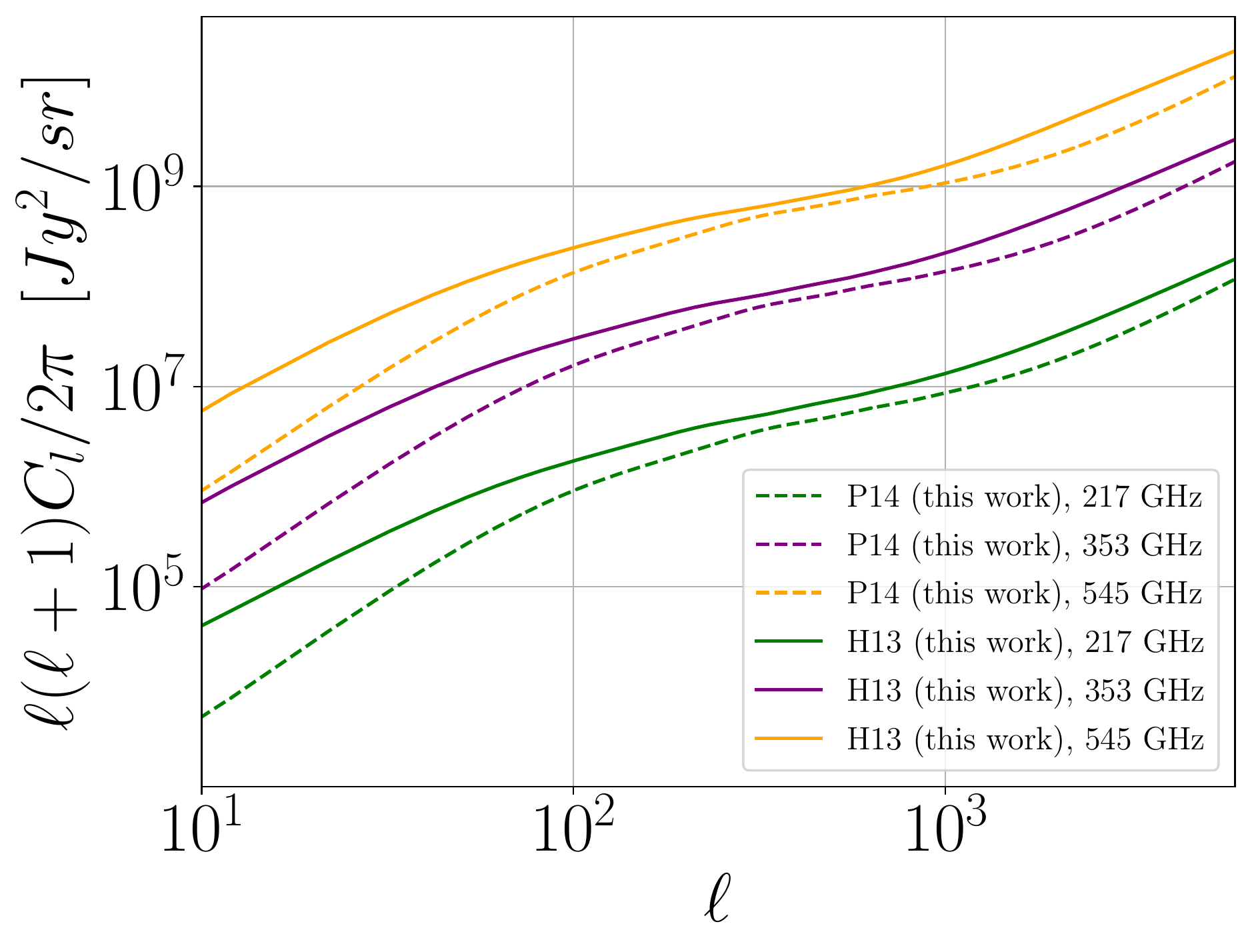}
    \caption{ {Comparison of the two CIB auto-correlation predictions (including shot noise) used in this work for the H13 CIB model (solid lines) and the P14 CIB model (dashed lines) at 217, 353, and 545 GHz. The notable difference on large scales between the two models arises from the fact that CIB measurements can have significant Galactic dust contamination, which can affect the derived modeling and interpretation. } }
    \label{fig:cib_compare}
\end{figure}

In Fig.~\ref{fig:cib_compare} (left panel), we show  the H13 CIB auto-correlation predictions computed with \verb|class_sz| using the modeling choices discussed in this section (Table~\ref{table:cib_params_websky}) at selected frequencies. In Fig.~\ref{fig:corr_cib_g} below, we present the correlation coefficients for the H13 CIB model (left panels) and each of the \emph{unWISE} samples for our predictions at each of the frequencies used in this work. 

\subsection{P14 CIB model}
\label{subsec:cib_planck}

The second CIB model we consider in our work is from the \emph{Planck} 2014 CIB paper \cite{Planck:2013cib}. The authors of Ref.~\cite{Planck:2013cib} fit the \emph{Planck} nominal mission CIB power spectrum results to the Shang \emph{et al.}~(2012) CIB model \cite{Shang_2012}, presented in Appendix~\ref{subsec:cib_hm}. Their constrained CIB model was further used in the literature, \textit{e.g.}, in Ref.~\cite{McCarthy_2021} to forecast improvements on CIB modeling by including CMB lensing data, or in Ref.~\cite{Sabyr_2022} to model the inverse-Compton scattering of the CIB, analogous to the tSZ effect. We refer to this CIB model as P14. 

We summarize the parameters of the P14 CIB model in Table~\ref{table:cib_params_websky}, which we use to predict the CIB auto- and cross-correlations in this work. {We validate our analytical predictions for the P14 model against the CIB power spectra from McCarthy \emph{et al.} \cite{McCarthy_2021} that uses the same \emph{Planck} 2014 model (curves obtained from the authors)}. Similarly to the H13 CIB model, we re-fit $L_0$, the overall free normalization of the \emph{L--M} relation, to mitigate differences coming from a different choice of the HMF or the mass definition than the original \emph{Planck} 2014 CIB paper \cite{Planck:2013cib} (which uses the HMF from~\cite{Tinker_2010}, the concentration-mass relation from~\cite{Duffy_2008}, and $\Delta = 200 \bar{\rho}_m$ mass definition, as well as a slightly different \emph{Planck} 2013 cosmology). In fact, Ref.~\cite{Planck:2013cib} does not quote an exact $L_0$ value, while Ref.~\cite{McCarthy_2021} uses $L_0$ = 6.4 $\times 10^{-8}$
Jy Mpc$^2$/$M_\odot$/Hz.  In this work, we use $L_0 = 7.0 \times 10^{-8}$
Jy Mpc$^2$/$M_\odot$/Hz for the P14 model. We use the same CIB shot noise and flux cut values as for the H13 CIB model, given in Tables~\ref{table:cib_shotnoise} and \ref{table:fluxcut}, respectively.

\subsection{Battaglia \textit{et al.} (2012) Compton-$y$}
\label{subsec:battaglia_y}

In this subsection, we present the comparison of our prediction of the Compton-$y$ (tSZ) power spectrum with the Battaglia \textit{et al.} (2012)~\cite{Battaglia_2012} (hereafter B12) pressure profile and data.  As described in Appendix~\ref{sec:spectra}, we use the B12 pressure profile, which was also used in the \emph{WebSky} simulation suite \cite{websky_Stein_2020}. The Compton-$y$ power spectra are computed within the halo model, as described in Appendix~\ref{subsec:tsz_hm}, using the default parameters. Specifically, we assume the AGN feedback B12 pressure profile parameters (Table~1 in Ref.~\cite{Battaglia_2012}), which we summarize in Table~\ref{table:pp_b12}.
\begin{table}[t]
\setlength{\tabcolsep}{10pt}
\renewcommand{\arraystretch}{1.5}
    \begin{tabular}{|c|c|c|c|c|}
        \hline
         Parameter &  Parameter description & $X_{\Delta}$ & $\alpha^y$ & $\omega^y$ \\
         \hline \hline
         $P_0$ &   Amplitude of the pressure profile &18.1 & 0.154 & -0.758 \\
         \hline
          $x_c$ & Core scale of the pressure profile& 0.497 &  -0.00865 & 0.731 \\
        \hline
        $\beta^y$ & Shape of the pressure profile & 4.35 & 0.0393 & 0.415 \\
         \hline
    \end{tabular}
    \caption{Values of the AGN feedback B12 \cite{Battaglia_2012} pressure profile parameters used in this work, described in \S \ref{subsec:tsz_hm}. All B12 parameters assumed in this work take their default for the $\Delta=200 \rho_c$ mass definition values as presented in Table~1 in Ref.~\cite{Battaglia_2012}.}
    \label{table:pp_b12}
\end{table}

\begin{figure}[t]
    \centering
    \includegraphics[scale=0.6]{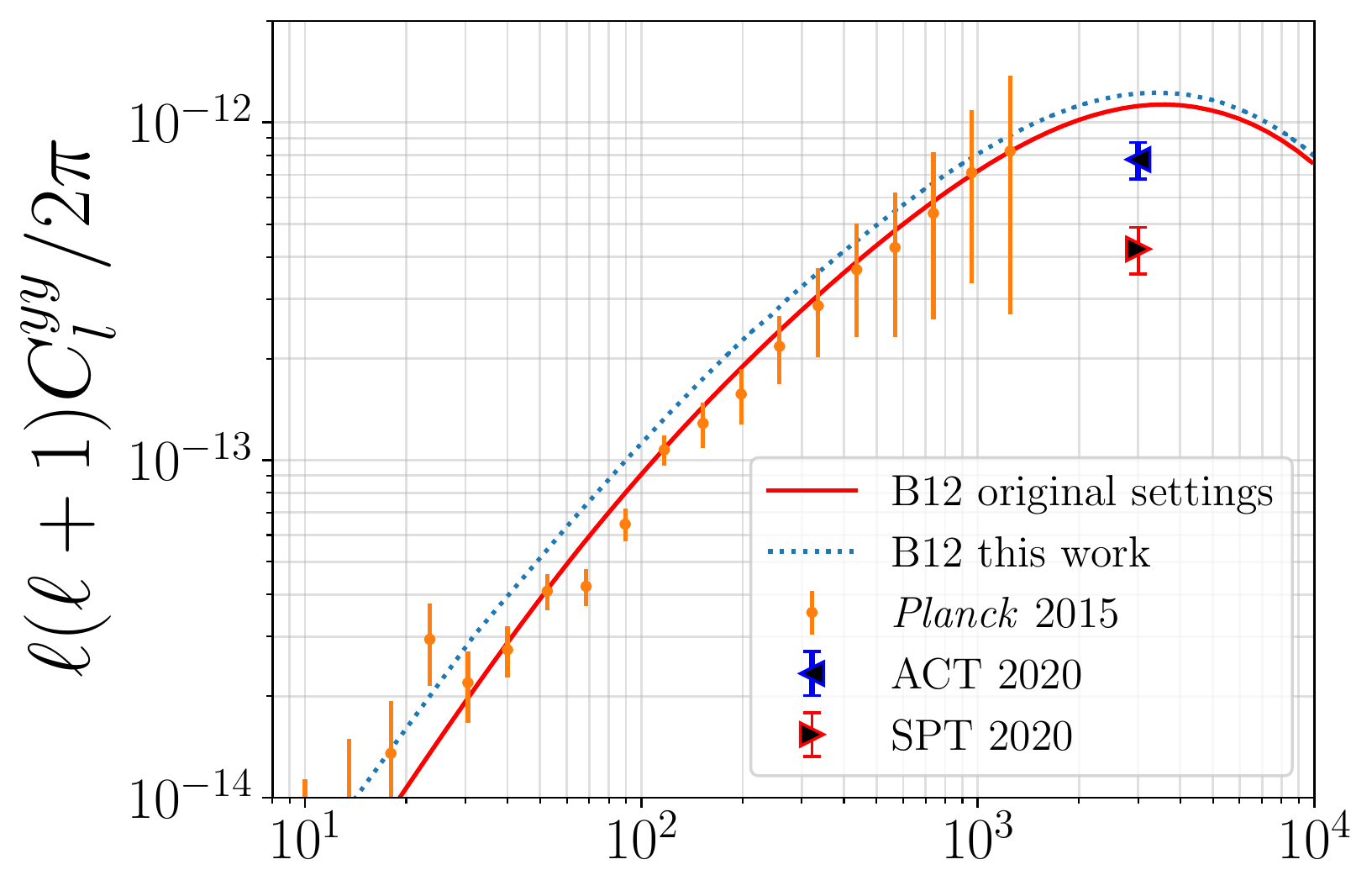}
    \caption{ Comparison of the Compton-$y$ auto-correlation measurements from \emph{Planck} 2015 \cite{planck2015_tsz} (orange dots with error bars), ACT 2020 \cite{Choi_2020} (at $\ell=3000$), and SPT 2021 \cite{Reichardt_2021} (at $\ell=3000$), along with the theoretical predictions computed using the Battaglia \emph{et al.}~(2012) \cite{Battaglia_2012} pressure profile (solid lines) used in this work. See Appendix~\ref{subsec:tsz_hm} and Appendix~\ref{subsec:battaglia_y} for details.}
    \label{fig:yy}
\end{figure}

In Fig.~\ref{fig:yy}, we compare our predictions for the Compton-$y$ auto-power spectrum to data. We show the Compton-$y$ measurements from \emph{Planck} 2015 \cite{planck2015_tsz} (orange dots with errorbars), ACT 2020 \cite{Choi_2020} at $\ell=3000$ (blue triangle), and the South Pole Telescope (SPT) 2021 \cite{Reichardt_2021} at $\ell=3000$ (red triangle). It is a well-known issue that the B12 pressure profile tends to overpredict the power at higher $\ell$ for a CMB-preferred cosmology, as can be seen in Fig.~\ref{fig:yy} for the ACT and SPT measurements (particularly the latter).

\begin{figure}[t]
    \centering
    \includegraphics[scale=0.4]{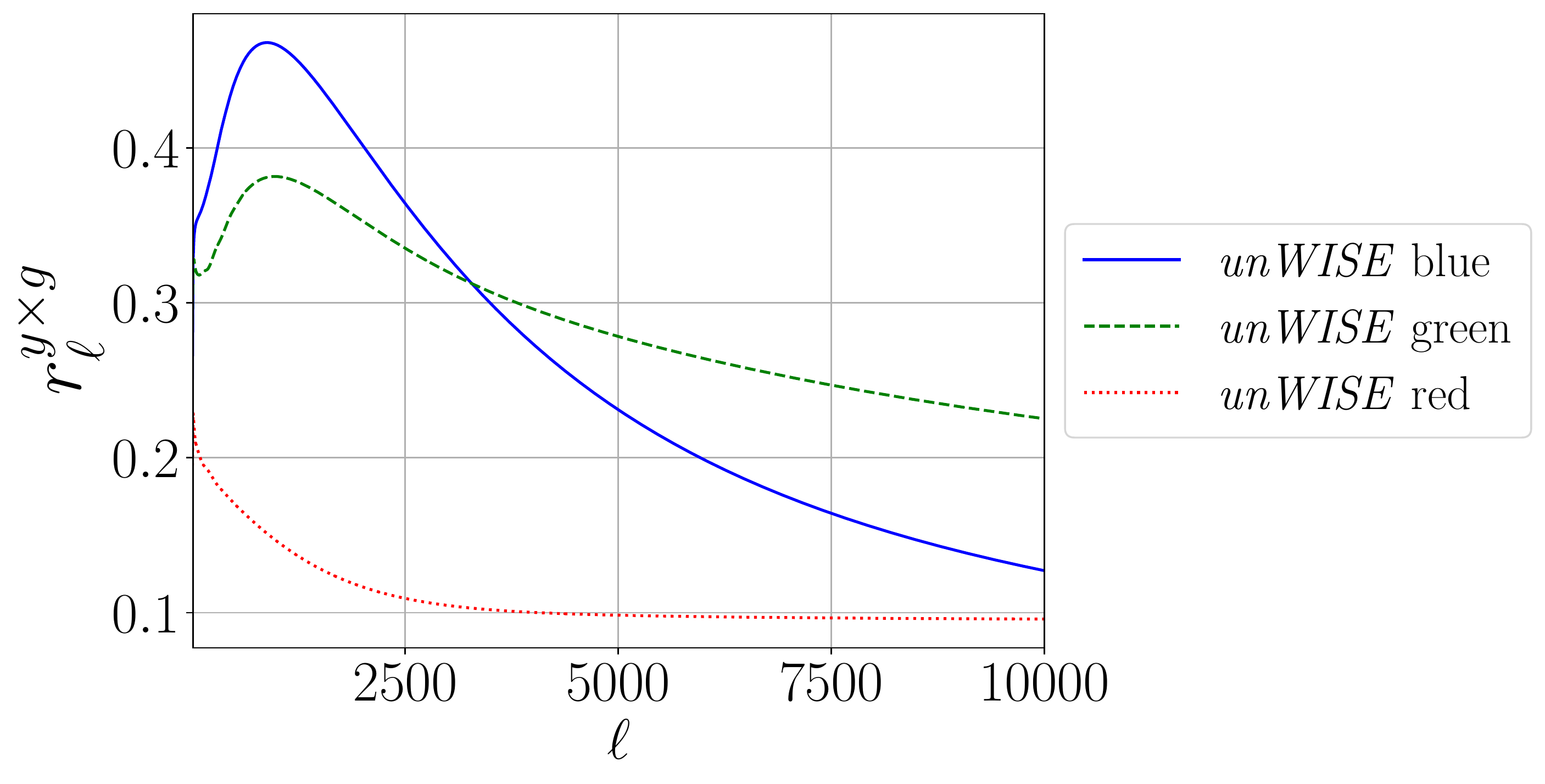}
    \includegraphics[scale=0.4]{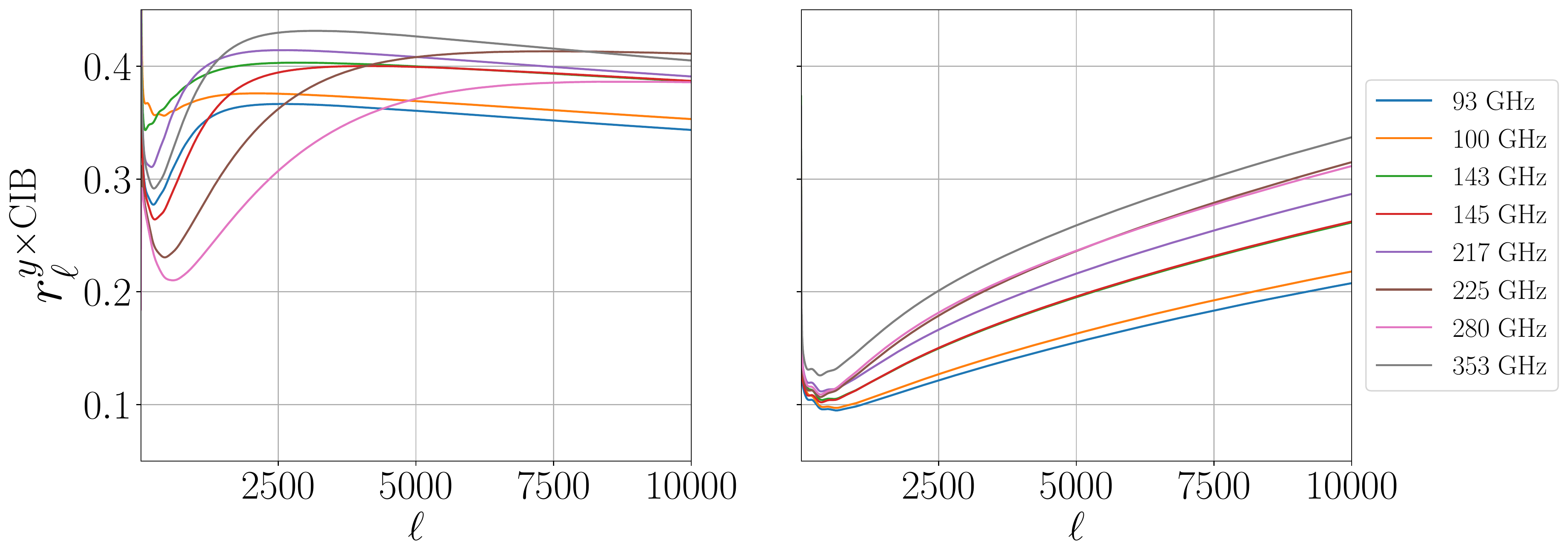}
    \caption{ \emph{Top}: Correlation coefficients of the Compton-$y$ field and the \emph{unWISE} galaxy samples, blue (solid blue), green (dashed green), and red (dotted red). \emph{Bottom}: Correlation coefficients of the Compton-$y$ field and the two CIB models, H13 (left) and P14 (right) for the eight frequency channels considered in this work. All curves are computed with the analytical halo model predictions for the auto- and cross-correlations presented in Appendix~\ref{subsec:hm_predictions} using {\tt class\_sz}. Both the CIB and galaxy auto-correlations include shot noise contributions (given in Tables~\ref{table:cib_shotnoise} and~\ref{table:gg_shotnoise}).  }
    \label{fig:corr_tsz_g_cib} 
\end{figure}

Note that the galaxy -- and CIB -- tSZ cross-correlations are also included in our theoretical modeling.  These play an important role in the de-(CIB+tSZ) method, where the LSS tracers are used to remove both CIB and tSZ contamination.  In the top panel of Fig.~\ref{fig:corr_tsz_g_cib}, we show correlation coefficients between the Compton-$y$ field and each \emph{unWISE} sample. The galaxy -- galaxy auto-correlations include the shot noise contributions (given in Tables~\ref{table:gg_shotnoise}). As expected, the blue sample yields the largest correlation coefficient with the tSZ effect (reaching about 50\%), which matches our qualitative expectation, as the tSZ redshift kernel peaks at redshift $z < 1 $, which is also where most of the blue sample galaxies are distributed (see Fig.~\ref{fig:dndz_COSMOS}). The correlation coefficient for the red sample and the Compton-$y$ field is the lowest, reaching only about 20\%, as the two fields do not overlap in redshift significantly.  While the galaxy -- tSZ correlations do not reach values as high as those for the galaxy -- CIB, they are non-negligible, illustrating that the \emph{unWISE} galaxies do partially trace out the tSZ field.  Including samples with higher number density at low redshifts, where much of the tSZ signal originates, would help to increase these correlations. 

In the bottom panel of Fig.~\ref{fig:corr_tsz_g_cib}, we show correlation coefficients between the Compton-$y$ field and the CIB emission for the two CIB models considered in this work, H13 (left) and P14 (right), for the eight frequency channels from 93 GHz to 353 GHz.  The correlations for the H13 model on average reach higher values (over 40\%) than for the P14 model. For the P14 model, the $y$ and CIB fields are fairly independent, except at high $\ell$ values.

\subsection{CIB \texorpdfstring{$\times$}~\textit{unWISE} with Lenz \textit{et al.}~(2019) data}
\label{subsec:lenz_cibg}

\begin{figure}[htb]
    \centering
    \includegraphics[scale=0.4]{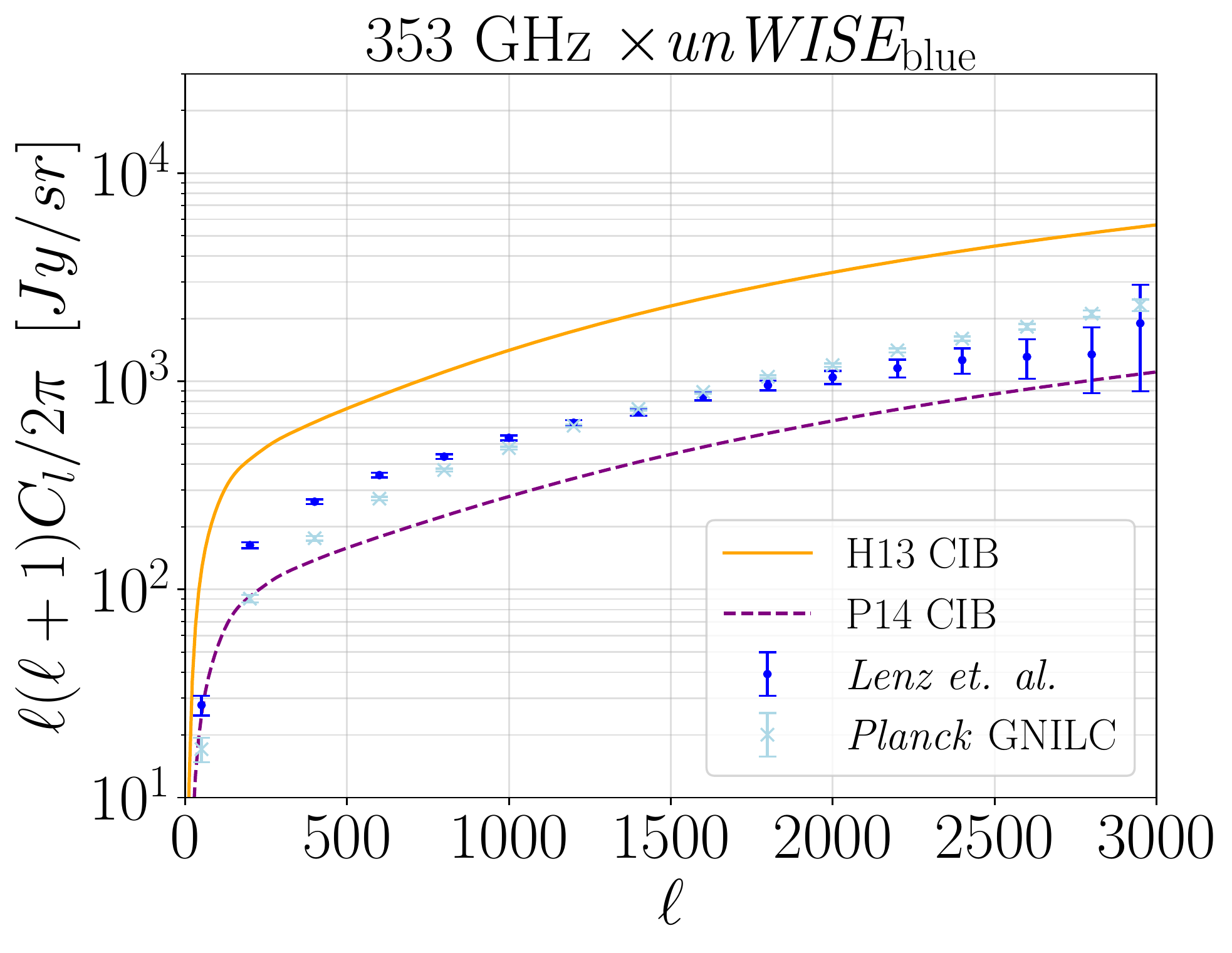}
    \includegraphics[scale=0.4]{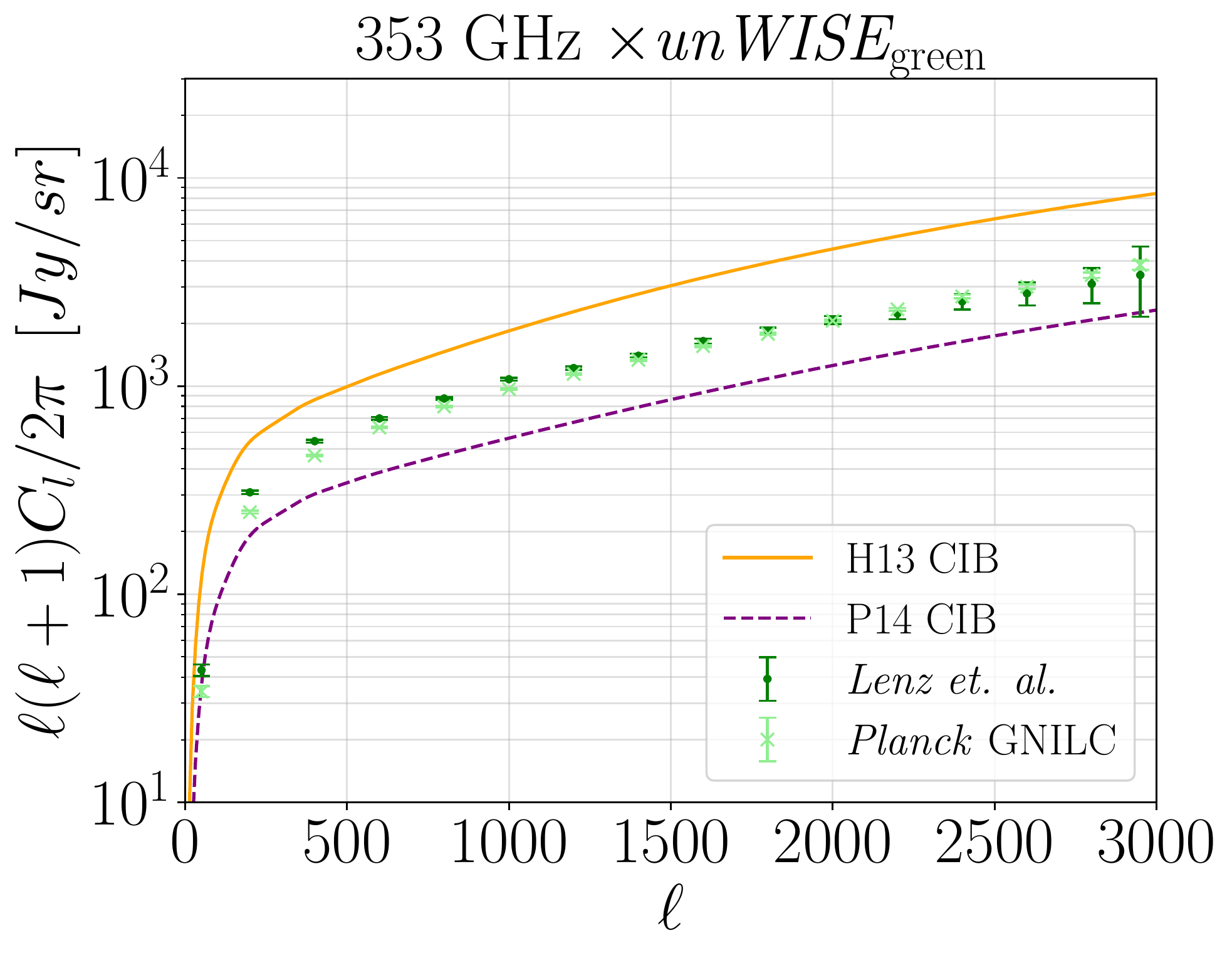}
    \includegraphics[scale=0.4]{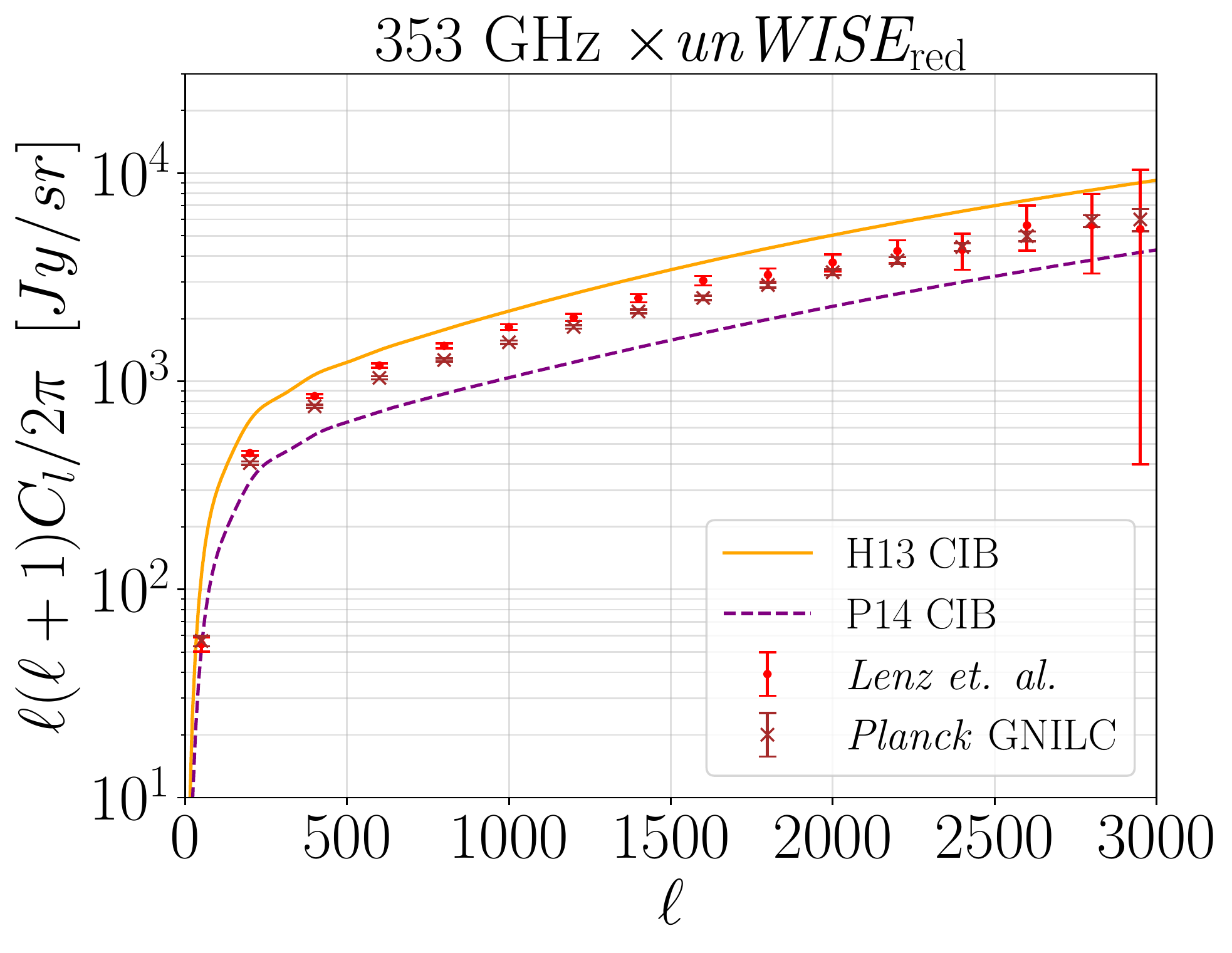}
    \caption{ Comparison of the CIB -- galaxy cross-power spectrum measurements using the \emph{unWISE} galaxies and CIB maps from Lenz \textit{et al.}~(2019) \cite{Lenz_2019} (dots) and the \emph{Planck} GNILC CIB maps (crosses) at 353 GHz, along with the theoretical predictions for these cross-correlations computed using the \emph{unWISE} HOD constrained in this work (Appendix~\ref{subsubsec:unwise_hod}) and the two CIB models considered, H13 and P14 (solid and dashed curves, respectively). Clockwise from top left: results for the \emph{unWISE} blue, green, and red samples. The measurements are encompassed by the two CIB models, which validates our choice to consider both in our work.  The differences in the data points at low $\ell$ are likely due to differences in residual Galactic contamination in the two CIB maps. See Fig.~\ref{fig:cibXg_data545} for the same comparison at 545 GHz.  }
    \label{fig:cibXg_data}
\end{figure}

To assess the validity of our assumptions for the two CIB model choices used in this work, we compare the \verb|class_sz| predictions for the cross-correlations of CIB with \emph{unWISE} galaxies with measurements made using the CIB maps from Ref.~\cite{Lenz_2019}. We compute the CIB -- \emph{unWISE} cross-power spectra with \verb|NaMaster| \cite{namaster} using the \emph{unWISE} galaxy overdensity maps described in \S \ref{subsec:unwise} and the Lenz \textit{et al.}~(2019) \cite{Lenz_2019} CIB maps with $f_{\rm{sky}} = 18.7 \%$, that is, the $N_{H I} < 2.5 \times 10^{20} \,\, {\rm cm}^{-2}$ threshold (see Fig.~15 in \cite{Lenz_2019}).  We consider the CIB maps at 353 and 545 GHz (note that we do not use the 545 GHz channel in our actual analysis in this paper, but due to the lack of CIB maps at lower frequencies, we consider it here as a useful sanity check). We also compute the cross-correlation with the \emph{Planck} GNILC CIB maps at the same frequency channels, downloaded from the {\emph{Planck} Legacy Archive}\footnote{\url{https://irsa.ipac.caltech.edu/data/Planck/release_2/all-sky-maps/foregrounds.html}}, although the method of removing Galactic dust may be more robust in Ref.~\cite{Lenz_2019}.

The results of this analysis at 353 GHz are shown in Fig.~\ref{fig:cibXg_data} for both CIB maps, along with the \verb|class_sz| predictions computed with the CIB models considered in this work and the \emph{unWISE} HOD presented in Appendix~\ref{subsubsec:unwise_hod} (an analogous plot for 545 GHz is shown in Fig.~\ref{fig:cibXg_data545}). 

First, we note that the difference in the two measurements using the Lenz \textit{et al.}~(2019) CIB maps and the \emph{Planck} GNILC CIB maps is effectively an estimate of the residual foreground contamination in these two CIB maps. The deviations are largest at low $\ell$, which is most likely due to residual Galactic dust, which correlates with large-scale systematics in the \emph{unWISE} maps (\textit{e.g.}, contamination from stars in the Milky Way, as well as fluctuations in the selection function of the galaxies due to Galactic dust). Second, the two predictions computed with the H13 and P14 CIB models fully encompass the measurements, with H13 (P14) predicting a cross-correlation larger (smaller) than that observed.  This validates our assumptions to consider both models in this work, so as to bracket the range of possibilities. To obtain close agreement between these measurements and the theoretical model, one would have to re-fit the measurements to constrain the CIB and \emph{unWISE} halo model parameters (simultaneously with the auto-spectra); we leave this detailed analysis for future work.

Note that in principle the CIB -- \emph{unWISE} cross-correlation should also include the shot noise contribution, as both signals are sourced (partially) by the same discrete objects (galaxies). However, there is no assessment of this contribution to date, and the two CIB models considered in this work, as seen in Fig.~\ref{fig:cibXg_data}, encompass the observations with a reasonable margin.  Thus we ignore this contribution to the CIB $\times$ \emph{unWISE} model for the purposes of this work.  To obtain an estimate of the shot noise contribution in the CIB $\times$ \emph{unWISE} cross-correlation, one would have to carefully fit those measurements to a theoretical prediction, which is beyond the scope of this work. 

\begin{figure}[t]
    \centering
    \includegraphics[scale=0.3]{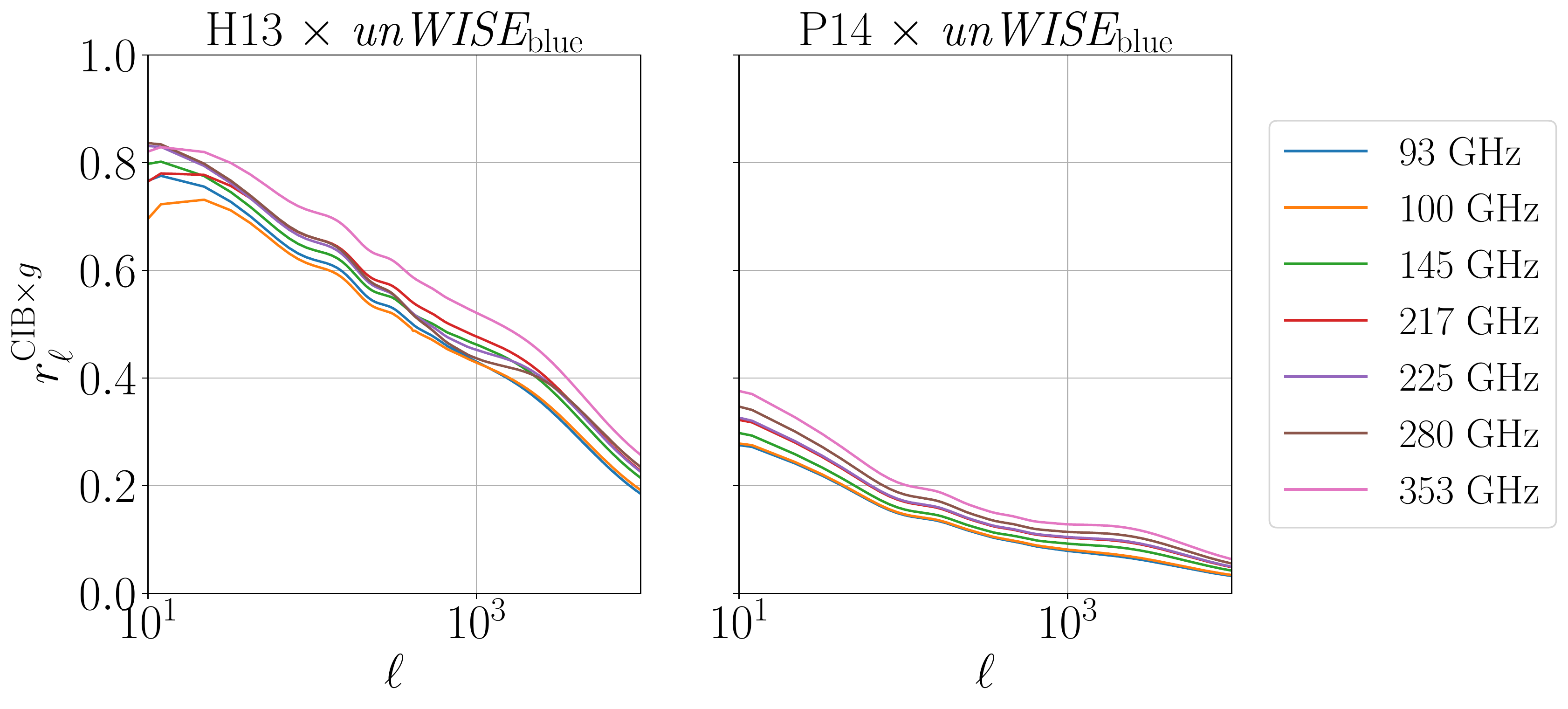}
    \includegraphics[scale=0.3]{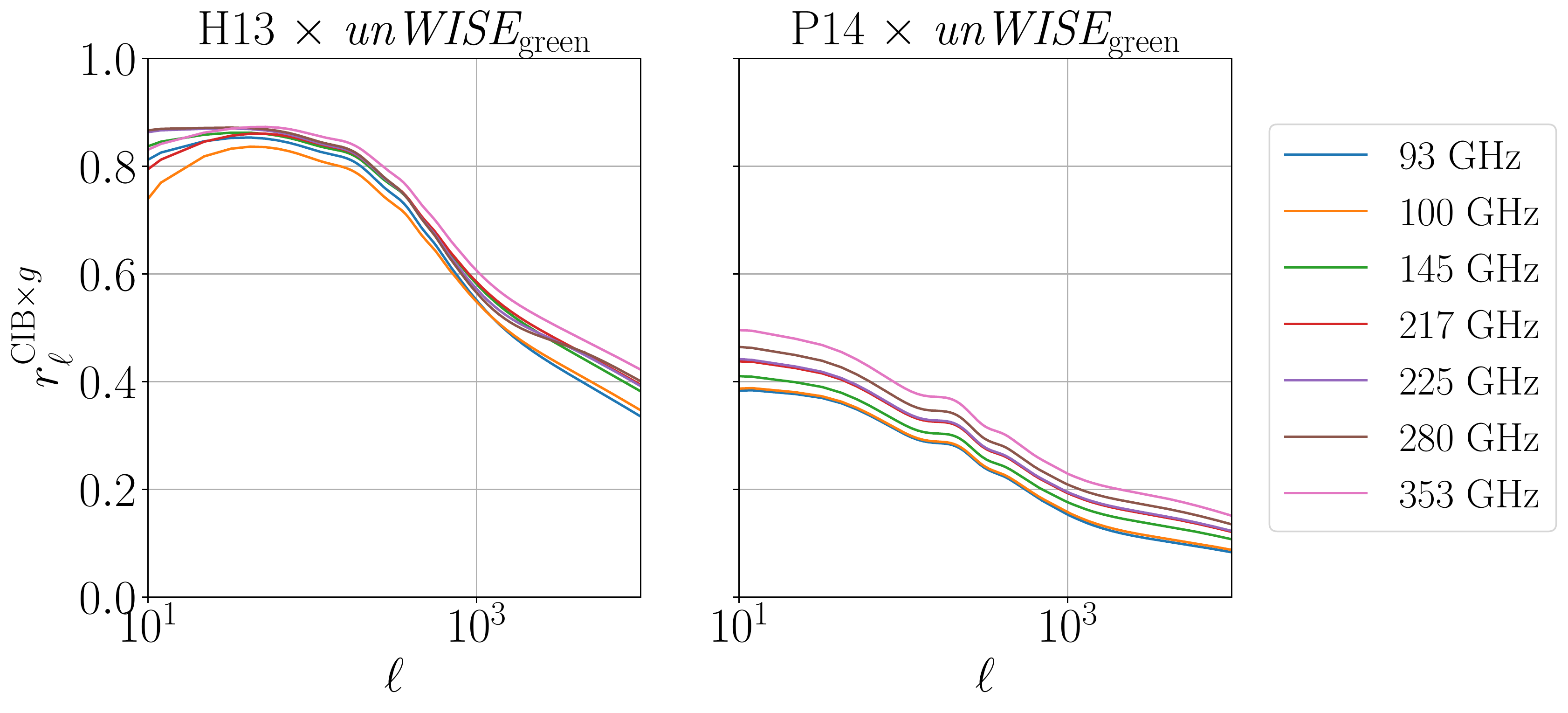}
    \includegraphics[scale=0.3]{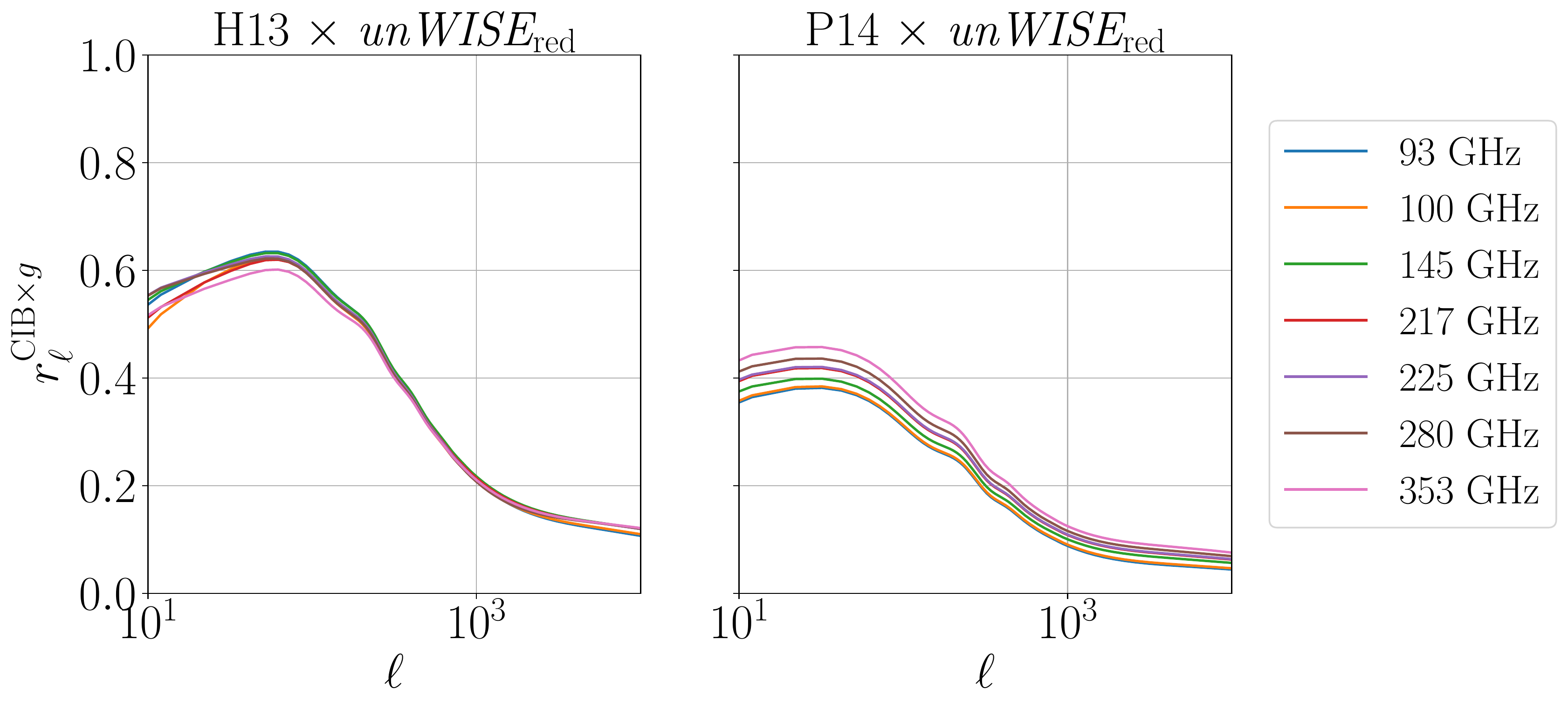}
    \caption{ Correlation coefficients of the CIB field and the \emph{unWISE} galaxy samples for the two CIB models considered in this work, H13 (left) and P14 (right) (see Appendix~\ref{subsec:cib_hermes} and Appendix~\ref{subsec:cib_planck}) for (from top to bottom) \emph{unWISE} blue, green, and red. All curves are computed with the analytical halo model predictions for the auto- and cross-correlations presented in Appendix~\ref{subsec:hm_predictions} using {\tt class\_sz}. Both the CIB and galaxy auto-correlations include shot noise contributions (given in Tables~\ref{table:cib_shotnoise} and~\ref{table:gg_shotnoise}, respectively). Results at 143 GHz are omitted for clarity.  }
    \label{fig:corr_cib_g}
\end{figure}

In Fig.~\ref{fig:corr_cib_g}, we show galaxy -- CIB correlation coefficients for the \emph{unWISE} blue, green, and red samples, where all components are computed with the analytical halo model predictions for the auto- and cross-correlations presented in Appendix~\ref{subsec:hm_predictions} using \verb|class_sz| for the two CIB models considered in this work, H13 and P14. The CIB -- CIB and galaxy -- galaxy auto-correlations include the shot noise contributions (given in Tables~\ref{table:cib_shotnoise} and~\ref{table:gg_shotnoise}, respectively). From Fig.~\ref{fig:corr_cib_g}, we see that the H13 (left) CIB model predicts a higher correlation coefficient between the CIB and the \emph{unWISE} galaxies, reaching almost 90\% correlation at low $\ell$, while the P14 CIB model predicts a maximal correlation of about 50\%. We thus conclude that the H13 CIB model is more optimistic than P14.

\subsection{Detector and Atmospheric Noise}
\label{subsec:noise}
\begin{figure}[t]
    \centering
    \includegraphics[scale=0.6]{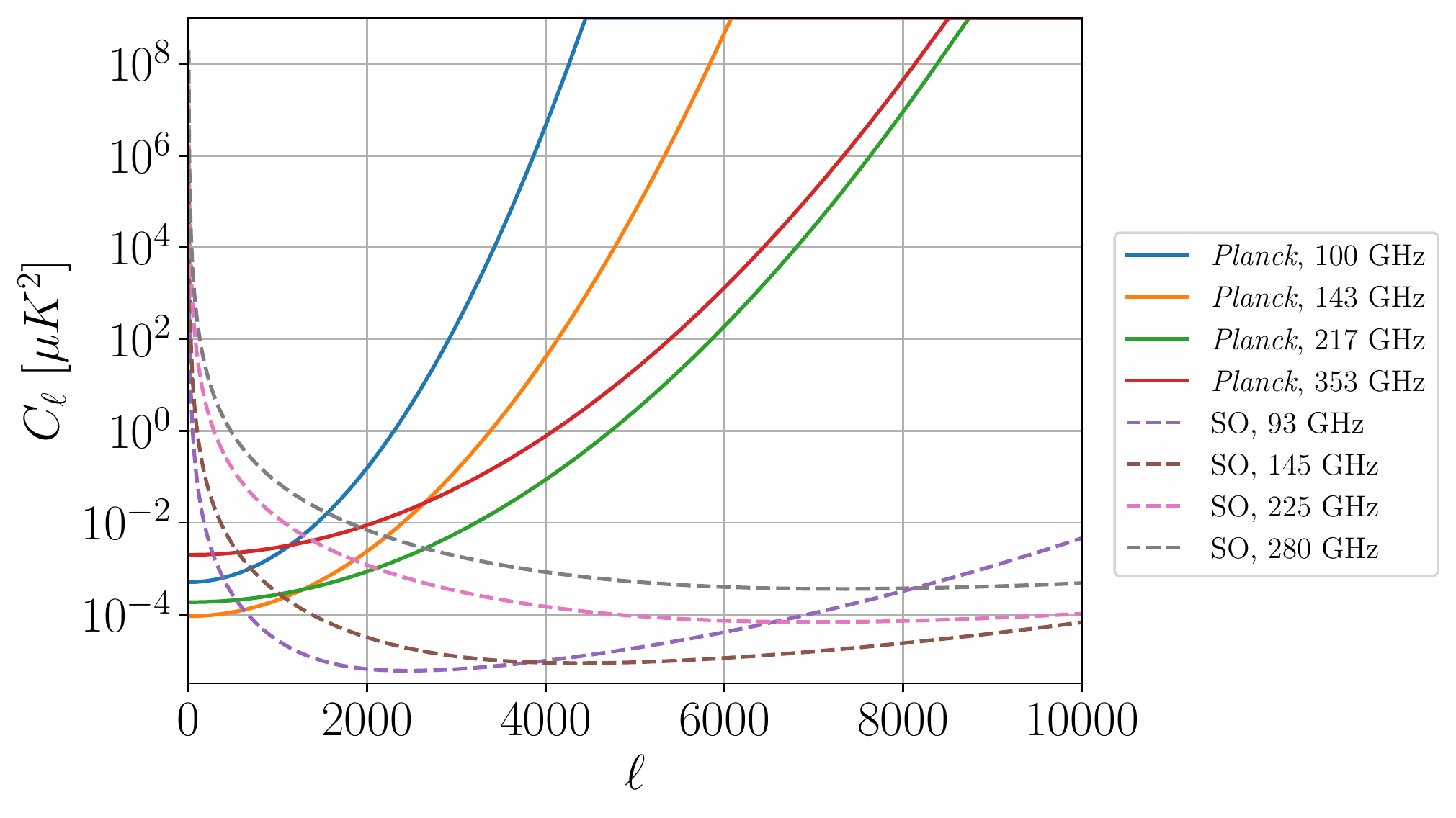}
    \caption{Noise power spectra from \textit{Planck} (solid curves) and SO (dashed curves) used in our calculations. For the \textit{Planck} noise curves, a cap is applied at high $\ell$ to avoid numerical overflow due to very high noise. For SO, atmospheric noise is present at low $\ell$. For simplicity, we only include noise auto-spectra and ignore cross-correlations due to atmospheric noise for SO. }
    \label{fig:noise_curves}
\end{figure}

We consider auto- and cross-frequency power spectra for the 93, 145, 225, and 280 GHz channels from Simons Observatory\footnote{\url{https://github.com/simonsobs/so_noise_models}} (SO) \cite{SO2019} and the 100, 143, 217, and 353 GHz channels from \emph{Planck} (see Table 7 of Ref.~\cite{Planck:2015LFI}, Table 6 of Ref.~\cite{Planck:2015HFI}, and Eq.~(2.32) in Ref.~\cite{Errard:2015}). The noise power spectrum for each frequency channel is shown in Fig.~\ref{fig:noise_curves}, and the parameters for the white noise contributions are given in Table \ref{table:noise}.  We model the \emph{Planck} noise as uncorrelated white noise.  For SO, we take into account both white noise at high multipoles (for the SO ``goal'' noise levels) and the atmospheric contribution at low multipoles, which is non-negligible (using the official SO noise model). The combination of SO and \emph{Planck} frequency channels allows us to obtain high SNR at both low and high $\ell$: for a ground-based telescope such as SO, atmospheric noise is present, dominating at low $\ell$, while for the \emph{Planck} satellite, the noise dominates at high $\ell$ due to the coarse angular resolution of the instrument. We note that we only include frequency-frequency noise auto-spectra in our models, ignoring frequency-frequency cross-correlations due to atmospheric noise (for SO).

\begin{table}[t]
    \centering
    \setlength{\tabcolsep}{8pt}
    \renewcommand{\arraystretch}{1.2}
    \begin{tabular}{ |c|c|c| } 
    \hline
      Frequency & White Noise [$\mu \mathrm{K} \cdot$ arcmin] &  Beam FWHM [arcmin] \\  
    \hline\hline
    \emph{Planck}, 100 GHz &77.4 &9.69  \\
    \hline
    \emph{Planck}, 143 GHz &33.0 &7.30 \\
    \hline
    \emph{Planck}, 217 GHz &46.8 &5.02 \\
    \hline
    \emph{Planck}, 353 GHz &153.6 &4.94\\
    \hline
    SO, 93 GHz &5.8 &2.2\\
    \hline
    SO, 145 GHz &6.3 &1.4\\
    \hline
    SO, 225 GHz &14.9 &1.0\\
    \hline
    SO, 280 GHz &37.2 &0.9\\
    \hline
    \end{tabular}
    \caption{Properties of the detector noise at each frequency used in our sky model. We note that atmospheric noise is also present for SO, which leads to a significant non-white component at low multipoles.}
    \label{table:noise}
\end{table}

\section{Additional Plots for the H13 CIB Model}
\label{app:additional_plots}

\begin{figure}[t]
    \centering
    \includegraphics[scale=0.55]{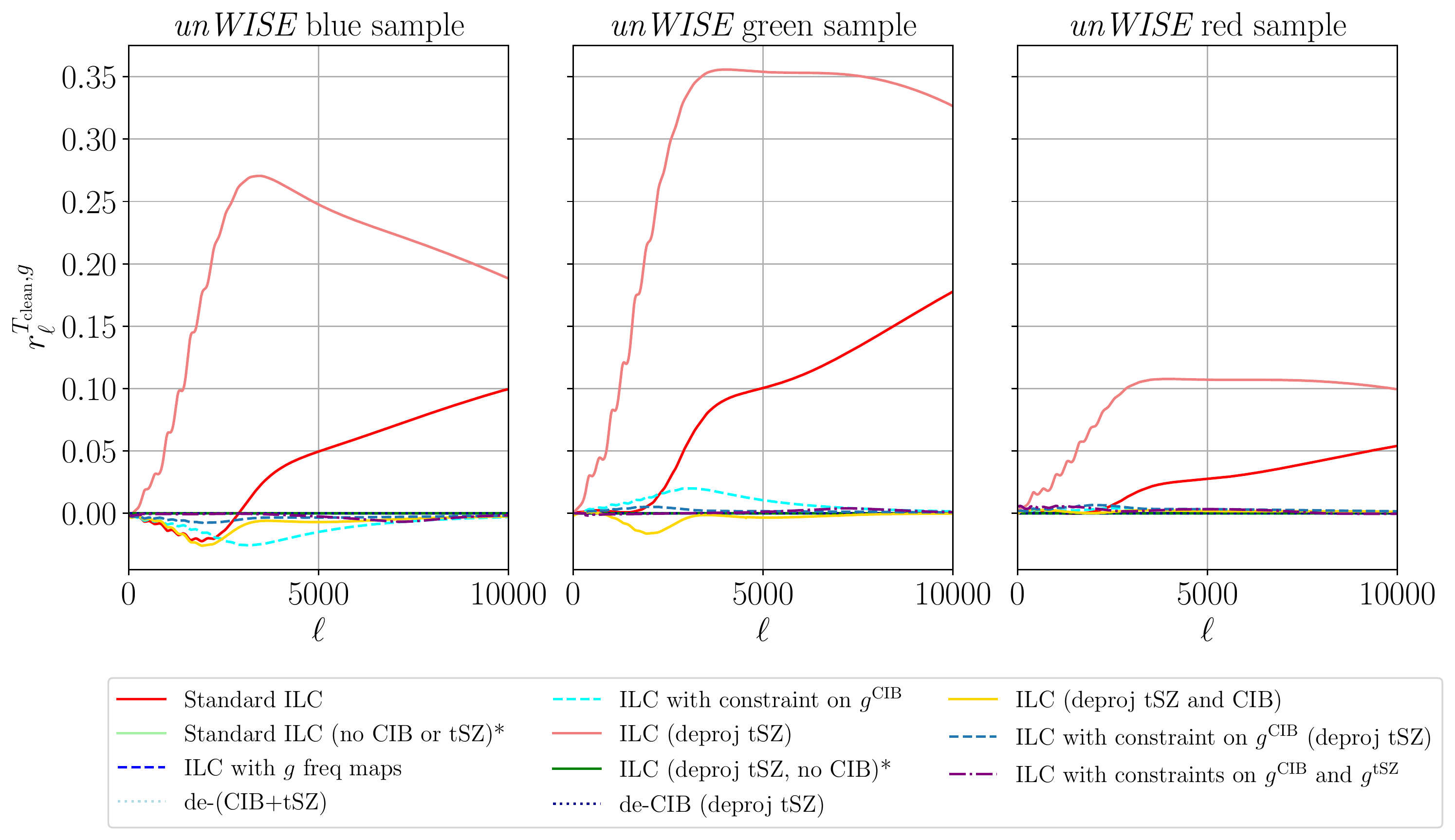}
    \caption{Correlation coefficient of the cleaned map with the \emph{unWISE} blue, green, and red samples, for each method considered in this work, using the H13 CIB model. ``Baseline" methods are shown as solid curves in shades of red and yellow. ``Idealized" methods are shown as solid curves in shades of green. Our new methods are shown as dashed or dotted curves in shades of blue and purple. The asterisk (*) indicates an idealized method that cannot be applied to actual data and is included in this work only for comparison purposes.}
    \label{fig:Tcleanxg}
\end{figure}

\begin{figure}[t]
    \centering
    \includegraphics[scale=0.6]{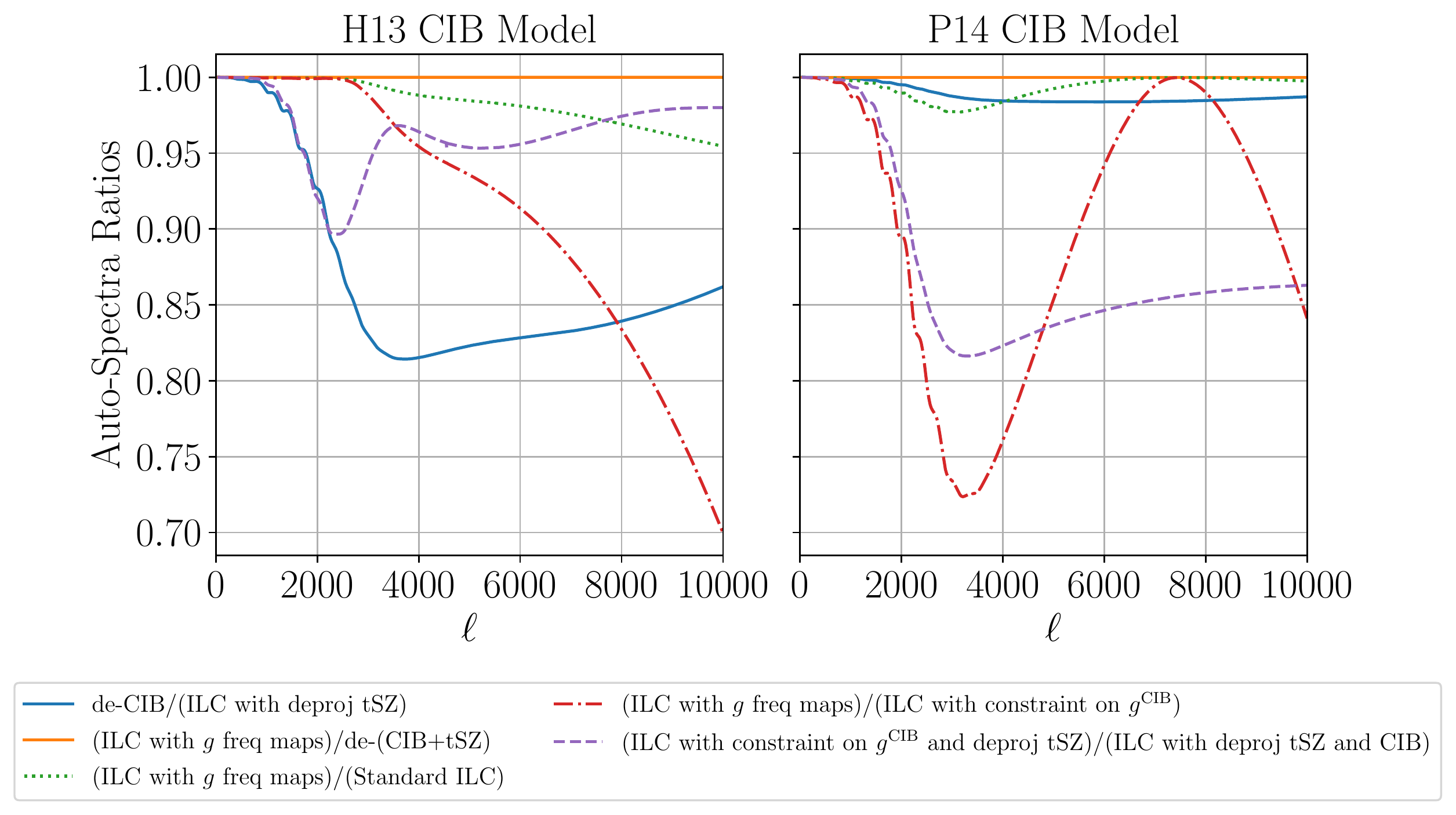}
    \caption{Ratios of auto-spectra of final cleaned maps from Fig.~\ref{fig:Tclean_auto_spectra} for both the H13 and P14 CIB models. We consider the ratio of the cleaned map auto-spectra from the de-CIB method (applied to a tSZ-deprojected ILC map) and the tSZ-deprojected ILC without de-CIBing [solid blue]; an ILC with the tracers $g$ as additional ``frequency'' maps and the de-(CIB+tSZ) method applied to a standard ILC map [solid orange]; an ILC with the tracers $g$ as additional ``frequency'' maps and standard ILC [dotted green]; an ILC with the tracers $g$ as additional ``frequency'' maps and an ILC with an additional constraint requiring zero cross-correlation of the cleaned map with $g^{\mathrm{CIB}}$ [dash-dot red]; and an ILC with deprojected tSZ signal and an additional constraint requiring zero cross-correlation of the cleaned map with $g^{\mathrm{CIB}}$ and an ILC with both the CIB and tSZ signals deprojected [dashed purple].}
    \label{fig:Tclean_auto_spectra_ratios}
\end{figure}

In this appendix, we provide additional plots using the H13 CIB model. First, to verify the removal of tracer-correlation from the cleaned maps and to assess the relative cleaning of each tracer sample, Fig.~\ref{fig:Tcleanxg} shows the correlation coefficients of the cleaned maps from various methods with the \emph{unWISE} blue, green, and red samples, \textit{i.e.},
$$r_\ell^{T^{\mathrm{clean}},g^a} \equiv C_{\ell}^{T^\mathrm{clean}, g^a} / \sqrt{C_{\ell}^{g^a g^a} C_{\ell}^{T^\mathrm{clean} T^\mathrm{clean}}}, $$
where $g^a$ represents one of the \emph{unWISE} samples (blue, green, or red). We note that this plot assumes the H13 CIB model. For all three samples, the baseline methods of standard ILC and tSZ-deprojected ILC, shown as solid curves in shades of red, display significant correlation with the \emph{unWISE} overdensity field, as the tracers are not used in those methods and there is no tailored optimization for CIB or tSZ contaminant removal. Notably, there is also a small dip in the correlation of the blue and green samples with the cleaned map from an ILC that deprojects both the CIB and tSZ signals. This is because we have assumed some effective CIB SED for this method, but in reality, the CIB SED is not well-defined since the CIB decorrelates across frequencies. The idealized (solid curves in shades of green) methods exhibit near-zero correlation with each of the tracer samples, as expected since the CIB and tSZ contaminants have been removed, so there is no remaining signal that is correlated with the tracers.

Our new methods (dashed and dotted curves in shades of blue and purple), for the most part, display near-zero correlation with the \emph{unWISE} overdensity field. However, for the ILC with zero-tracer-correlation constraint using $g^{\mathrm{CIB}}$ (dashed cyan curves), there is residual correlation of the cleaned map with the blue and green samples. This is because we have optimized for CIB removal but not for tSZ signal removal in this method, and the tSZ signal is significantly correlated with the blue and green samples, as shown in Fig.~\ref{fig:corr_tsz_g_cib}. However, each individual galaxy overdensity sample is not fully cleaned since, at each frequency, we remove an optimally weighted linear combination of the samples to remove the most CIB contamination at that frequency.

\begin{figure}[t]
    \centering
    \includegraphics[scale=0.73]{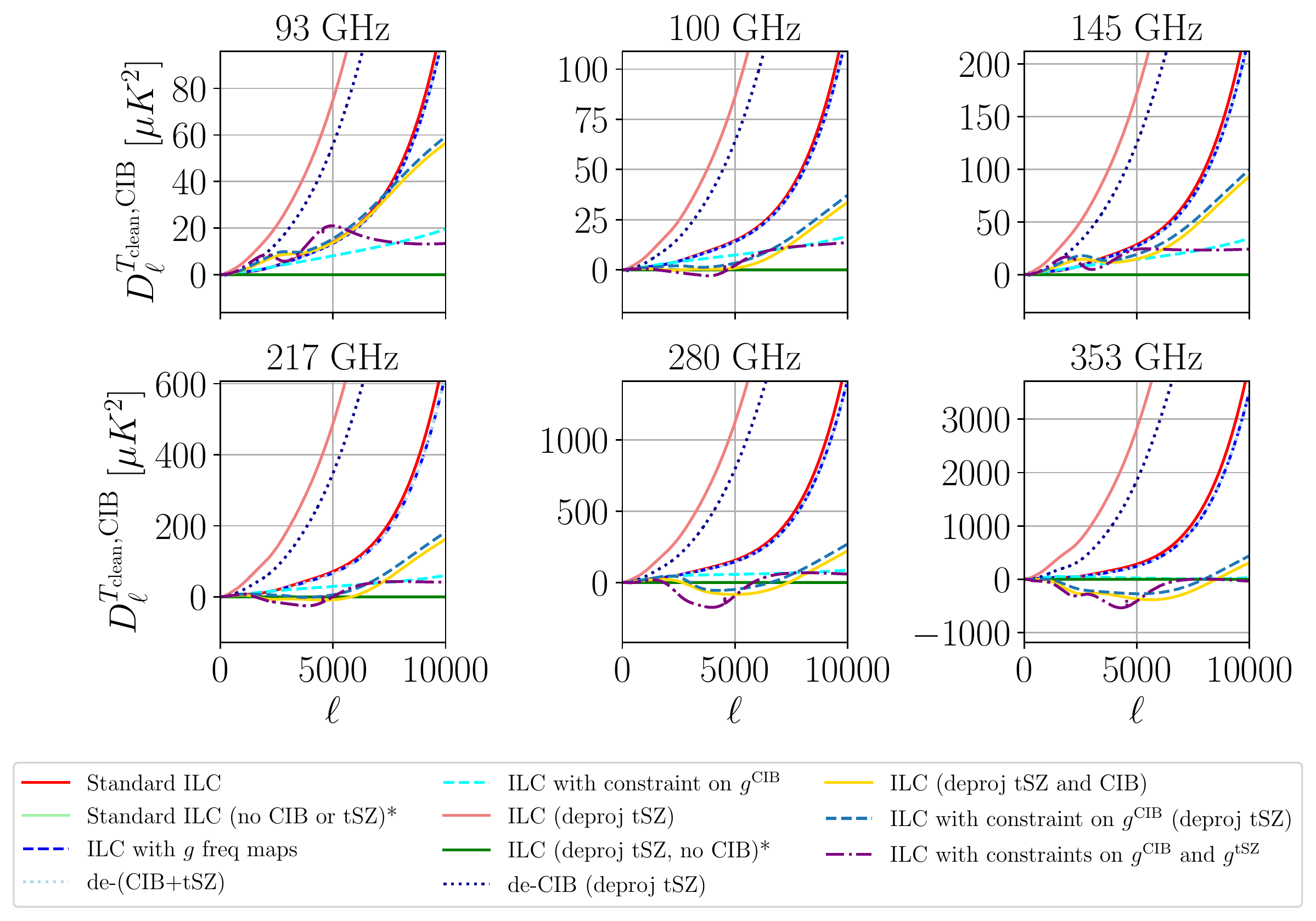}
    \caption{Cross-power spectra (in $\mu \mathrm{K}^2$) of the cleaned maps with the H13 CIB at six frequencies, for each cleaning method, plotted as $D_\ell^{T_{\rm clean}, \rm CIB} = \ell(\ell+1)C_\ell^{T_{\rm clean}, \rm CIB} / (2\pi)$. The asterisk (*) indicates an idealized method that cannot be applied to actual data and is included in this work only for comparison purposes. Our new methods (dashed and dotted curves in shades of blue and purple) decrease the cross-spectra of the cleaned maps with the CIB from those of the baseline methods (solid curves in shades of red and yellow) toward those of the idealized methods (solid curves in shades of green).  Results for 143 and 225 GHz are omitted for concision.}
    \label{fig:TcleanxCIB_spectra}
\end{figure}

Additionally, we show ratios of the auto-spectra of a few of our methods in Fig.~\ref{fig:Tclean_auto_spectra_ratios}. These ratios are discussed in detail in \S \ref{sec:results}. Finally, we show cross-spectra of the cleaned maps with the CIB in Fig.~\ref{fig:TcleanxCIB_spectra}.

\section{Results for P14 CIB Model}
\label{app:planck_results}

\begin{figure}[t]
    \centering
    \includegraphics[scale=0.55]{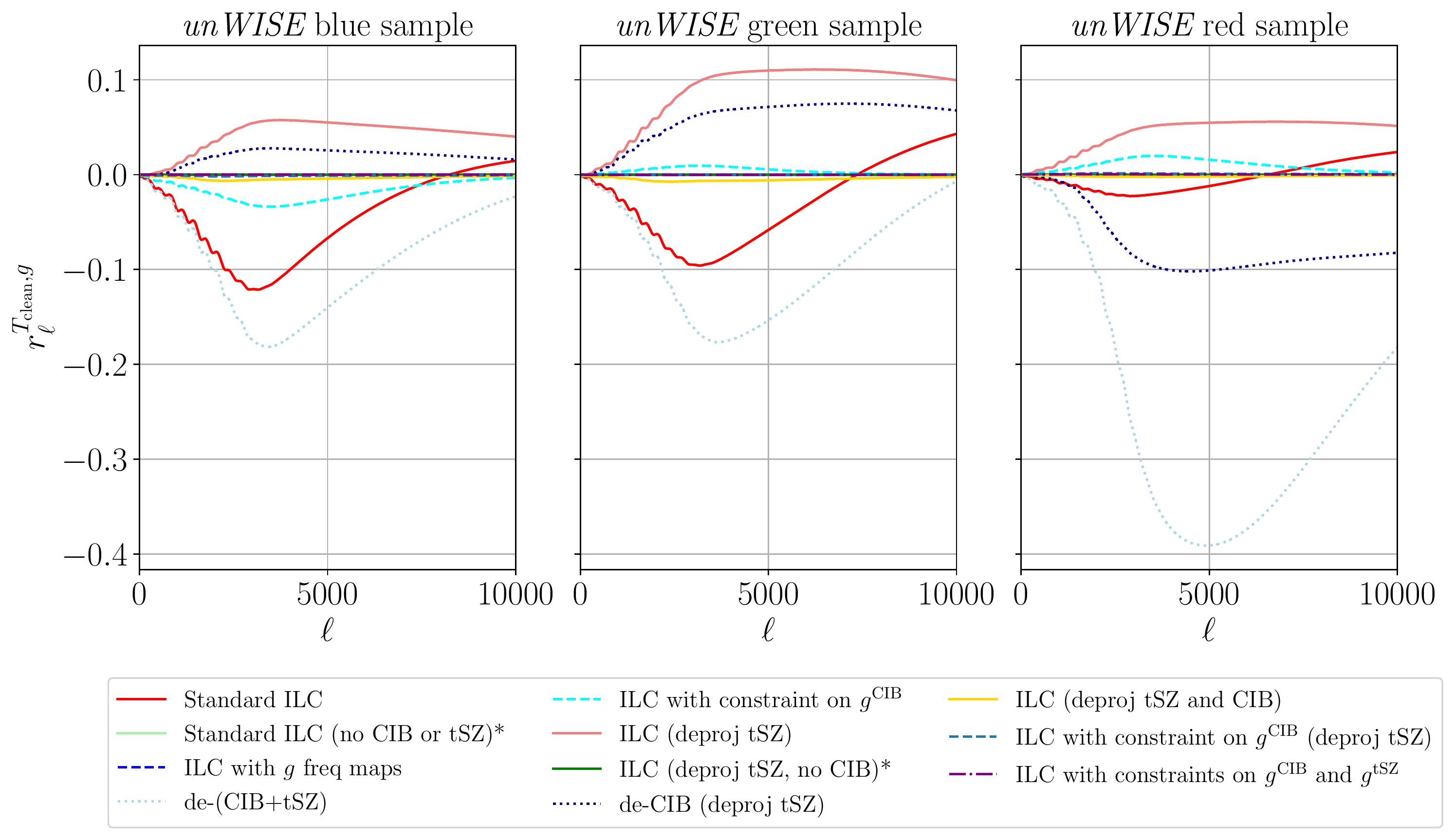}
    \caption{Correlation coefficient of the cleaned map with the \emph{unWISE} blue, green, and red samples, for each method considered in this work using the P14 CIB model. The asterisk (*) indicates an idealized method that cannot be applied to actual data and is included in this work only for comparison purposes.}
    \label{fig:Tcleanxg_planck}
\end{figure}

{In this appendix, we present the same evaluation metrics as in \S \ref{sec:results} but for input spectra generated using the P14 CIB model. Fig.~\ref{fig:Tcleanxg_planck} shows the correlation coefficient of the cleaned map from each method with the \emph{unWISE} blue, green, and red samples.}

\begin{figure}[t]
    \centering
    \includegraphics[scale=0.7]{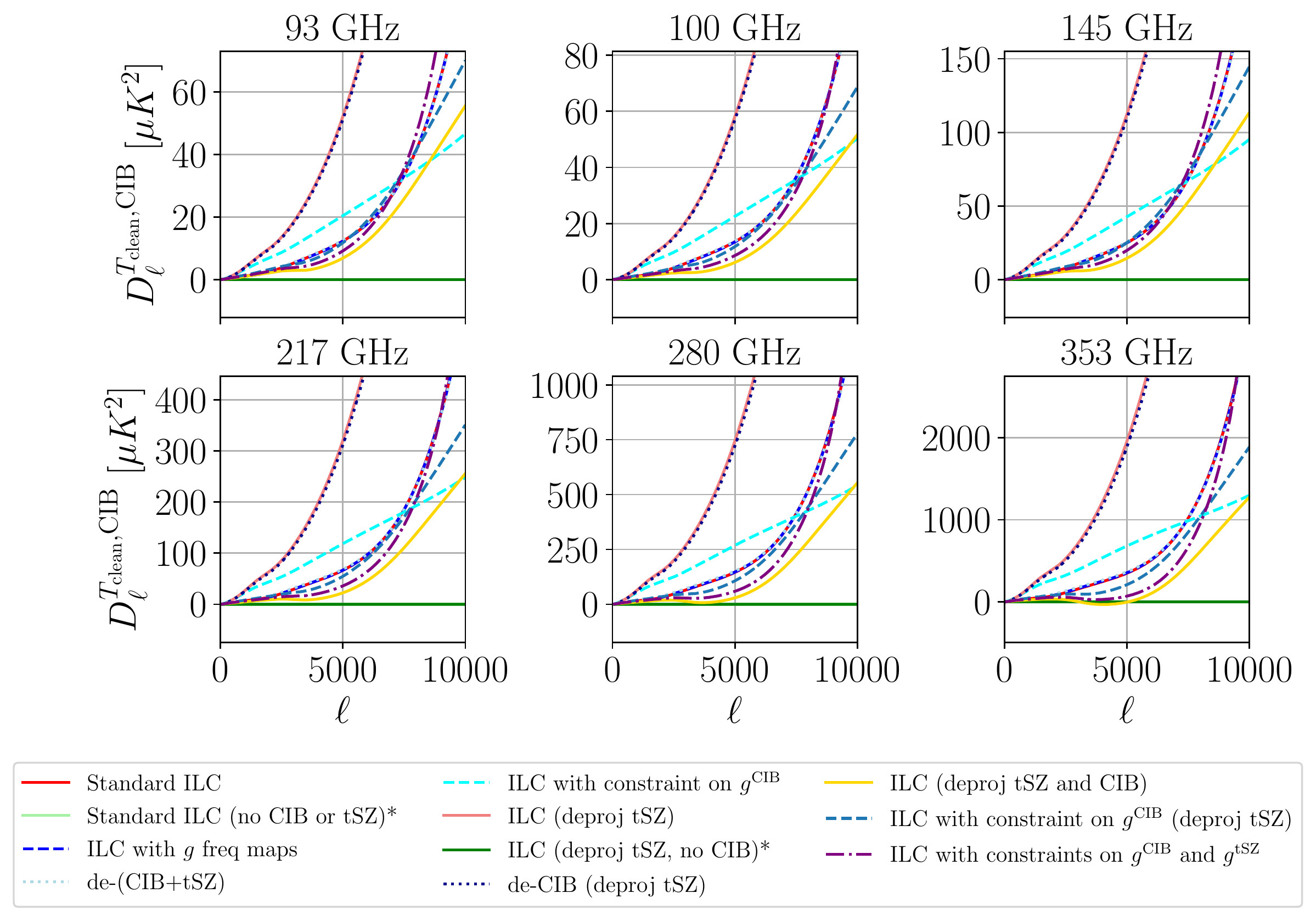}
    \caption{Cross-spectra (in $\mu \mathrm{K}^2$) of the cleaned map with the CIB at six frequencies, for each method as labeled, for the P14 CIB model. The spectra are plotted as $D_\ell^{T_{\rm clean}, \rm CIB} = \ell(\ell+1)C_\ell^{T_{\rm clean}, \rm CIB} / (2\pi)$. For clarity, we do not show results for 143 and 225 GHz, which are very similar to those for 145 and 217 GHz, respectively. The asterisk (*) indicates an idealized method that cannot be applied to actual data and is included in this work only for comparison purposes.}
    \label{fig:TcleanxCIB_spectra_planck}
\end{figure}

\begin{figure}[t]
    \centering
    \includegraphics[scale=0.7]{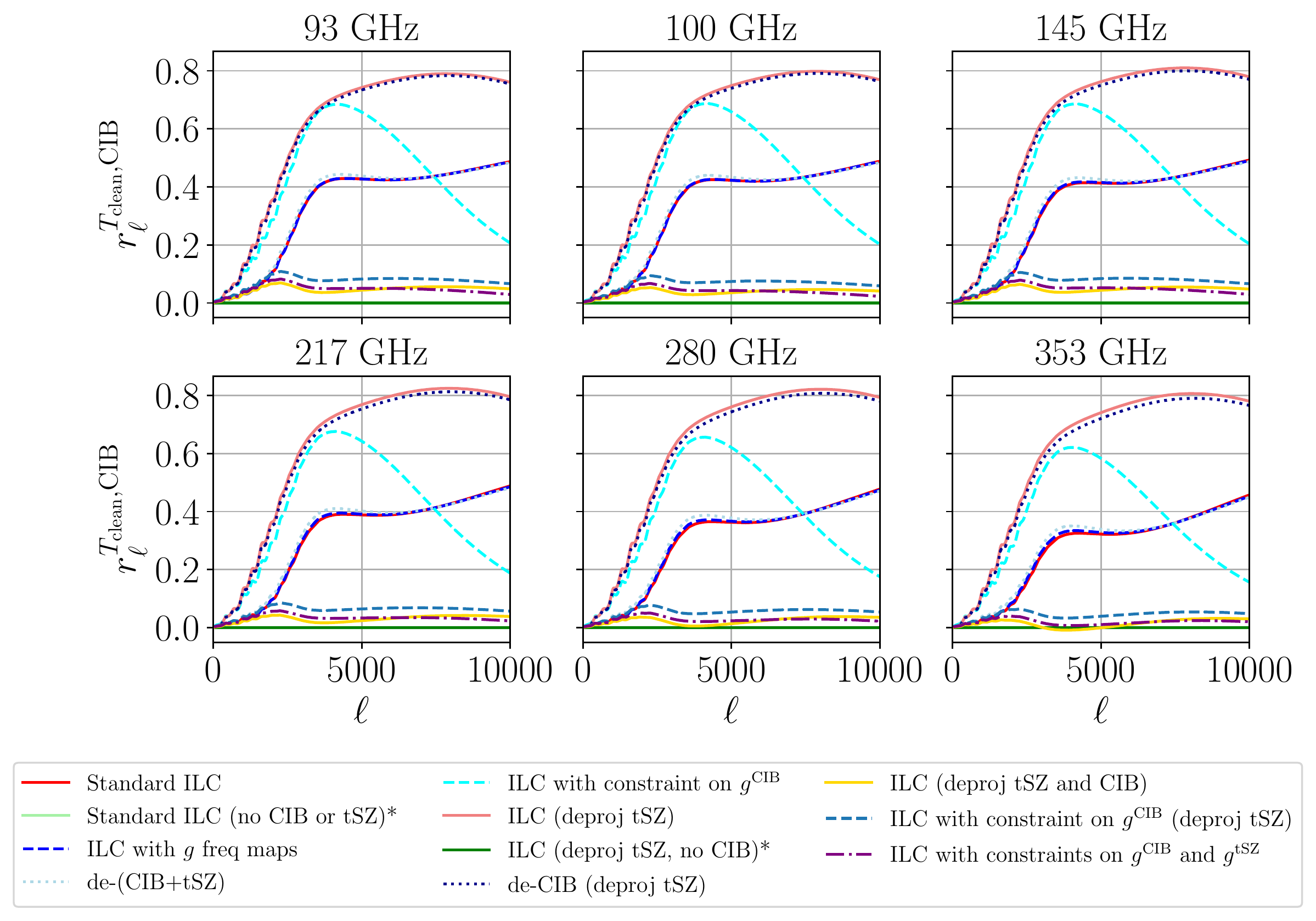}
    \caption{Correlation coefficient of the cleaned map with the CIB at six frequencies, for each method as labeled, for the P14 CIB model. For clarity, we do not show results for 143 and 225 GHz, which are very similar to those for 145 and 217 GHz, respectively. The asterisk (*) indicates an idealized method that cannot be applied to actual data and is included in this work only for comparison purposes.}
    \label{fig:TcleanxCIB_corr_planck}
\end{figure}

\begin{figure}[t]
    \centering
    \includegraphics[scale=0.72]{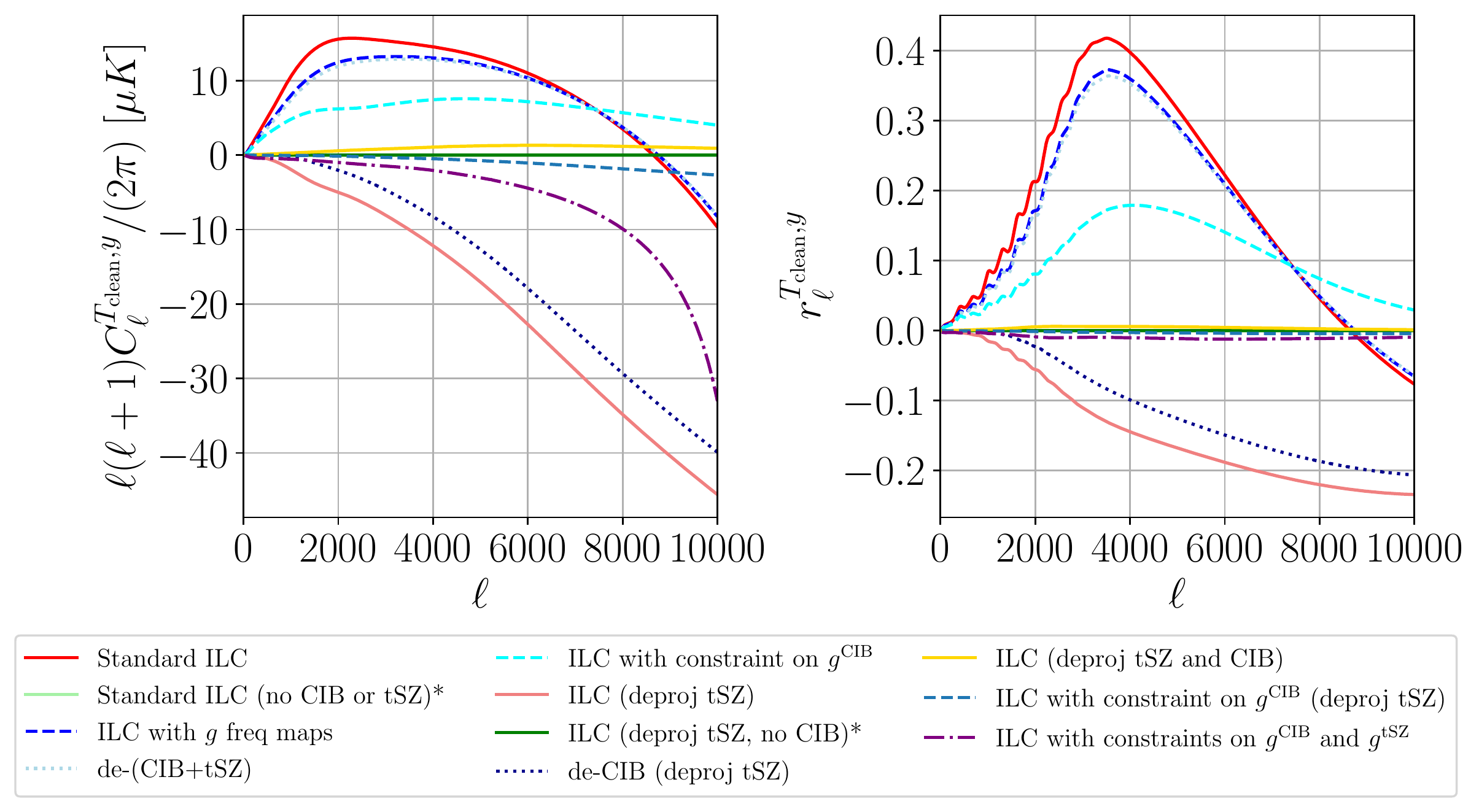}
    \caption{\textbf{Left}: Cross-spectra of the cleaned map with the tSZ field (in dimensionless Compton-$y$ units), plotted as $\ell(\ell+1)C_{\ell}^{T_{\rm clean},y}/(2\pi)$ (in $\mu \mathrm{K}$). \textbf{Right}: Correlation coefficient of the cleaned map with the tSZ field. The asterisk (*) indicates an idealized method that cannot be applied to actual data and is included in this work only for comparison purposes. Both plots assume the P14 CIB model. }
    \label{fig:TcleanxtSZ_planck}
\end{figure}

To assess how well the various methods perform in cleaning the tSZ and CIB signals, Fig.~\ref{fig:TcleanxCIB_spectra_planck} shows the cross-spectra of the cleaned map with the CIB at each frequency, and Fig.~\ref{fig:TcleanxCIB_corr_planck} shows the correlation coefficient of the cleaned map with the CIB at each frequency. Similarly, Fig.~\ref{fig:TcleanxtSZ_planck} shows the cross-spectra and correlation coefficient of the cleaned map with the tSZ signal (in dimensionless Compton-$y$ units).

\begin{figure}[t]
    \centering
    \includegraphics[scale=0.72]{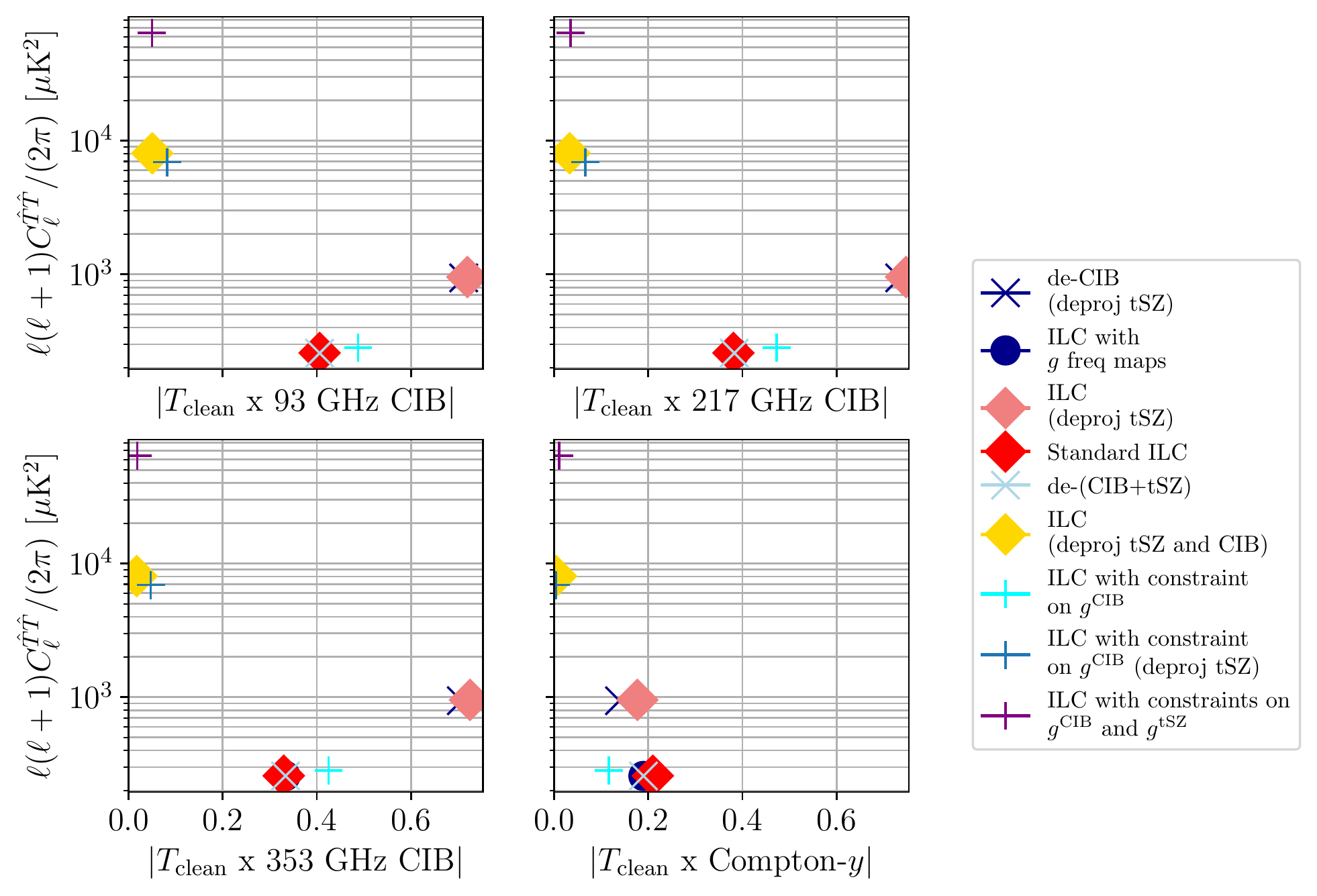}
    \caption{{Bias-variance trade-off for different methods, as assessed in terms of trade-offs among the noise penalty, CIB correlation, and tSZ correlation, shown for the P14 CIB model. The mean $D_\ell^{\hat{T}\hat{T}}$ over multipoles $2000 \leq \ell \leq 10000$ is plotted against the mean absolute $T_{\rm clean} \times \mathrm{CIB}$ or $T_{\rm clean} \times \mathrm{Compton-}y$ correlation coefficient, also computed over multipoles $2000 \leq \ell \leq 10000$. Correlation coefficients are shown for three select CIB frequencies as well as the Compton-$y$ field.}}
    \label{fig:tradeoff_planck}
\end{figure}

\section{\emph{unWISE} pixel window function treatment}
\label{app:pwf}
In this appendix, we describe the treatment of the pixel window function in the \emph{unWISE} galaxy clustering measurements used in this work to constrain the HOD. 

The \verb|NaMaster|{-obtained} \emph{unWISE} $C_\ell^{gg,\rm{data}}$ data points used in the analysis in \cite{krolewski_2020,krolewski2021cosmological,Kusiak_2022} are already divided by the pixel window function (PWF) of the $N_{\rm side}=2048$ HEALPix map, which was not accounted for in the previous HOD analysis in Ref.~\cite{Kusiak_2022} (this, however, would not have a significant effect on the results of that paper, as the considered scales are well below those where the PWF is significant). However, for our small-scale analysis, we need to include it in our modeling. The $ C_\ell^{gg,\rm{data}}$ galaxy -- galaxy data points that we have on hand and which were used in \cite{Kusiak_2022} were divided by the square of the PWF, after subtracting the fiducial, $\frac{1}{ \bar{n}_g}$ value of the shot noise (which should not be corrected by the PWF~\cite{Jing2005}), and then adding back that $\frac{1}{ \bar{n}_g}$ value again. These  $C_\ell^{gg,\rm{data}} $ data points are thus of the form 
\begin{equation}
    C_\ell^{gg,\rm{data}}  = \left( C_\ell^{gg,\rm{measured}} - \frac{1}{ \bar{n}_g } \right) P_{\ell}^{-2} + \frac{1}{ \bar{n}_g }, 
\end{equation}
where $C_\ell^{gg,\rm{measured}}$ are the power spectra measured directly from the $N_{\rm side}=2048$ \emph{unWISE} maps and $P_{\ell}$ is the pixel window function for maps of $N_{\rm side}=2048$ computed with \verb|healpy| \cite{healpy_paper1, healpy_paper2}. Given that the measured power spectra should be modeled as the sum of the shot noise contribution and the theoretical power spectra prediction multiplied by the square of the pixel window function, the $C_\ell^{gg,\rm{data}}$ that we have on hand should be written as 
\begin{equation}
    C_\ell^{gg,\rm{data}} = C_\ell^{gg,\rm{theory}} +  A_\mathrm{SN} P_{\ell}^{-2} + \frac{1}{ \bar{n}_g } \left( 1 - P_{\ell}^{-2} \right),
\end{equation}
where $C_\ell^{gg,\rm{theory}}$ are the theory predictions for the galaxy -- galaxy power spectra (including the lensing magnification contribution), here computed with \verb|class_sz|, and $A_\mathrm{SN}$ is the shot noise amplitude, which for us is a free parameter. This correction was the only change between the likelihood used in \cite{Kusiak_2022} and the one in this work used to re-fit the \emph{unWISE} $C_\ell^{gg,\rm{data}}$ data points.

\section{CIB -- Galaxy Cross-Correlation}
\label{app:plots}
\begin{figure}[t]
    \centering
    \includegraphics[scale=0.4]{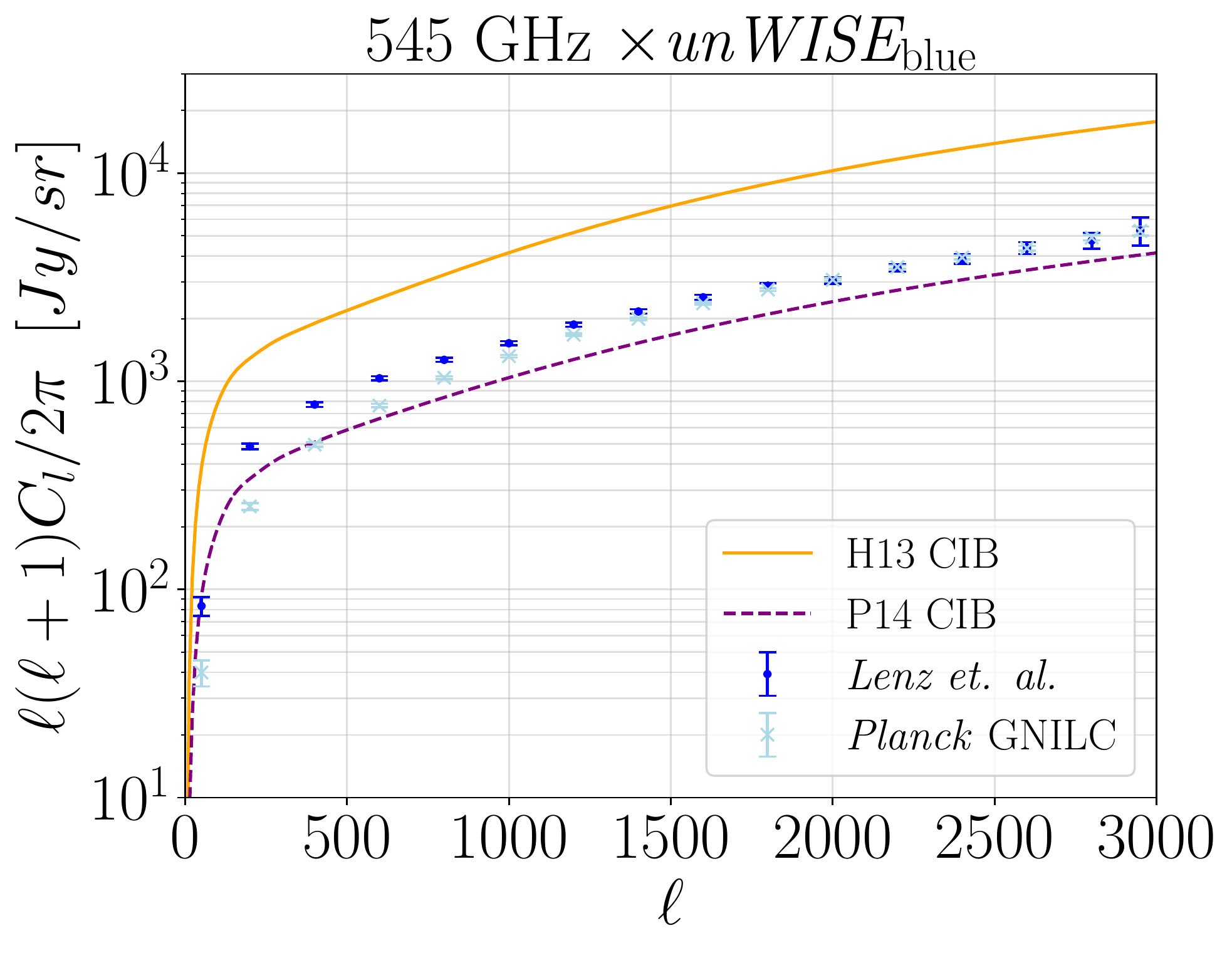}
    \includegraphics[scale=0.4]{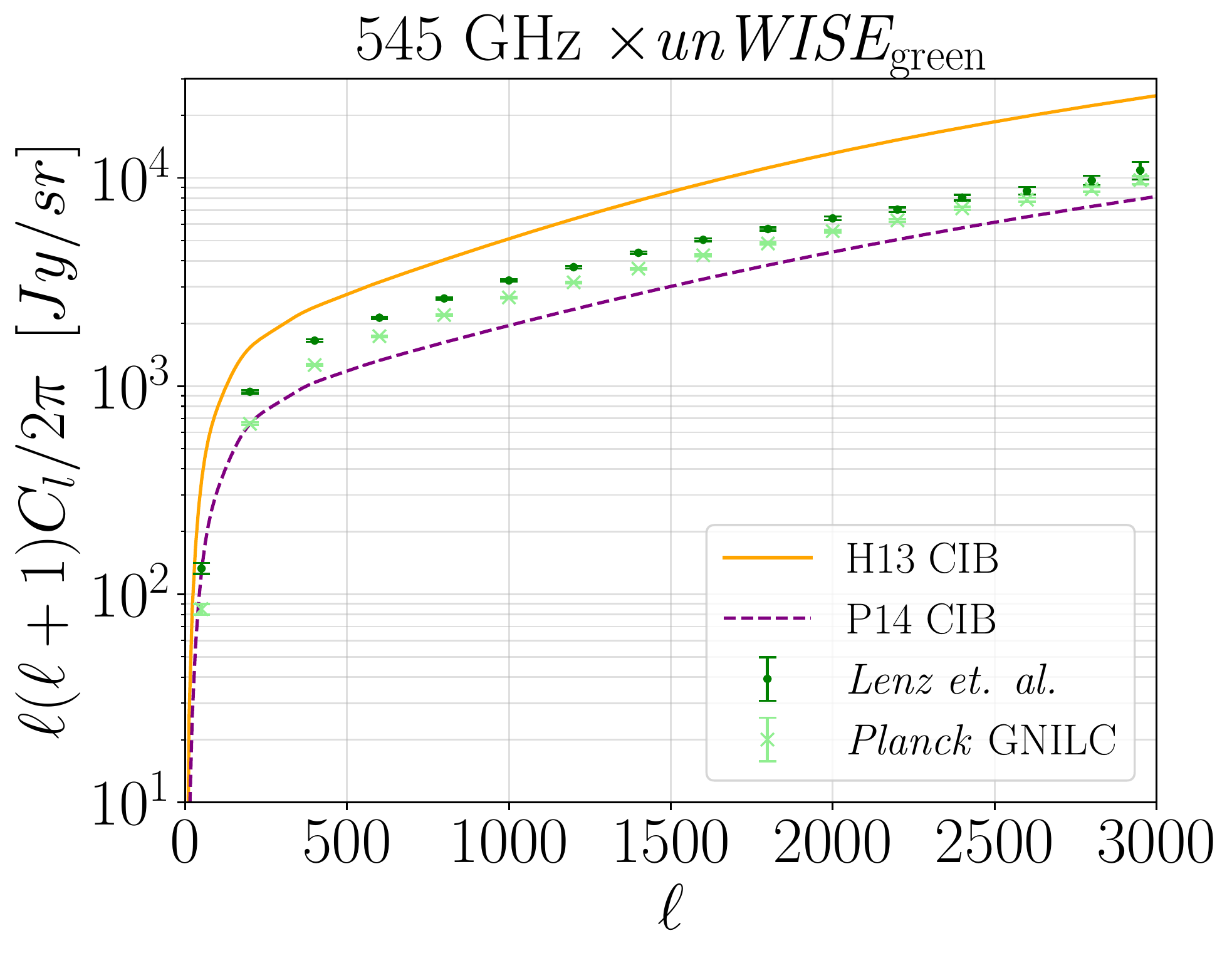}
    \includegraphics[scale=0.4]{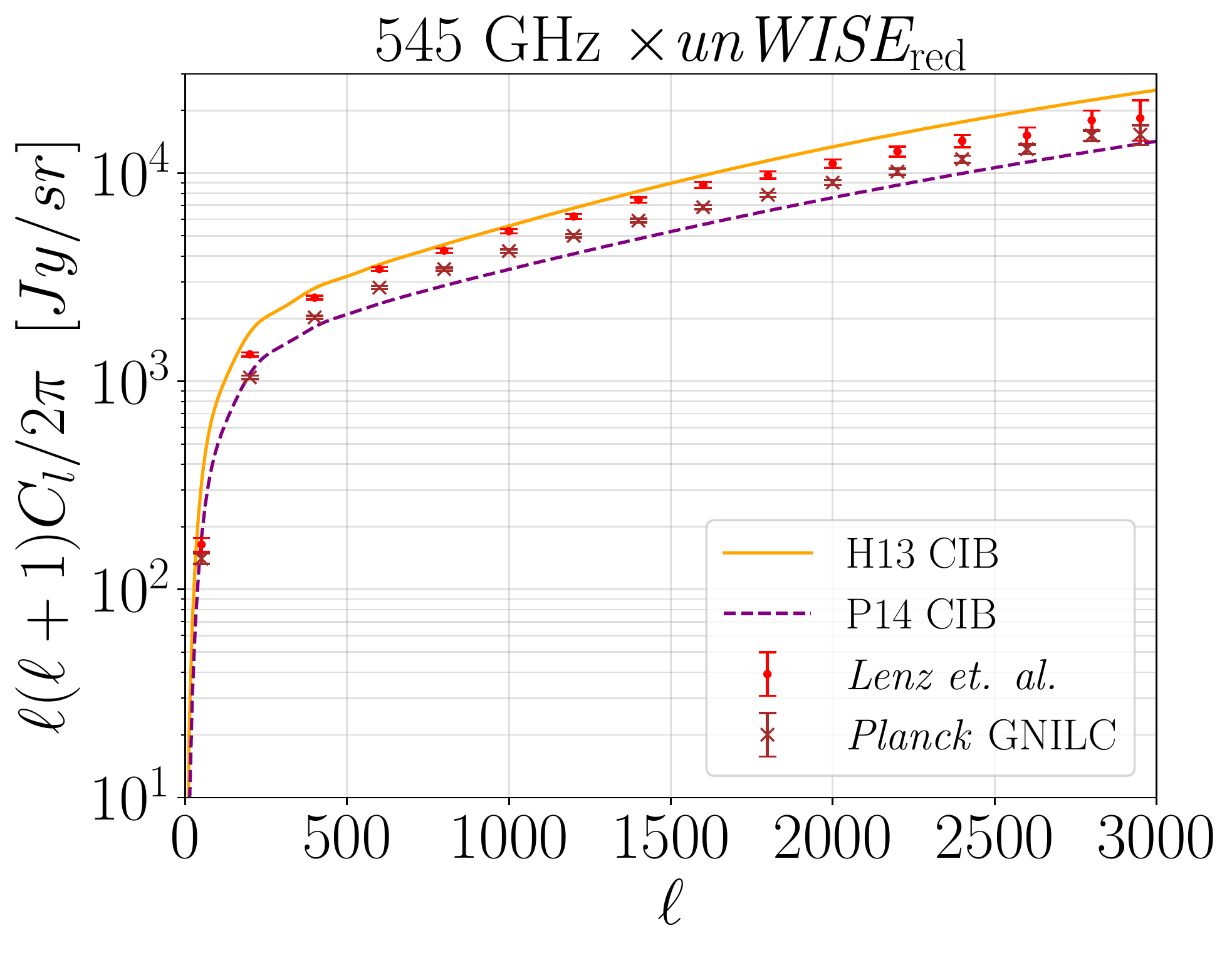}
    \caption{ Same as Fig.~\ref{fig:cibXg_data}, but for the CIB at 545 GHz. Comparison of the CIB -- galaxy cross-power spectrum measurements using the \emph{unWISE} galaxies and the CIB maps from \cite{Lenz_2019} (dots) and the \emph{Planck} GNILC CIB maps (crosses) at 545 GHz, along with the theoretical predictions for these correlations computed using the \emph{unWISE} HOD constrained in this work (Appendix~\ref{subsubsec:unwise_hod}) and the two CIB models (see Appendices~\ref{subsec:cib_planck} and \ref{subsec:cib_hermes}), P14 and H13 (solid and dashed curves, respectively). Clockwise from top left: results for the \emph{unWISE} blue, green, and red samples. The measurements are encompassed by the two CIB models, which validates our choice to consider both in our work. Note that the 545 GHz frequency channel is not used in any of the CIB-cleaning methods presented in this work.}
    \label{fig:cibXg_data545}
\end{figure}

In Fig.~\ref{fig:cibXg_data545}, we show the comparison of the measured CIB -- \emph{unWISE} cross-correlations for CIB maps from \cite{Lenz_2019} (dots) and from \emph{Planck} GNILC CIB maps (crosses) at 545 GHz with the theoretical predictions for this cross-correlation computed with the \emph{unWISE} HOD (see Appendix~\ref{subsubsec:unwise_hod}) and the two CIB models considered in this analysis, H13 (solid orange lines) and P14 (dashed purple) (this is the same as Fig.~\ref{fig:cibXg_data}, but for 545 GHz). The 545 GHz frequency channel is not included in any of the CIB removal methods presented in this paper, but we show it here as a useful check of the validity of our assumed CIB and \emph{unWISE} modeling.

\section{CIB -- CIB Correlation Coefficients}
\label{app:corr_coeff}

\begin{figure}[t]
    \centering
    \includegraphics[scale=0.6]{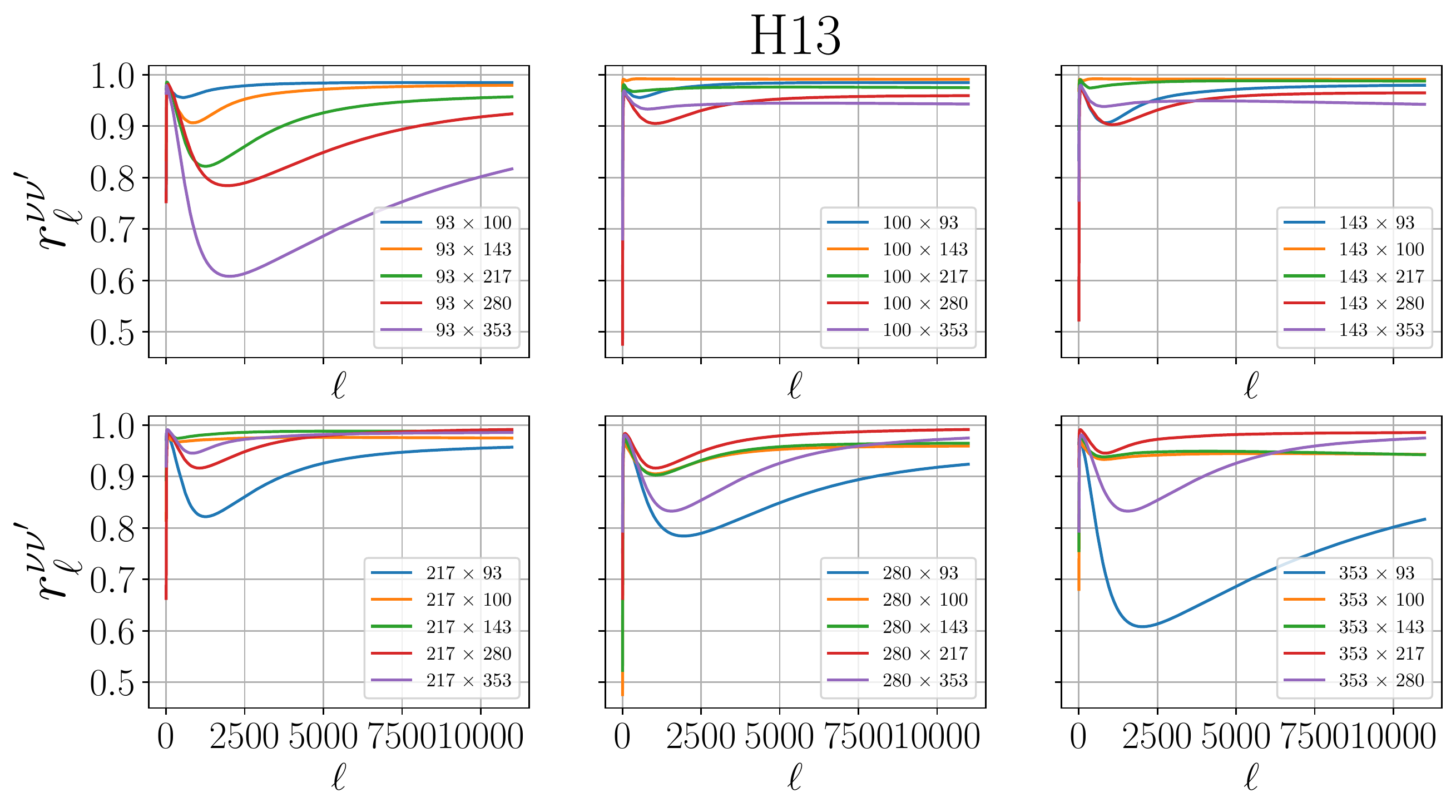}
    \includegraphics[scale=0.6]{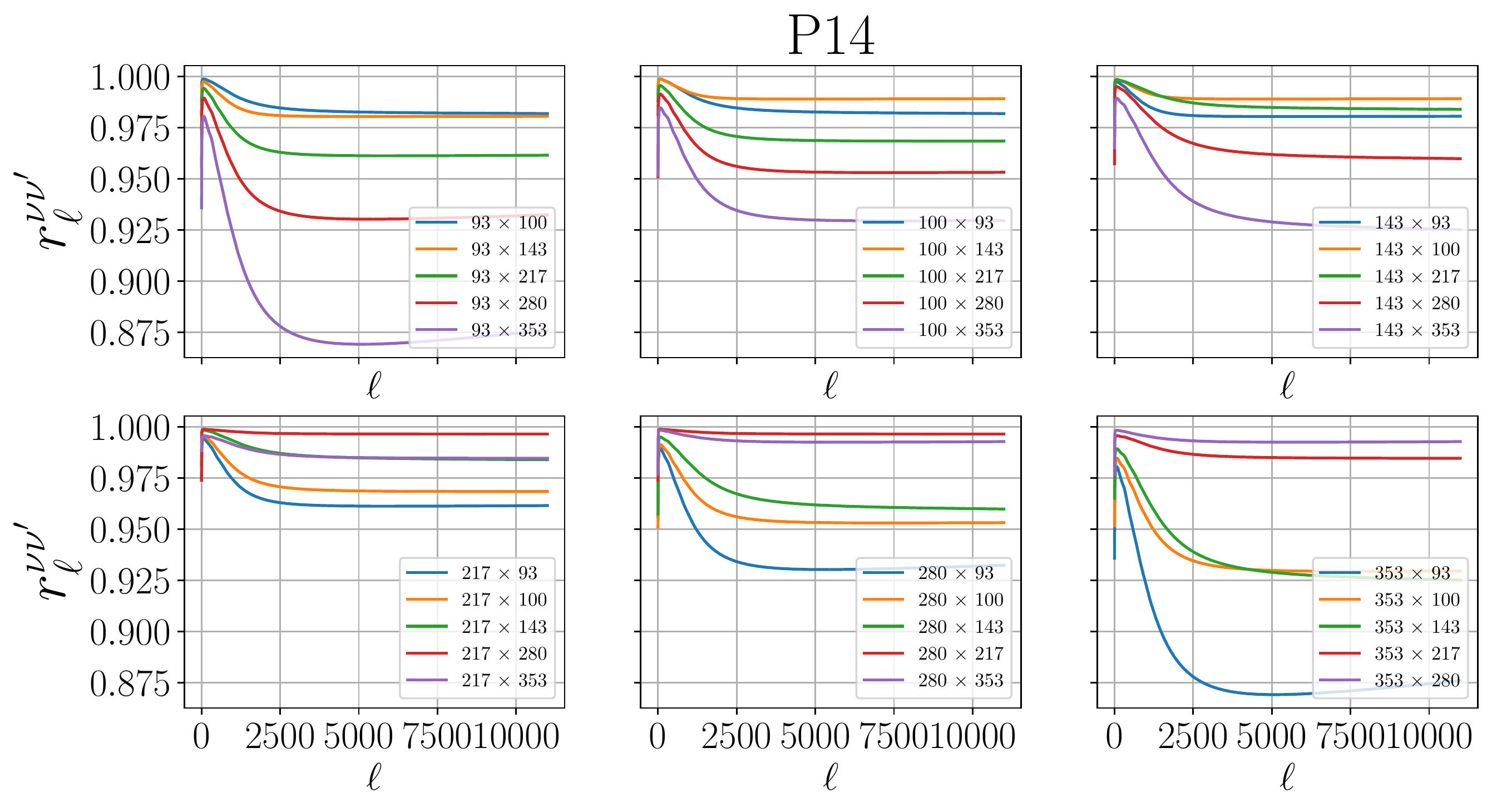}

    \caption{Correlation coefficients of the CIB field between six frequencies considered in this work (we omit plots for 145 GHz and 225 GHz) for the two CIB models, H13 (top) and P14 (bottom) (see Appendix~\ref{subsec:cib_hermes} and~\ref{subsec:cib_planck}). The H13 model is highly sensitive to the flux cut values, which are significantly lower for SO frequencies than for the \emph{Planck} frequencies (see Appendix~\ref{subsec:hm_predictions} and Table~\ref{table:fluxcut}); compare, \textit{e.g.}, 353 GHz $\times$ 100 GHz (\emph{Planck} only frequencies) and 353 GHz $\times$ 93 GHz (\emph{Planck} $\times$ SO frequency). This sensitivity to flux cut values is not as evident for the P14 model, which likely implies P14 predicts fewer bright sources. All curves are computed with the analytical halo model predictions for the auto- and cross-correlations presented in Appendix~\ref{subsec:hm_predictions} using {\tt class\_sz}, and include shot noise contributions (given in Table~\ref{table:cib_shotnoise}).  }
    \label{fig:cib_corr}
\end{figure}

In this appendix, we present the correlation coefficients of the CIB at different frequencies. In this work we include four frequency channels from \emph{Planck} (100, 143, 217, 353 GHz) and four from SO (93, 145, 225, 280 GHz). In Fig.~\ref{fig:cib_corr}, we show the CIB correlation coefficients for six of those frequencies (we omit plots for 143 GHz and 225 GHz) for the two CIB models, H13 (top) and P14 (bottom).

As expected~\cite{Planck:2013cib,Mak2017,Lenz_2019}, the correlation coefficients between the CIB field at different frequencies are very high, often close to one, especially for neighboring frequencies.  These high correlation coefficients allow us to approximate the CIB as a component with a well-defined SED, which underlies some of our cleaning methods (see Appendix~\ref{app:mbb}). However, from Fig.~\ref{fig:cib_corr}, we can also conclude that the H13 CIB model is more sensitive to the flux cut values, which are significantly lower for SO frequencies than for \emph{Planck} frequencies (see Appendix~\ref{subsec:hm_predictions} and Table~\ref{table:fluxcut} for discussion and flux cut values) --- compare, \textit{e.g.}, 353 GHz $\times$ 100 GHz (\emph{Planck} only frequencies) and 353 GHz $\times$ 93 GHz (\emph{Planck} $\times$ SO frequency).  The P14 model (bottom plot) is not as sensitive to flux cut values, which likely implies P14 predicts fewer bright sources than H13.

\section{CIB SED as a Modified Blackbody}
\label{app:mbb}

In this appendix, we estimate the true SED of the two CIB models presented in this work, H13 and P14. 

First, we give the conversion from specific intensity units, which are used for the CIB emission in \verb|class_sz|, to CMB thermodynamic temperature units, in $\mathrm{K}_\mathrm{CMB}$ or $\mu \mathrm{K}_\mathrm{CMB}$. To convert from specific intensity  units to thermodynamic temperature units, we take a derivative of the Planck formula with respect to temperature:
\begin{equation}
    d B_{\nu}(T) = \frac{2h^2 \nu^4 e^x}{c^2 k_{\rm B} T^2(e^x -1)^2} dT \,,
    \label{eq:planck_formula}
\end{equation}
where 
\begin{equation}
    x = \frac{h_{\rm{P}}\nu}{k_{\rm{B}}T_{\rm{CMB}}} \,.
\end{equation}
In Table~\ref{table:units}, we list the exact conversion factors from [Jy/sr] to [$\mu$K] for the frequencies considered in this work. 

\begin{table}[ht]
\setlength{\tabcolsep}{10pt}
\renewcommand{\arraystretch}{1.0}
    \begin{tabular}{|c|c|}
        \hline
         Frequency & Conversion Factor\\
         $\left[ \mathrm{GHz} \right]$ &  [Jy/sr $\mu$K$^{-1}$]  \\
         \hline 
         \hline
          93 & 213.55 \\
         \hline
          100 & 238.81\\
          \hline
         143 & 380.00\\
         \hline
        145& 385.47\\
         \hline
         217 & 483.82\\
         \hline
         225  & 482.90\\
         \hline
          280& 429.38\\
          \hline
          353 & 296.88\\
          \hline
       
    \end{tabular}
    \caption{Conversion factors from [Jy/sr] to [$\mu$K] for the frequencies considered in this work, computed using Eq.~\eqref{eq:planck_formula}.}
    \label{table:units}
\end{table}

Second, we want to approximate the CIB SED as a modified blackbody, to use in constrained ILC calculations. Following Ref.~\cite{Madhavacheril_2020}, 
the CIB SED as a modified blackbody can be written as 
\begin{equation}
    f_{\rm{CIB}} (\nu) \propto \frac{\nu^{3+\beta_{\rm{CIB}}}}{e^{h_{\rm{P}}\nu / (k_{\rm{B}}T_{\rm{CIB}})} -1} \left ( \left.\frac{ d B_{\nu}(T)}{d T} \right\vert_{T=T_{\mathrm{CMB}}} \right ) ^{-1} ,
    \label{eq:mbb}
\end{equation}
where $\beta_{\mathrm{CIB}}$, the spectral index, and $T_{\mathrm{CIB}}$, the effective dust temperature, are free parameters.  As in Ref.~\cite{Madhavacheril_2020}, we take $\beta_{\mathrm{CIB}}=1.2$ and $T_{\mathrm{CIB}}=24$ K (note that these do not correspond to modified blackbody parameters describing any individual CIB source, but rather an approximation of the SED of the full CIB field). Then the full CIB power spectrum at some frequency can be approximated by\footnote{This assumes that the decorrelation of CIB across frequencies is very small.} 
\begin{equation}
    \mathrm{CIB}(\nu, \hat{n}) = f_{\rm{CIB}} (\nu) a^{\mathrm{CIB}}(\hat{n}),
\end{equation}
where $a^{\mathrm{CIB}}(\hat{n})$ is to the CIB what the Compton-$y$ field is to the tSZ field, allowing the CIB to be approximately deprojected in the ILC procedure similar to the tSZ field. 

\begin{figure}[t]
    \centering
    \includegraphics[scale=0.4]{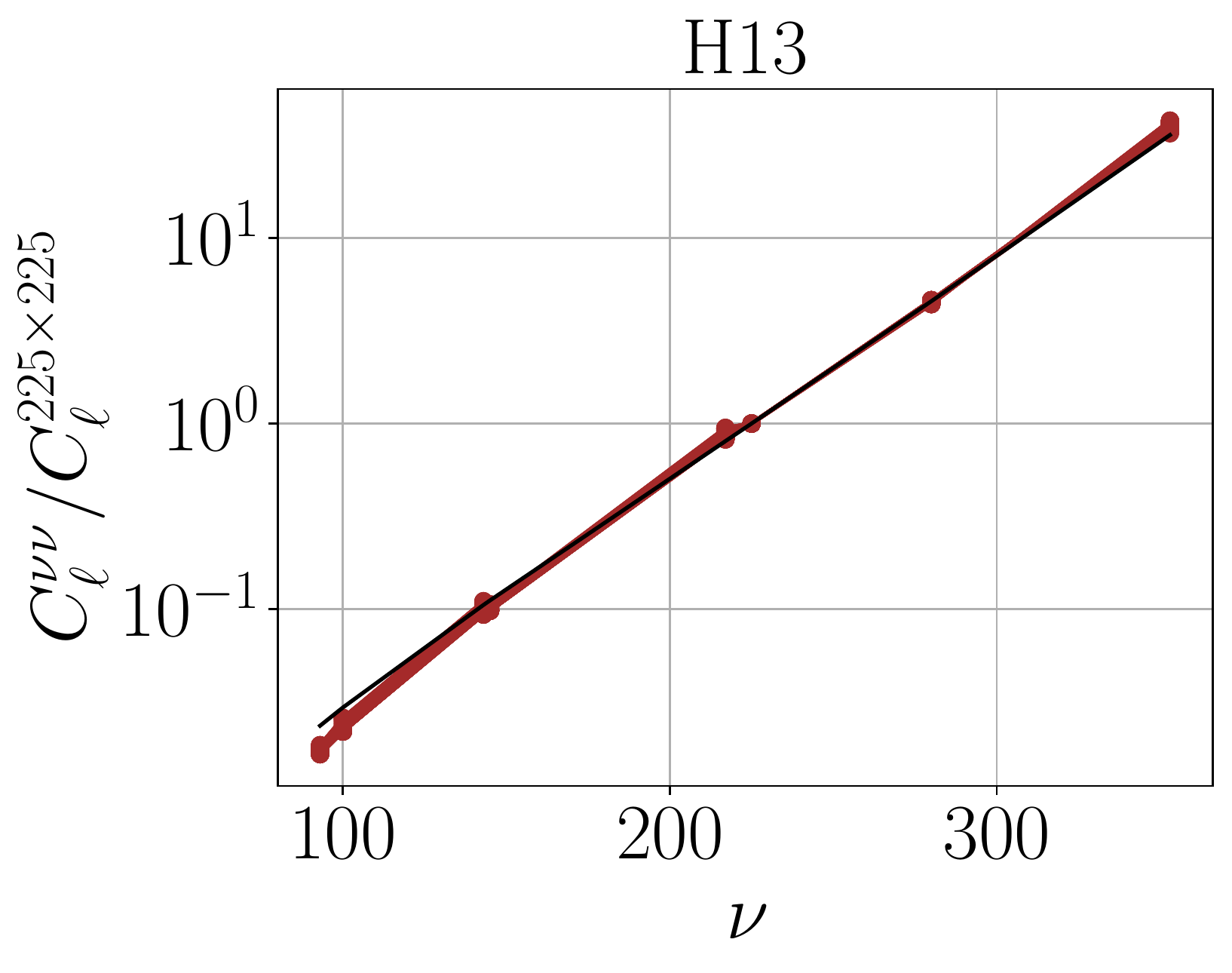}
    \includegraphics[scale=0.4]{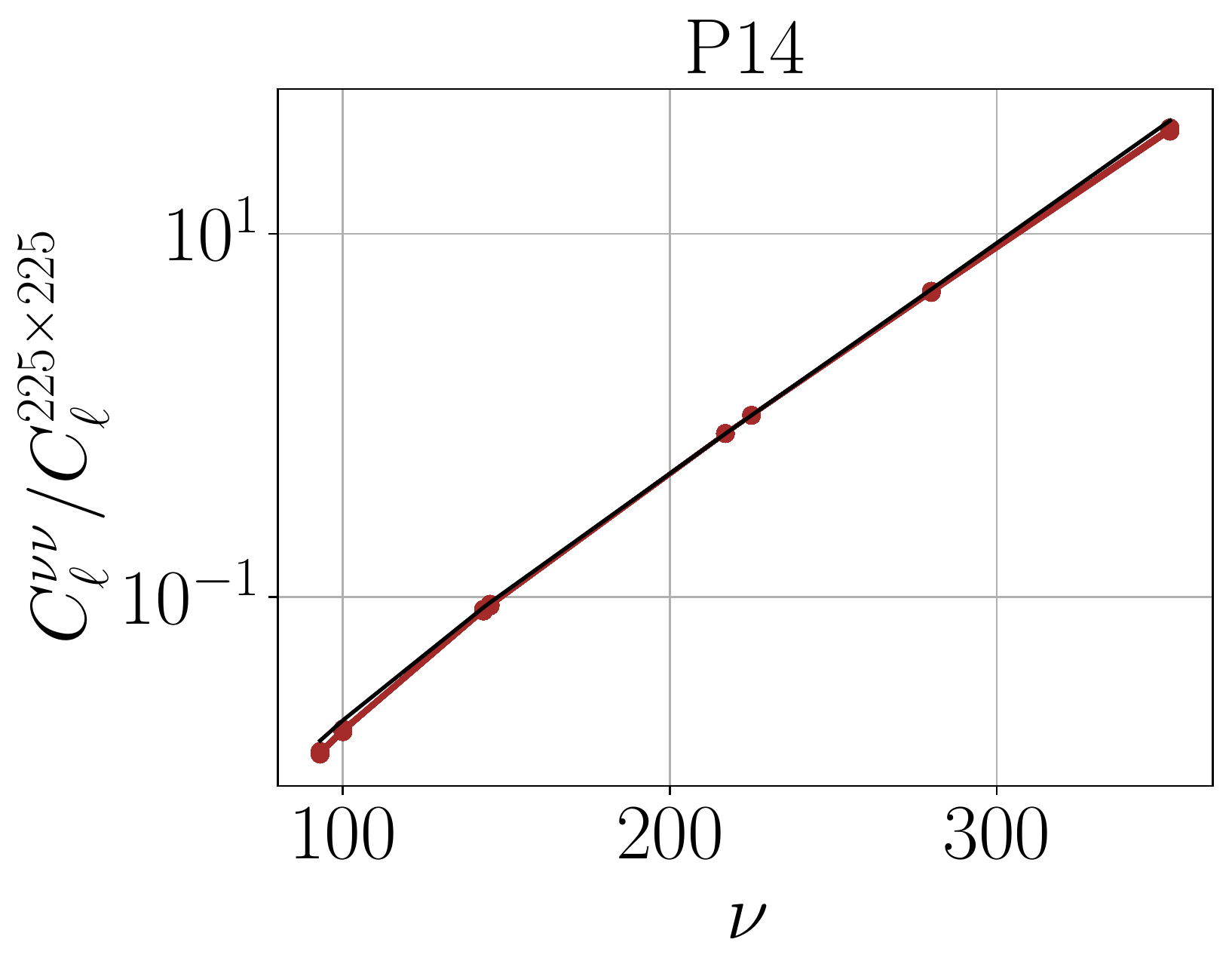}
    \caption{ Modified blackbody SED estimation for the two CIB models presented in this work, H13 (left) and P14 (right) (see Appendix~\ref{subsec:cib_hermes} and \ref{subsec:cib_planck}). We plot CIB auto-power spectra normalized at 225 GHz  $C_{\ell}^{\nu\nu} $/ $C_{\ell}^{225 \times 225}$ versus frequency $\nu$ for the eight frequency channels used in this work, for various $\ell$ values (brown curves, with dots), and compare it with a modified blackbody SED (also normalized at 225 GHz) computed following Eq.~\eqref{eq:mbb} with the choice of parameters discussed in Appendix \ref{app:mbb} for the two models (black solid curve).  The points for different $\ell$ values are nearly coincident, illustrating that the SED is not strongly scale-dependent in these models.}
    \label{fig:mbb}
\end{figure}

To estimate the CIB SED, in Fig.~\ref{fig:mbb} we plot a normalized CIB auto-power spectrum versus frequency $\nu$ for the eight frequency channels considered in this work for various $\ell$ values (brown curves with dots), and compare it with the modified blackbody used in the CIB deprojections computed following Eq.~\eqref{eq:mbb} (black solid curve). We normalize all the auto-power spectra at 225 GHz, that is,
$C_{\ell}^{\nu\nu} $/ $C_{\ell}^{225 \times 225}$.  For the H13 CIB model, the assumed parameter values of the modified blackbody SED from Eq.~\eqref{eq:mbb} ($\beta_{\mathrm{CIB}}=1.2$ and $T_{\mathrm{CIB}}=24$ K) agree quite well with the model predictions, and we use those values in the CIB deprojections for the H13 model in the ILC.  For the P14 model, there is a significant offset between the model predictions and the assumed parameter values of the modified blackbody SED from Eq.~\eqref{eq:mbb}; therefore we re-fit those parameter values by hand and obtain $\beta_{\mathrm{CIB}}=1.45$ and $T_{\mathrm{CIB}}=20$ K. We proceed with those values for the ILC CIB deprojections for the P14 model.

\end{appendices}
\bibliographystyle{apsrev}
\bibliography{unWISE}
\end{document}